\documentclass[12pt]{article}
\usepackage[scale={0.75,0.8}]{geometry}

\pdfoutput=1
\usepackage[utf8]{inputenc}
\usepackage{graphicx}
\usepackage{subfig}
\usepackage{hyperref}
\usepackage{amsmath}
\usepackage{color}
\usepackage{epstopdf} 
\usepackage{float}
\usepackage{upgreek}
\usepackage{amssymb}
\usepackage{pdflscape}
\usepackage[nice]{nicefrac}
\usepackage[dvipsnames]{xcolor}
\usepackage[figuresleft]{rotating}
\usepackage[numbers, sort&compress]{natbib}
\usepackage{makecell}
\usepackage[normalem]{ulem}
\usepackage{comment}

\newcommand{\hh}{\kappa}
\newcommand{\hchi}{h}

\newcommand{\rhoend}{\rho_{\rm end}}
\newcommand{\rhoreh}{\rho_{\rm re}}

\newcommand{\wrehbar}{\bar{w}_{\rm re}} 
\newcommand{\wreh}{w_{\rm re}}
\newcommand{\Nreh}{N_{\rm re}}
\newcommand{\X}{\mathcal{X}}

\newcommand{\upmutobealpha}{\alpha}
\newcommand{\upalphatosigma}{\sigma}
\newcommand{\Vend}{\mathcal{V}_{\rm end}}

\newcommand{\V}{\mathcal{V}}

\newcommand{\Treh}{T_{\rm re}}
\newcommand{\x}{x}
\newcommand{\n}{ j}
\newcommand{\DD}{D}
\newcommand{\corr}[1]{\textcolor{black}{#1}}

\newcommand{\GG}{\Gamma}
\newcommand{\Nk}{N_{\rm k}}
\newcommand{\SM}{\mathsf{a}}
\newcommand{\vv}{\mathsf{v}}
\newcommand{\g}{\mathsf{g}}
\renewcommand{\x}{x}
\newcommand{\sigmar}{\sigma_r}
\newcommand{\sigmans}{\sigma_{n_s}}
\newcommand{\nsbar}{\bar{n}_s}
\newcommand{\rbar}{\bar{r}}
\newcommand{\xx}{a}
\newcommand{\yy}{b}
\newcommand{\khor}{k_{\rm hor}}
\newcommand{\Nhor}{N_{\rm hor}}
\newcommand{\Ninf}{N_{\rm inf}}

\newcommand{\logy}{\x}
\newcommand{\logY}{\logy}

\newcommand{\M}{M}

\begin{document}

\title{ 
\begin{flushright}
{\scriptsize 
}
\end{flushright}
{\bf \boldmath 
LiteBIRD and CMB-S4 Sensitivities to Reheating in Plateau Models of Inflation 
} \\[8mm]}

\author{
Marco Drewes$^{a}$\footnote{marco.drewes@uclouvain.be}, Lei Ming$^{b,a}$\footnote{minglei@mail.sysu.edu.cn}, Isabel Oldengott$^{a}$\footnote{isabel.oldengott@uclouvain.be} \\
\\
{\normalsize \it $^a$Centre for Cosmology, Particle Physics and Phenomenology,}\\ {\normalsize \it Universit\'{e} catholique de Louvain, Louvain-la-Neuve B-1348, Belgium}\\ 
{\normalsize \it $^b$School of Physics, Sun Yat-Sen University,
Guangzhou 510275, China}
\vspace{-0.5cm}
}
\date{}

\maketitle

\begin{abstract}
    \noindent We study the sensitivity of LiteBIRD and CMB-S4 to the reheating temperature and the inflaton coupling in three types of plateau-potential models of inflation, namely mutated hilltop inflation, radion gauge inflation, and $\alpha$-attractor T models. We first find relations between model parameters and CMB observables in all models. We then perform Monte Carlo Markov Chain based forecasts to quantify the information gain on the reheating temperature, the inflaton coupling, and the scale of inflation that can be achieved with LiteBIRD and CMB-S4, assuming a fiducial tensor-to-scalar ratio $\bar{r} \sim 0.02$ and neglecting foreground contamination of the B-mode polarization spectrum. We compare the results of the forecasts to those obtained from a  recently proposed simple analytic method. We find that both LiteBIRD and CMB-S4 can simultaneously constrain the scale of inflation and the reheating temperature in all three types of models. They can for the first time obtain both an upper and lower bound on the latter, comprising the first ever measurement of the big bang temperature. In the mutated hilltop inflation and radion gauge inflation models this can be translated into a measurement of the inflaton coupling in parts of the parameter space. Constraining this microphysical parameter will help to understand how these models of inflation may be embedded into a more fundamental theory of particle physics.   
\end{abstract}

\newpage

\tableofcontents

\newpage
\section{Introduction}

\emph{Cosmic inflation} \cite{Starobinsky:1979ty,Starobinsky:1980te,Guth:1980zm,Linde:1981mu} continues to be the most popular explanation for the overall homogeneity and isotropy of the observable universe. It can also explain the properties of the small perturbations that are visible in the Cosmic Microwave Background (CMB), and which formed the seeds for galaxy formation. 
However, very little is known about the mechanism that may have driven the accelerated expansion. A wide range of theoretical models of inflation exist, but the observational data is not sufficient to clearly single out any of them. Moreover, even less is known about how a given model of inflation should be embedded into a more fundamental microphysical theory of nature. Understanding this connection would be highly desirable from the viewpoints of both cosmology and particle physics. 

It has been known for a long time that CMB anisotropies carry information about the temperature of the primordial plasma  
at the onset of the radiation dominated epoch $\Treh$    
\cite{Martin:2010kz,Adshead:2010mc,Easther:2011yq}, known as the \emph{reheating temperature}.
Since the reheating process that determined $\Treh$ is ultimately driven by fundamental interactions, this also implies a sensitivity to microphysical parameters that connect inflation to particle physics \cite{Drewes:2015coa}. 
However, current observational data is not sufficient to measure $\Treh$ or microphysical parameters \cite{Drewes:2022nhu}.
 In the next decade the observational situation will change drastically \cite{Martin:2014rqa}. Upgrades at the South Pole Observatory \cite{Moncelsi:2020ppj}
and the Simons Observatory \cite{SimonsObservatory:2018koc} aim at pushing the uncertainty in the scalar-to-tensor ratio $r$ down to $\sigma_r \sim 3\times 10^{-3}$. 
In the 2030s JAXA's LiteBIRD satellite \cite{Sugai:2020pjw} and the ground-based  CMB Stage 4 (CMB-S4) program \cite{CMB-S4:2020lpa} can further reduce this to $\sigma_r \sim  1 \times 10^{-3}$ and $\sigma_r \sim  0.5 \times 10^{-3}$, respectively.
It has been estimated in \cite{Drewes:2019rxn} that this accuracy should be sufficient to for the first time measure\footnote{Here and in the following we use the term \emph{measurement} if the one-$\sigma$ interval of the posterior is smaller than the range of values that a given parameter can take, i.e., data imposes both an upper and a lower limit on that parameter.} 
$\Treh$ as well as 
individual 
microphysical
coupling constants in models with a plateau potential, a class of models that is currently favoured by observations.
In \cite{Drewes:2022nhu} a simple semi-analytic Bayesian approach was used to confirm this.  
This calls for a more detailed evaluation of the perspectives to determine the inflaton coupling from CMB data in different models of inflation.

 In the present work we investigate the possibility to obtain information on the reheating epoch from future CMB observations with a combination of analytic methods and Monte Carlo Markov Chain (MCMC) based forecasts. 
We focus on three types of 
models in which inflation is effectively driven by a single scalar field $\Phi$ -- 
namely mutated hilltop inflation (MHI) \cite{Pal:2009sd,Pal:2017bmd}, 
 radion gauge inflation (RGI) model~\cite{Fairbairn:2003yx,Martin:2013nzq} and $\alpha$-attractor T-models ($\alpha$-T)~\cite{Kallosh:2013maa,Carrasco:2015pla,Carrasco:2015rva} 
-- and assess the capacity
of future CMB observations to constrain the scale of inflation, the reheating temperature, and the  
couplings that connect $\Phi$ to other fields in an agnostic way. Here \emph{agnostic} means that one can obtain a constraint on the inflaton coupling constants 
within a given model of inflation
without having to specify details of the complete particle physics theory in which this model is embedded.
The conditions under which this is possible have been laid out in \cite{Drewes:2019rxn}. 
We perform MCMC based forecasts using 
{\sc MontePython}
\cite{Audren:2012wb,Brinckmann:2018cvx} and compare them to the semi-analytic approach pursued in \cite{Drewes:2022nhu}. We thereby assume a fiducial tensor-to scalar ratio $\Bar{r}\sim 0.02$ and neglect contamination of B-mode polarization due to foregrounds.
We also include the observational uncertainty in the amplitude of spectral perturbations $A_s$, which had been fixed to its best fit value in \cite{Drewes:2022nhu}. 
In addition to studying $\g$, we also present constraints on the reheating temperature $\Treh$ in all three models.\footnote{
Here $\Treh$ refers to the temperature at the onset of the radiation dominated epoch, which is in general not the highest temperature of the radiation in the history of the universe \cite{Giudice:2000ex}. 
}  
To the best of our knowledge, these are the first forecasts for the sensitivities of LiteBIRD and CMB-S4 to the reheating $\Treh$ and $\g$  in these classes of models.

This article is organised as follows. 
In Sec.~\ref{Sec:OlleKamellen} we review the general formalism that we use to constrain $\Treh$ and the inflaton coupling.
In Sec.~\ref{sec:Benchmark models and analytic estimates} we introduce the three classes of benchmark models that we consider and analytically study the relations between the model parameters and observables. 
In Sec.~\ref{Sec:Methods} we outline two methods for estimating the LiteBIRD  and CMB-S4 sensitivities to the reheating temperature and the inflaton coupling. 
In Sec.~\ref{ChorPasseMole} we present the results obtained with these methods and discuss them.
In Sec.~\ref{Sec:Conclusion} we conclude.
In the appendices we comment on the choice of the pivot scale (appendix \ref{Sec:PivotScale}), justify the neglect of next-to-leading order corrections in the slow-roll parameters throughout this work (appendix~\ref{app:Next-to-leading}), review the conditions under which the inflaton coupling can be constrained from CMB data without having to specify all details of the underlying particle physics models (appendix \ref{Sec:FeedbaclEffects}), comment on the difficulty to fulfil those conditions in $E$-type $\alpha$-attractor models (appendix \ref{JinURevisited}), and present an extended set of plots to illustrate the relations between various parameters and observables in the models we consider (appendix \ref{Sec:AppendixWithAllPlots}).

\begin{figure}
\centering
\subfloat[]{
\includegraphics[width=0.49\linewidth]{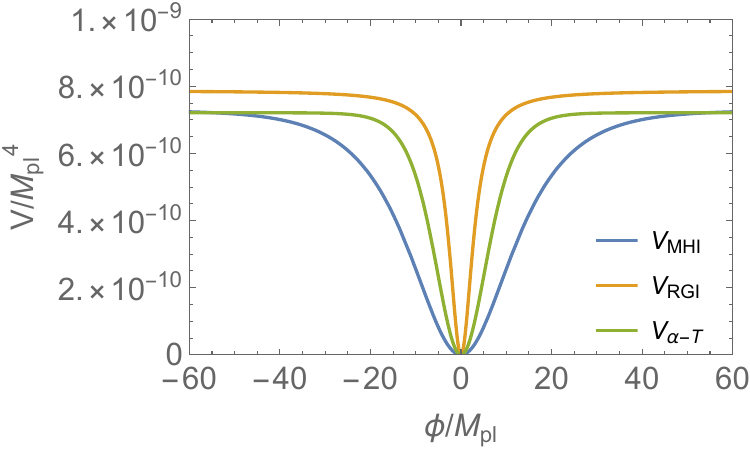}}
\subfloat[]{
\includegraphics[width=0.49\linewidth]{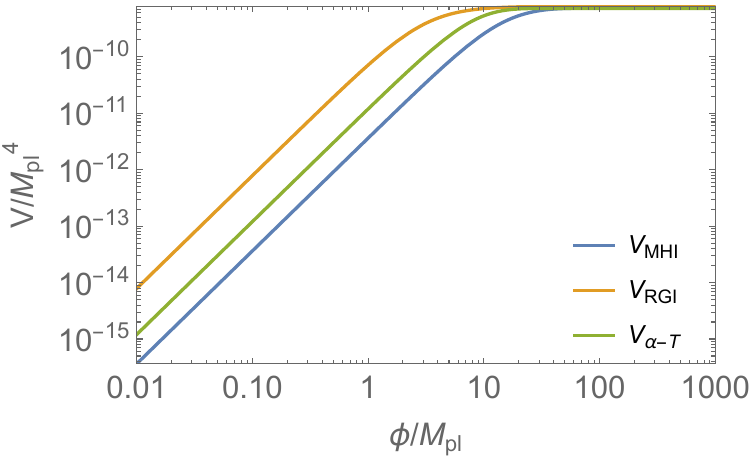}}	
\caption{The shape of the potential for the 
MHI) model, RGI model and the $\alpha$-T model
with their potential given in (\ref{MHI V}), (\ref{RGI V}) and (\ref{alpha V}), and the parameters as used in Sec.~\ref{ChorPasseMole}: in the MHI model $\alpha=3.4$ and $M=5.19\times 10^{-3}M_{pl}$, in 
 the RGI model $\alpha=19$ and $M=5.29\times 10^{-3}M_{pl}$, in the $\alpha$-T model $\alpha=6$ and $M=5.18\times 10^{-3}M_{pl}$. The values of $M$ are chosen according to the fiducial values used in Sec.~\ref{ChorPasseMole} and summarised in Tab.~\ref{tab:priors}.}
\label{VVV}
\end{figure} 

\section{Review of the main ingredients}\label{Sec:OlleKamellen}
Leaving aside the notable exception of Higgs inflation \cite{Bezrukov:2007ep}, 
it is generally believed that inflation was driven by new degrees of freedom beyond the Standard Model (SM) of particle physics. It is clear that these new degrees of freedom must couple to the known particles of the SM in some way (even if only indirectly through a chain of mediators) to enable  \emph{cosmic reheating}~\cite{Albrecht:1982mp,Dolgov:1989us,Traschen:1990sw,Shtanov:1994ce,Kofman:1994rk,Boyanovsky:1996sq,Kofman:1997yn}, i.e., 
 the dissipative transfer of energy from the inflationary sector to other degrees of freedom that filled the universe with particles and set the stage for the \emph{hot big bang}.
Hence, gaining information about reheating would  shed light on the connection between inflation and particle physics models beyond the SM. 
Moreover, reheating is an interesting process on its own because it sets the initial conditions for the radiation dominated epoch. 
This includes the reheating temperature $\Treh$,
which is an important parameter for the production of thermal relics (such as Dark Matter \cite{Cooley:2022ufh}) and many mechanisms to explain the observed matter-antimatter asymmetry of the universe \cite{Canetti:2012zc}.

\subsection{Goals of this work}

 The only known direct messenger from the reheating epoch would be gravitational waves (GWs), cf.~\cite{Caprini:2018mtu} and references therein.\footnote{Thermal emission of gravitons provides unavoidable GW background \cite{Ghiglieri:2015nfa} that can in principle be used to probe $\Treh$ \cite{Ghiglieri:2020mhm,Ringwald:2020ist}, though this is very challenging in practice \cite{Drewes:2023oxg}. }
However, it can be studied indirectly 
through the impact that the post-inflationary expansion history has on the redshifting of CMB perturbations \cite{Martin:2010kz,Adshead:2010mc,Easther:2011yq}. 
This has been used to impose constraints on the expansion history of the universe in a wide range of inflationary models,\footnote{
Examples include 
Starobinski inflation \cite{Cook:2015vqa},
$\alpha$-attractor inflation \cite{Ueno:2016dim,Nozari:2017rta,DiMarco:2017zek,Drewes:2017fmn,Maity:2018dgy,Rashidi:2018ois,German:2020cbw,Mishra:2021wkm,Ellis:2021kad},
power law \cite{Cai:2015soa,DiMarco:2018bnw,Maity:2018qhi,Maity:2019ltu,Antusch:2020iyq} and polynomial potentials \cite{Dai:2014jja,Cook:2015vqa,Domcke:2015iaa,Dalianis:2016wpu},
natural inflation \cite{Munoz:2014eqa,Cook:2015vqa,Wu:2018vuj,Stein:2021uge},
Higgs inflation \cite{Gong:2015qha,Cook:2015vqa,Cai:2015soa},
curvaton models \cite{Hardwick:2016whe},
inflection point inflation \cite{Choi:2016eif},
hilltop type inflation \cite{Cook:2015vqa,Cai:2015soa},
axion inflation \cite{Cai:2015soa,Takahashi:2019qmh} (also in the context of the SMASH model \cite{Ballesteros:2016xej,Ballesteros:2016euj,Ringwald:2022xif}),
fiber inflation \cite{Cabella:2017zsa},
 tachyon inflation \cite{Nautiyal:2018lyq}, 
K\"ahler moduli inflation  \cite{Kabir:2016kdh,Bhattacharya:2017ysa},
and other SUSY models \cite{Cai:2015soa,Dalianis:2018afb}.
}
in particular on the duration of the reheating epoch, which are commonly translated into bounds on the reheating temperature $\Treh$ defined in~\eqref{TR}. 
It has further been pointed out that these constraints can be translated into bounds of microphysical parameters \cite{Drewes:2015coa} if certain conditions are fulfilled \cite{Drewes:2019rxn}.
This possibility has primarily been studied in the context of $\alpha$-attractor models \cite{Ueno:2016dim,Drewes:2017fmn,Ellis:2021kad,Drewes:2022nhu}.
In the present work we go beyond the existing literature in two ways.
Firstly, we for the first time use fully fleshed MCMC forecasts to estimate the foreground-free sensitivity of LiteBIRD and CMB-S4 to $\Treh$ in all three classes of models under consideration.
Secondly, we use those forecasts to critically assess claims in the previous literature regarding the possibility to constrain microphysical parameters with these experiments.

\subsubsection{Theoretical background and parameter classification}

We focus on scenarios in which the dynamics during inflation and reheating can effectively be described by a single scalar field $\Phi$.
We assume that this description holds during both inflation and reheating\footnote{\label{MultifieldFootnote}It is well-known that the single field description may break down during reheating even if it holds during inflation. An important example are instabilities due to the metric in field space \cite{Renaux-Petel:2015mga} that are e.g.~present in 
Higgs inflation \cite{Ema:2016dny,Sfakianakis:2018lzf} as well as 
popular multi-field embeddings of $\alpha$-attractor type potentials \cite{Krajewski:2018moi,Iarygina:2018kee,Iarygina:2020dwe}. In these scenarios our results for $\g$ can be applied if effects that cannot be described within the effective single field framework remain sub-dominant. 
The results for $\Treh$, on the other hand, are more general.}
and distinguish three types of microphysical parameters.
\begin{itemize}
\item A \emph{model of inflation} is defined by specifying the effective potential $\V(\varphi)$. 
Ignoring radiative corrections and the running couplings, 
specifying $\V(\varphi)$ fixes the set of coefficients $\{\vv_i\}$ of all operators in the  action that can be constructed from $\Phi$ alone. In the models considered here all coefficients $\{\vv_i\}$ are fixed by two parameters $M$ and $\alpha$. 
\item The coupling constants (or Wilson coefficients) associated with operators that are constructed from $\Phi$ and other fields form the set of \emph{inflaton couplings} $\{\g_i\}$. 
\item  A complete \emph{particle physics model} typically contains a much larger number of parameters than the combined sets $\{\vv_i\}$ and $\{\g_i\}$. 
These include the masses of the particles produced during reheating as well as their interactions amongst each other and with fields other than $\Phi$. 
We refer to the set of all parameters that are not already contained in $\{\vv_i\}$ and $\{\g_i\}$  as $\{\SM_i\}$. This set e.g.~contains the parameters of the SM.
\end{itemize}

In the models we consider here there are two free parameters in each potential. One of them, $M$, indicates the height of the plateaus in Fig.~\ref{VVV}.
The second one, $\alpha$, characterises the width of the valley in which reheating takes place, and thereby determines the inflaton mass $m_\phi$ for fixed $M$.
Hence, the two microphysical parameters $M$ and $\alpha$ fix two relevant physical scales, namely the scale of inflation (roughly identified with $M$) and the inflaton mass $m_\phi$. 
The third relevant scale is the energy density at the beginning of the radiation dominated epoch, conventionally parameterised in terms of $\Treh$, which necessarily depends on the inflaton couplings $\{\g_i\}$, as they govern the efficiency of the reheating process.
We assume that one interaction with coupling constant $\g\in\{\g_i\}$ dominates the reheating process. 
Our goal is then to investigate if constraints on the scales $\{M, m_\phi,\Treh \}$ as well as 
the microphysical parameters $\{M,\alpha,\g\}$ can be obtained with CMB-S4~\cite{CMB-S4:2020lpa} and LiteBIRD~ \cite{LiteBIRD:2022cnt}. 
We restrict ourselves to the observables $\{n_s,A_s,r\}$, the relation of which to $\Treh$ is well-described by the equations presented below, and which can readily be implemented in forecasts.\footnote{In the future, further observables will provide additional information \cite{Achucarro:2022qrl,Komatsu:2022nvu}.
For instance, the running of the scalar spectral index $n_s$ \cite{Adshead:2010mc} can be constrained by combining CMB observations with data from optical, infrared and 21cm surveys \cite{Mao:2008ug,Kohri:2013mxa,Munoz:2016owz,Munoz:2016owz} 
and can provide information on reheating \cite{Martin:2016oyk}, especially if $r$ is small \cite{Easther:2021eje}.  
We comment on this point in appendix \ref{app:Next-to-leading}.
Also non-Gaussianities which will be probed with CMB observations, galaxy surveys and 21cm observations  \cite{Renaux-Petel:2015bja,Sekiguchi:2018kqe,Achucarro:2022qrl}.}
Here $A_s$ is the amplitude of scalar perturbations, $n_s$ their spectral index, and $r$ the scalar-to-tensor ratios; the precise definitions of these quantities are given in appendix \ref{Sec:PivotScale}.

\subsubsection{Observational perspectives}

As discussed in Sec.~\ref{sec:Benchmark models and analytic estimates}, the ability both of CMB-S4 and LiteBIRD to constrain $\g$ is primarily owed to their sensitivity on the tensor-to-scalar ratio $r$. While  Planck's \cite{Planck:2018vyg} constraints on $r$ come from the large scales of the CMB temperature anisotropy spectrum, the sensitivity on $r$ of CMB-S4 and LiteBIRD (and BICEP \cite{BICEP:2021xfz}) comes from measuring the B-mode polarization spectrum of the CMB. The primordial B-mode spectrum (generated by tensor perturbations) gets contaminated by CMB lensing which converts E-modes (generated by scalar perturbations) into B-modes. This brings out two  observable windows for B-modes, namely the reionization bump at the largest scales, $\ell \lesssim 10$, and the recombination bump centered at $\ell \approx 80$. At smaller scales, $\ell \gtrsim 100$, the primordial tensor modes are already sub-horizon during recombination implying the B-modes to quickly drop off while the lensing contaminant becomes entirely dominant. Even though both experiments -- CMB-S4 and LiteBIRD -- have as one of their main science goals the measurement of primordial B-modes, they are considered being complimentary as they aim to measure them on different scales. LiteBIRD targets to measure the reionization bump at the very large scales where the lensing contamination is expected to be very small. The satellite mission is therefore designed with very good sensitivity but modest resolution. CMB-S4 in contrast aims to measure the recombination bump which requires removal of the lensing induced B-modes. This is aimed to be done via a procedure called \textit{delensing}: The lensing induced B-mode signal is predicted from the measurements of the lensing potential and the E-mode spectrum and then subtracted from the measured B-modes, the remains should be the primordial ones. CMB-S4 will consist of many ground-based detectors providing very good sensitivity and resolution (required for successful delensing) but only small sky coverage.

\subsection{General formalism}\label{Sec:GeneralFormalism}
We  assume that the time evolution of the field expectation value $\varphi\equiv \langle \Phi\rangle$ follows an equation of motion of the form\footnote{It is in general not obvious that such a simple Markovian equation can be used during reheating, but for the range of parameters in which $\g$ can be measured its use for our purpose can be justified, as discussed in appendix C in \cite{Drewes:2019rxn}. }
\begin{equation}\label{EOM}
    \ddot{\varphi} + (3H + \GG)\dot{\varphi} + \partial_\varphi \V(\varphi)=0.
\end{equation}
Here $\V(\varphi)$ is an effective potential, $\Gamma$ an effective damping rate for $\varphi$, and
$H=\dot{a}/a$ is the Hubble rate, with $a$ the scale factor.
Inflation ends when the equation of state exceeds $w>-1/3$, i.e., the universe stops accelerating ($\ddot{a} \leq 0$).
For practical convenience we here define the end of inflation (and thus the beginning of reheating) in terms of the slow-roll parameters
\begin{equation}
    \epsilon=\frac{M^2_{pl}}{2}\left(\frac{\partial_\varphi \V}{\V}\right)^2\ , \quad 
    \eta=M^2_{pl}\frac{\partial^2_\varphi \V}{\V}.
\label{slowrollpara}
\end{equation}
At leading order in these parameters inflation ends when 
 $\epsilon|_{\varphi_{\rm end}}=1$.\footnote{
Throughout our analysis we work at leading order in the slow roll parameters. 
Expressions at next-to-leading order are given in appendix \ref{app:Next-to-leading}, where we confirm that these do not affect our results.
} 
The reheating epoch ends when the energy density of radiation $\rho_R$ exceeds the energy density $\rho_\varphi\simeq \dot{\varphi}^2/2 + \V $ of the condensate $\varphi$ and the universe becomes radiation dominated ($w=1/3$). This moment in good approximation coincides with the moment when $\Gamma=H$, and we in the following use the latter equality to define the end of reheating.

For a given model of inflation the power spectrum of cosmological perturbations at the end of inflation is determined by the parameters $\{\vv_i\}$ in the potential $\V(\varphi)$. 
The impact that the reheating epoch has on the redshifting of these perturbations is primarily sensitive to the duration of the reheating epoch in terms of $e$-folds $N_{\rm re}$ and the averaged equation of state $\wrehbar$ during reheating, 
\begin{equation}
\wrehbar= \frac{1}{N_{\rm re}}\int_0^{N_{\rm re}} w(N) dN.
\end{equation}
The latter is in good approximation fixed by the $\{\vv_i\}$, as the inflaton condensate $\varphi$ dominates the energy density and pressure during reheating (and hence $\wrehbar$). 
Hence, $\Nreh$ is the only parameter that cannot be fixed by specifying the potential $\V(\varphi)$ (i.e., the parameters $\{\vv_i\}$).

To establish a connection to observables we follow the approach taken in section 2.2 of ref.~\cite{Drewes:2017fmn}, which is based on the derivation given in \cite{Ueno:2016dim}.  
$\Nreh$ can be expressed as 
\begin{equation}
\label{Nre}
    N_{\rm re}=\frac{4}{3\Bar{w}_{\rm re}-1}\left[N_k+{\rm ln}\left(\frac{k}{a_0 T_0}\right)+\frac{1}{4}{\rm ln}\left(\frac{40}{\pi^2g_*}\right)+\frac{1}{3}{\rm ln}\left(\frac{11g_{s*}}{43}\right)-\frac{1}{2}{\rm ln}\left(\frac{\pi^2M^2_{ pl}r A_s}{2\sqrt{\Vend}}\right)\right]~,
\end{equation}
where $M_{pl}=2.435\times 10^{18}~{\rm GeV}$ is the reduced Planck mass, and $g_*$ and $g_{s*}$ are the effective number of degrees of freedom contributing to the energy- and entropy-densities of the primordial plasma, respectively.  
We use $g_*=g_{s*}=10^2$ in this work.\footnote{\label{WhatWeNeglect}By treating $g_*$ and $g_{s*}$ as constant, we neglect mild dependencies such as $\sim g_*^{1/4}$. We further neglect late time and foreground effects, cf.~section 2.2 in \cite{Drewes:2019rxn}.}
$N_k$  is the e-folding number between
the horizon crossing of a perturbation with wave number $k$ and the end of inflation,
\begin{equation}\label{Nk}
    N_k={\rm ln}\left(\frac{a_{\rm end}}{a_k}\right)=\int_{\varphi_k}^{\varphi_{\rm end}}\frac{H d\varphi}{\dot{\varphi}}
    \approx\frac{1}{M^2_{pl}}\int_{\varphi_{\rm end}}^{\varphi_k}d\varphi\frac{\V }{\partial_\varphi \V }~,
\end{equation}
with $a_0$ being the scale factor at  present time and $T_0=2.725~{\rm K}$ the temperature of the CMB at the present time.
We introduce the subscript notation $H_k, \varphi_k, \epsilon_k, \eta_k$ to indicate the values of the respective quantities  $H, \varphi, \epsilon, \eta$ in the moment when the mode $k$ crosses the horizon.
In this work we pick $k_p/a_0=0.05 {\rm Mpc}^{-1}$ as this pivot scale of the CMB data.
In the past, sometimes the pivot scale $k/a_0=0.002 {\rm Mpc}^{-1}$ has been used to determine the scalar-to-tensor ratio, which we will denote by $r_{0.002}$. 
We comment on the relation between $r$ and $r_{0.002}$ in appendix \ref{Sec:PivotScale}.
$\varphi_k$ can be expressed in terms of the CMB parameters, namely the spectral index $n_s$ and the tensor-to-scalar ratio $r$, through the relation (at linear order in the slow-roll parameters)
\begin{equation}
    n_s=1-6\epsilon_k+2\eta_k~, \quad r=16\epsilon_k~.
\label{nANDr}    
\end{equation}
From the slow-roll approximation, we obtain 
\begin{equation}
\label{H_k}
	H^2_k=\frac{\V(\varphi_k)}{3M_{pl}^2}~
		=\pi^2 M_{pl}^2\frac{r A_s}{2}.
\end{equation}
In the present work we expand to leading order in the slow roll parameters
and consider three observables $\{n_s,A_s,r\}$. 
A priori this is questionable, as next generation CMB observatories are in general sensitive to higher order corrections and e.g.~the running of $n_s$.
We checked that, for the models and model parameter choices under consideration, next-to-leading order contributions in the slow-roll parameters can be neglected for the purpose of constraining $\Treh$ and $\g$. This also justifies the neglect of the running of the spectral index throughout this work, for a detailed discussion see appendix~\ref{app:Next-to-leading}. 
From \eqref{nANDr} we obtain
\begin{equation}
    \epsilon_k=\frac{r}{16}~,\quad \eta_k=\frac{n_s-1+3r/8}{2}.
\end{equation}
Combining this with \eqref{slowrollpara} we find
\begin{eqnarray}\label{TakaTukaUltras}
&&\frac{\partial_\varphi \V(\varphi)}{\V(\varphi)}\Bigl|_{\varphi_k}=\sqrt{\frac{r}{8M_{pl}^2}} \ , \quad
\frac{\partial^2_\varphi \V(\varphi)}{\V(\varphi)}\Bigl|_{\varphi_k}=\frac{n_s-1+3r/8}{2M_{pl}^2}.
\end{eqnarray}
Together with \eqref{H_k} this provides three equations that relate the effective potential and its derivatives to the three observables $\{n_s,A_s,r\}$. 
This at least in principle permits to determine three microphysical parameters from observation, which we typically choose as the overall normalisation of the potential $\M$, the inflaton coupling $\g$, and one more parameter $\alpha$ in the potential that determines the inflaton mass $m_\phi$. 
The overall normalisation $\M$ of $\V(\varphi)$ has to be obtained from \eqref{H_k}, then the two equations in \eqref{TakaTukaUltras} permit to identify one more parameter in the potential (which is typically necessary to determine the inflaton mass) and $\Nreh$.
For the models considered in this work the solution for the set of equations relating $\{M,\alpha,\g\}$ to $\{n_s,A_s,r\}$ is always invertible within the observationally allowed ranges. However, the current error bar on $n_s$ is too large to fit all three parameters, as we show in detail in sec.~\ref{MHI model}.

\subsubsection{Constraining the reheating temperature}\label{sec:TRconstraints} 
The expansion of the universe is governed by the Friedmann equation 
\begin{equation}\label{Friedmann}
H^2=\rho/(3M^2_{pl}). 
\end{equation}
Knowledge of $\Nreh$ and $\wreh$ permits to compute 
the energy density $\rhoreh$ at the end of reheating by using 
redshifting relation $\rho(N)=\rhoend\exp[-3\int_0^N(1+w(N'))dN']$,
\begin{equation}\label{ThisIsTheEndMyOnlyFriendTheEnd}
    \rhoreh = \rhoend\exp(- 3\Nreh(1 + \wrehbar)).
\end{equation}
The energy density at the end of inflation $\rhoend$ at leading order in the slow roll parameters \eqref{slowrollpara}  can be expressed in terms of the parameters of the inflationary model 
\begin{eqnarray}\label{RhoEndApprox}
\rho_{\rm end} \simeq \frac{4}{3}\Vend ~,
\end{eqnarray}
where $\Vend$ is the potential at the end of inflation. 
The energy density $\rhoreh$ is often expressed in terms of an effective reheating temperature 
defined by the relation
\begin{eqnarray}\label{TR}
    \frac{\pi^2 g_*}{30}\Treh^4\equiv 
    \rhoreh.
\end{eqnarray}
$\Treh$ can be identified with a physical temperature in the thermodynamic sense if the decay products reach local thermal equilibrium quickly, cf. refs.~\cite{Mazumdar:2013gya,Harigaya:2013vwa,Harigaya:2014waa,Mukaida:2015ria} for a discussion, otherwise it should be interpreted as an effective parametrisation of the energy density $\rho_{\rm re}$.
Assuming that reheating happens instantaneously when $\GG = H$,\footnote{The impact that this assumption has on $\wrehbar$ has been estimated in appendix A of \cite{Martin:2010kz}, where it was found that it holds in reasonable approximation.} 
one obtains the standard relation
\begin{eqnarray}
T_{\rm re}=\sqrt{\Gamma M_{pl} }\left(\frac{90}{\pi^2g_*}\right)^{1/4}\Big|_{\Gamma=H}.\label{TRsimple}
\end{eqnarray}
Solving \eqref{TR} and \eqref{ThisIsTheEndMyOnlyFriendTheEnd}  for $\Treh$ gives
\begin{equation}\label{Tre}
	\Treh=\exp\left[-\frac{3(1+\bar{w}_{\rm re})}{4}N_{\rm re}\right]\left(\frac{40 \Vend}{g_*\pi^2}\right)^{1/4}.
\end{equation}
Leaving aside the mild dependence on $g_*$,
$\Nreh$ is the only parameter on the RHS of \eqref{Tre} that is not determined by fixing the parameters $\{\vv_i\}$ in $\V $.
Hence, determining $\Treh$ boils down to establishing a relation between $\Nreh$ and observable quantities.
This is done by plugging \eqref{Nre} with \eqref{Nk} into \eqref{Tre}, 
where $\varphi_k$ is determined by solving \eqref{nANDr}, and $\Vend$ and $\varphi_{\rm end}$ are obtained by solving $\epsilon=1$ for $\varphi$.
This opens up the possibility to constrain the reheating temperature from observation.

Before moving on, we recall that $\Treh$ is already constrained by the good agreement between theoretical calculations of big bang nucleosynthesis (BBN) and the abundances of light elements in the intergalactic medium \cite{Cyburt:2015mya}, imposing a lower bound  
\begin{eqnarray}\label{TreTBBN}
    \Treh > T_{\rm BBN}.
\end{eqnarray}
The highest temperature that directly affects BBN is the freeze-out of neutrinos, which happens at roughly $1.4$ MeV (cf.~e.g. \cite{Bennett:2020zkv}). However, since different modes freeze out at different temperatures, the lower bound on the reheating temperature is in fact a bit higher \cite{Hasegawa:2019jsa}.\footnote{The inflaton decay affects BBN in different ways \cite{Kawasaki:1999na,Kawasaki:2000en}, and the precise value of the lower bound depends on the decay modes. If one, for instance, assumed no hadronic decay modes, it is a bit lower \cite{Hasegawa:2019jsa,deSalas:2015glj}. } 
 In the present work we use $T_{\rm BBN}=10$ MeV.\footnote{\label{BBNfootnote}It is widely believed that the universe in fact was much hotter. This believe is not only based on the theoretical prejudice that most models of single-field inflation involve rather high energy scales, but also on the observation that the present day universe practically consists only of matter (and not antimatter, cf.~e.g.~\cite{Canetti:2012zc}), which in the context of inflationary cosmology necessarily requires violation of the baryon number $B$ \cite{Sakharov:1967dj}. The only known source of $B$-violation in the SM are thermally induced electroweak sphaleron transitions \cite{Kuzmin:1985mm}, which are efficient above a temperature of $T_{\rm sph}\simeq 131$ GeV \cite{DOnofrio:2014rug}.
However, there are models of baryogenesis that can generate the asymmetry at lower temperature (cf.~e.g.~\cite{Barrow:2022gsu}), hence $\Treh > T_{\rm sph}$ would be a model-dependent requirement, which $\Treh > T_{\rm BBN}$ is a comparably hard observational bound.}

\subsubsection{Constraining the inflaton coupling}\label{Sec:InflatonCoupling}

Any information about $\Treh$  discussed in the previous section \ref{sec:TRconstraints} 
is already of great interest not only for cosmology, but also from the viewpoint of particle physics because the thermal history of the universe determines the abundances of thermal relics (cf.~e.g.~\cite{Bernal:2017kxu,Garcia:2021iag}) and is crucial for the viability of baryogenesis scenarios \cite{Bodeker:2020ghk}. 
We go one step further and establish a more direct connection between CMB observables and fundamental interactions by converting the information on 
$\Nreh$ into constraints on microphysical parameters, in particular the inflaton coupling $\g$. 
This is possible because $\Nreh$ is primarily determined by the efficiency of the energy transfer from $\varphi$
to radiation, which we characterise by an effective dissipation rate $\GG$.
This transfer at a microscopic level is driven by fundamental interactions between elementary particles.
Hence, $\Nreh$ is in principle sensitive to the fundamental parameters that govern these interactions.

\begin{table}
\centering
\begin{tabular}{c c c c c}
interaction & 
$y\Phi\bar{\psi}\psi$ & 
$g\Phi\chi^2$ &
$\frac{\upalphatosigma}{\Lambda}\Phi F_{\mu\nu}\tilde{F}^{\mu\nu}$ & $\frac{\hchi}{3!}\Phi\chi^3$\\
\hline 
$\g$ & $y$ & $\tilde{g}=g/m_\phi$ & $\tilde{\upalphatosigma} = \upalphatosigma m_\phi/\Lambda$ & $h$\\
$\#$ & $8\pi$  & $8\pi$ & $4\pi$ & $3! 64(2\pi)^3$ \\
rescaling factor & $1$ & 1 & $\frac{1}{\sqrt{2}}$ & $8\sqrt{6}\pi$
\end{tabular}
\caption{\label{ConversionTable}
Values for the parameters in \eqref{GammaPerturbativeGeneric} for the different interactions  in \eqref{WecheselWirkungen}.
In the main text we primarily use $\g=y$, constraints on other couplings can be obtained by multiplying with the number in the last row. 
Note that this simple conversion is only possible when $\GG$ is expressed in the form \eqref{GammaPerturbativeGeneric}. In particular, it does not apply to the dimensionful couplings $g_\phi$, $g$ and $\upalphatosigma$, but only to the dimensionless ratios \eqref{DimensionlessCouplings}, i.e., $\g\in\{y, \tilde{g}, \tilde{\upalphatosigma}, h\}$. 
We use the notation $\g$ whenever we generically refer to an inflaton coupling and write $y$ when quoting a specific bound on the Yukawa coupling. We will also introduce the variable $\x = \log_{10}y$ later in \eqref{xDef} as this is the quantity that is practically used in the analysis in Sec.~\ref{Sec:Methods}.
}
\end{table}

We can use the Friedmann equation \eqref{Friedmann} with \eqref{ThisIsTheEndMyOnlyFriendTheEnd} and the knowledge that reheating ends when $\GG=H$ to obtain the identity
 \begin{eqnarray}
\GG|_{\GG=H}=\frac{1}{ M_{pl}}\left(\frac{\rhoend}{3}\right)^{1/2} e^{-3(1+\wrehbar) \Nreh/2}. \label{GammaConstraint}
\end{eqnarray}
Within a given model of inflation
all parameter on the RHS of \eqref{GammaConstraint} can
be expressed in terms of CMB observables $\{n_s, A_s, r\}$ by applying the procedure sketched in Sec.~\ref{Sec:GeneralFormalism}, while the LHS is in principle calculable in terms of microphysical parameters when the inflaton interactions are specified. 
$\GG$ necessarily depends on the $\{\g_i\}$ and $\{\vv_i\}$. For instance, for reheating through elementary particle decays, one typically finds 
\begin{equation}\label{GammaPerturbativeGeneric}
\GG = \g^2 m_\phi /{\#}, 
\end{equation}
with $\g\in \{\g_i\}$ a coupling constant
and $\#$ a numerical factor. 
The inflaton mass $m_\phi \in \{\vv_i\}$ is defined by the expansion
\begin{eqnarray}\label{TaylorV}
\V(\varphi)  
&=& 
\frac{1}{2}m_\phi^2 \varphi^2 + \frac{g_\phi}{3!} \varphi^3 + \frac{\lambda_\phi}{4!}\varphi^4 + \mathcal{O}[\varphi^5]
\end{eqnarray}
with $\vv_3 = g_\phi/\Lambda $ and $\vv_4 = \lambda_\phi$ in the notation used in appendix \ref{Sec:FeedbaclEffects}. 
In this work we consider the operators
\begin{eqnarray}\label{WecheselWirkungen}
g\Phi\chi^2 \ , \quad
\frac{\upalphatosigma}{\Lambda}\Phi F_{\mu\nu}\tilde{F}^{\mu\nu} \ , \quad
y\Phi\bar{\psi}\psi, \quad \frac{\hchi}{3!}\Phi\chi^3~.
\end{eqnarray}
They include interactions of $\Phi$ with other scalars $\chi$, fermions $\psi$ as well as U(1) gauge bosons with field strength tensor $F_{\mu\nu}$. 
The interactions \eqref{WecheselWirkungen}  yield $\GG$ of the form \eqref{GammaPerturbativeGeneric}.
It is convenient to introduce the dimensionless coupling constants,
\begin{eqnarray}\label{DimensionlessCouplings}
\tilde{g}\equiv\frac{g}{m_\phi} \quad , \quad 
\tilde{\upalphatosigma} \equiv \frac{m_\phi}{\Lambda}\upalphatosigma \quad , \quad \tilde{g}_\phi \equiv\frac{g_\phi}{m_\phi}
\end{eqnarray}
so that $\g\in\{\tilde{g}, \tilde{\upalphatosigma}, y, h\}$
and $\vv_i \in \{\tilde{g}_\phi, \lambda_\phi\}$.
In the main part of this article we primarily consider the Yukawa coupling $y$, constraints on other quantities can be obtained with table \ref{ConversionTable}.

\paragraph{Conditions on the parameter values} If the effective dissipation rate $\GG$ at the moment when it equals the Hubble rate $H$ is dominated by elementary decays, one can plug \eqref{GammaPerturbativeGeneric} into \eqref{GammaConstraint} and solve for $\g$ to translate a constraint on $\Nreh$ into one on $\g$.
This dependence has been explored in $\alpha$-attractor models in \cite{Ueno:2016dim,Drewes:2017fmn,Ellis:2021kad,Drewes:2022nhu}.
In general this procedure is hampered by the fact that feedback effects invalidate the usage of \eqref{GammaPerturbativeGeneric} in \eqref{GammaConstraint} and introduce a dependence on the $\{\SM_i\}$, making a determination of $\g$ from CMB data impossible.
This in particular happens when non-perturbative processes dominate the early phase of reheating, sometimes referred to as preheating.

Non-linear terms in the inflaton equation of motion
due to self-interactions
can efficiently transfer energy density of from the coherent condensate $\varphi$ into a plasma of inflaton particles $\phi$ through a self-resonance, a process also referred to as fragmentation. 
A very rough criterion to suppress such non-linearities reads
\begin{eqnarray}\label{phi cr}
\varphi_{\rm end} < \varphi_{\rm cr}, \tilde{\varphi}_{\rm cr} \quad {\rm with} \quad 
\varphi_{\rm cr}^2 = \frac{12}{\lambda_\phi} m_\phi^2 \ , \ \tilde{\varphi}_{\rm cr} = \frac{3m_\phi^2}{g_\phi}.
\end{eqnarray}
It turns out, however, that \eqref{phi cr} is not very accurate. 
The conditions on the potential
under which \eqref{GammaPerturbativeGeneric} can be used on the LHS of \eqref{GammaConstraint} 
 have been studied in \cite{Drewes:2019rxn} and are briefly sketched in appendix \ref{Sec:FeedbaclEffects}.
A conservative estimate reads\cite{Drewes:2019rxn}\footnote{Note that there are two numerical errors in the constraints on $g_\phi$ in equation (4.7) in \cite{Drewes:2019rxn}, one forgotten factor $1/3!$ and a typo in the decimal point.}
\begin{eqnarray}
\label{avoid self PR}
\begin{tabular}{c c c}
$\lambda_\phi\ll 8 \left(\frac{m_\phi}{\varphi_{\rm end}}\right)^2$ &, \ & $\lambda_\phi\ll 5.5 \left(\frac{m_\phi}{\varphi_{\rm end}}\right)^{3/2}\left(\frac{m_\phi}{M_{pl}}\right)^{1/2}$, \\ 
$\tilde{g}_\phi 
\ll \frac{1}{2}\frac{m_\phi}{\varphi_{\rm end}} $
&, \ &
$\tilde{g}_\phi
\ll 0.34 \left(\frac{m_\phi}{\varphi_{\rm end}}\frac{m_\phi}{M_{pl}}\right)^{1/2}$.
\end{tabular}
\end{eqnarray}
We restrict ourselves to values of the parameters $\{\vv_i\}$ in $\V(\varphi)$ for which the  conditions \eqref{avoid self PR}  are fulfilled, as otherwise no agnostic constraint on any microphysical parameter can be derived, irrespective of the value of $\g$.
They practically restrict us 
to field elongations $\varphi_{\rm end}$ at the end of inflation which are sufficiently small that the $\varphi$-oscillations are approximately harmonic with a frequency given by the inflaton mass $m_\phi$,\footnote{This conclusion may be too conservative, cf.~footnote \ref{MarcosFootnote}.
}
so that
\begin{equation}\label{EquationOfState}
\wrehbar = 0. 
\end{equation}

The applicability of \eqref{GammaPerturbativeGeneric} on the LHS of \eqref{GammaConstraint} is further restricted by the requirement to avoid feedback from produced particles on the expansion history. For the interactions in \eqref{WecheselWirkungen} considered here the resulting condition reads \cite{Drewes:2019rxn}
\begin{eqnarray}\label{UBoot}
    |\g| \ll {\rm min}\left( \left(\frac{m_\phi}{\varphi_{\rm end}}\frac{m_\phi}{M_{pl}}\right)^{1/2}, \ 
    \frac{m_\phi}{M_{pl}}
    \right),
\end{eqnarray}
which is obtained from Eq.~\eqref{CrocodileBatidaSpecial} in the appendix.

\paragraph{Simple estimates in plateau models}
Keeping in mind that the requirement \eqref{avoid self PR} 
restricts us to potentials that are roughly parabolic at the end of inflation, 
the inflaton mass $m_\phi$ is typically suppressed with respect to $M_{pl}$ as 
$m_\phi \sim  M^2/\varphi_{\rm end}$, with $M$ the typical energy scale during inflation.\footnote{This assumption holds in good approximation in the plateau models introduced in Sec.~\ref{sec:Benchmark models and analytic estimates}, on which we focus in this work (while it is less justified for power-law models, as the one studied in Sec.~\ref{ChaosAD}).}
This follows from $\V(\varphi_{\rm end}) \simeq \frac{1}{2}m_\phi^2\varphi_{\rm end}^2 \sim \V(\varphi_k) \sim M^4$, where the second relation assumed that $\V(\varphi_{\rm end})$ is of the same order of magnitude as $\V(\varphi_k)$ due to the flatness of the potential during inflation. Using  \eqref{H_k} one then finds
\begin{eqnarray}\label{InflatonMass}
m_\phi \sim  \frac{M_{pl}^2}{\varphi_{\rm end}} \pi \sqrt{3 r A_s} \ , \quad
M \sim M_{pl} \left(\frac{3\pi^2}{2} A_s r\right)^{1/4}.
\end{eqnarray}
For $\varphi_{\rm end}\sim M_{pl}$ the conditions \eqref{avoid self PR} and \eqref{UBoot} simplify to
\begin{eqnarray}\label{YippieYaYaySchweinebacke}
\lambda_\phi \ll \left(3\pi^2  r A_s\right) \ , \ 
\tilde{g}_\phi \ll \left(3\pi^2  r A_s\right)^{1/2} \ , \ 
\g \ll \left(3\pi^2  r A_s\right)^{1/2}   
\end{eqnarray}
Plugging in the upper bound $r=0.06$ from \cite{Planck:2018vyg} yields an upper bound of $4\times 10^{-5}$ on $\g$ for interactions that are linear in $\Phi$, which is larger than the electron Yukawa coupling in the 
SM of particle physics.
In the derivation of  \eqref{YippieYaYaySchweinebacke} we neglected various numerical prefactors that depend on the parameters $\{\vv_i\}$ in the potential, which limits its applicability to rough estimates. It is nevertheless a very useful criterion due to its remarkable simplicity, and it provides an explanation for the otherwise surprisingly similar numerical values that the upper bound \eqref{UBoot} yields when applied to the different models. 
Finally, we can use the estimate of $M$ in \eqref{InflatonMass} to rewrite \eqref{YippieYaYaySchweinebacke} as 
\begin{eqnarray}
\lambda_\phi \ll \left(M/M_{pl}\right)^{4} \ , \ 
\tilde{g}_\phi \ll \left(M/M_{pl}\right)^{2} \ , \ 
\g \ll \left(M/M_{pl}\right)^{2}
\end{eqnarray}
implying that models with a comparably low energy scale of inflation can avoid a feedback only for very small values of the inflaton coupling. 
Luckily these conditions are too conservative in many realistic scenarios. We briefly comment on this in appendix \ref{Sec:FeedbaclEffects}.
Note also that 
\eqref{avoid self PR} and \eqref{UBoot}
only restrict the translation of a bound on $\Nreh$ into a constraint on $\g$ 
via \eqref{GammaConstraint}. Imposing a bound on $\rhoreh$ or $\Treh$ via \eqref{Tre} is still perfectly possible in the regime where these conditions are violated.

\subsection{Basic Example: Chaotic inflation}\label{ChaosAD}
\begin{figure}
    \centering
    \includegraphics[width=0.7\textwidth]{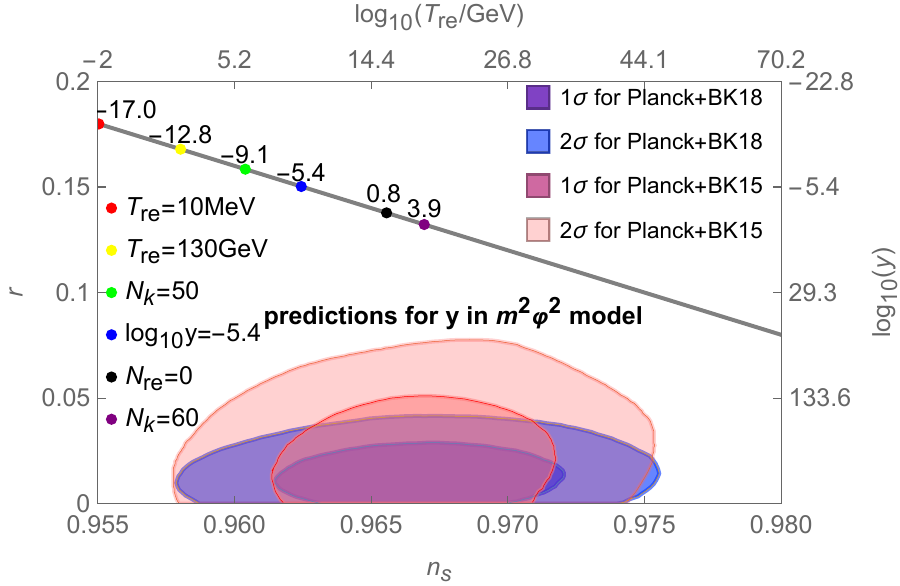}
    \caption{Impact of the reheating phase on the predictions of the chaotic inflation model 
    \eqref{m2phi2potential}
    for $n_s$ and $r$. The relation \eqref{FullHouse} fixes a line in the $n_s$, $r$ plane. 
    Along this line each point corresponds to a specific choice of values for $\g$ and $\Treh$. The dependence is monotonic, larger values of $\g$ (and hence $\Treh$) correspond to smaller values of $r$ and larger values of $n_s$. 
    Here we identify $\g$ with the Yukawa coupling $y$ in table \ref{ConversionTable}.
 	 The predictions for ${\rm log}_{10}y$ are given by the numbers above the discs. 
 The value ${\rm log}_{10}y=-17$ (red disc) corresponds to $\Treh=T_{\rm BBN}$, smaller values are ruled out by the condition \eqref{BBNsimple}. The yellow disc with ${\rm log}_{10}y=-12.8$ marks the point to the right of which $\Treh$ exceed the temperature of the electroweak sphaleron freezeout $T_{\rm sph} \simeq 131$ GeV, cf.~footnote \ref{BBNfootnote}. 
 The green and purple discs indicate the points where $N_k=50$ and $N_k=60$, respectively. 
 The red and blue areas display  marginalized joint confidence regions for $(n_s, r)$ at the $1\sigma$ $(68\%)$ CL (dark) and $2\sigma$ $(95\%)$ CL (light) from Planck+BICEP/Keck 2015 and 2018 data, respectively (Fig.~28 in~\cite{Planck:2018vyg} and Fig.~5 in~\cite{BICEP:2021xfz}, respectively). 
	} 
    \label{FLUSH}
\end{figure}

The present work is dedicated to plateau-type models of inflation, which are currently favoured by observational data. Before moving to the discussion of realistic models of this type we start with a brief discussion of the potential
\begin{equation}\label{m2phi2potential}
    \V(\varphi) = \frac{1}{2} m_\phi^2\varphi^2,
\end{equation}
the simplest example of a monomial potential \cite{Linde:1983gd}, which can also act as an effective description of string-inspired scenarios~\cite{Dimopoulos:2005ac,McAllister:2008hb}.
The effect of reheating in the model \eqref{m2phi2potential} has previously been studied in \cite{Creminelli:2014fca}.
Though it is by now severely disfavoured by the observational upper bound on $r$, cf.~Fig.~\ref{FLUSH}, its simplicity makes it suitable for illustrative purposes. 
For the potential \eqref{m2phi2potential} one finds $\epsilon = \eta = 2M_{pl}^2/\varphi^2$, which leads to \begin{equation}\label{FullHouse}
1-n_s  = \frac{r}{4}. 
\end{equation}
Solving $\epsilon|_{\varphi=\varphi_{\rm end}}=1$ in \eqref{slowrollpara} and $r=16\epsilon_k$ in \eqref{nANDr} respectively we find
\begin{equation}
    \varphi_{\rm end}=\sqrt{2}M_{pl}, \quad \varphi_k=4\sqrt{\frac{2}{r}}M_{pl}~.
\end{equation}
Inserting them into \eqref{Nk} we get a simple expression for $N_k$,
\begin{equation}
N_k=\frac{8}{r}-\frac{1}{2}~.    
\end{equation}
Note that the above relations are entirely independent of $m_\phi$ and also independent of $\g$. 
The fact that specifying the potential defines a line in the  $n_s$-$r$ plane is a general result that applies to all models under consideration here. 
The shape of this line is typically independent of $\g$. Fixing $\g$ then determines where on the line the model prediction lies, cf.~Fig.~\ref{FLUSH}. 
However, the fact that the position of the line defined by \eqref{FullHouse} is independent on the only parameter $m_\phi$ in the potential \eqref{m2phi2potential} is a particularity of this simple model. In general, choosing a type of potential defines a family or curves in the $n_s$-$r$ plane, and selecting a specific curve within this family requires fixing some parameters in the potential.

\begin{figure}
    \centering
    \includegraphics[width=0.7\textwidth]{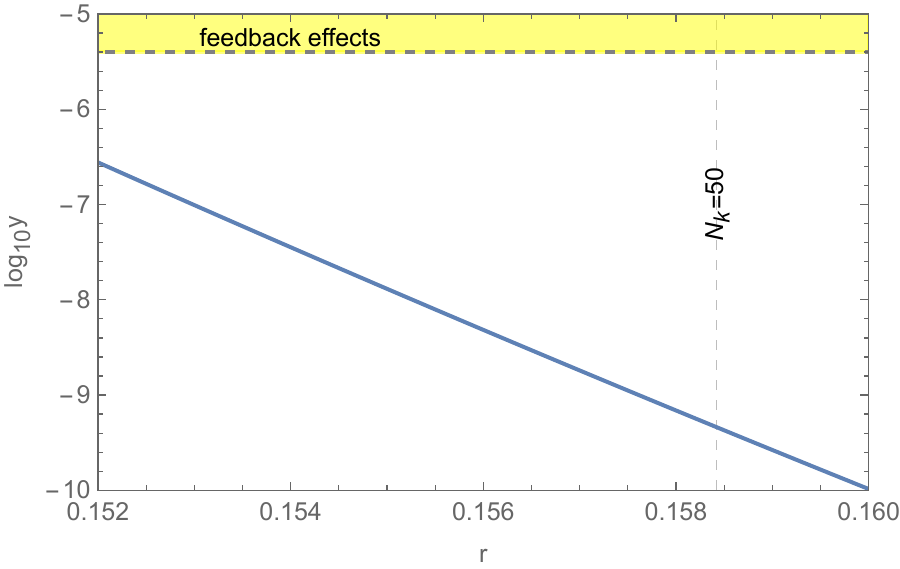}
    \caption{The logarithm of the Yukawa coupling $y$ as a function of the tensor-scalar ratio $r$ in the model \eqref{m2phi2potential}.
    }
    \label{STRAIGHT}
\end{figure}

Fixing $m_\phi$ by \eqref{H_k} requires knowledge of $A_s$ (but does not explicitly depend on $n_s$), 
\begin{equation}
    m_\phi=\sqrt{\frac{3}{32}}\pi
    \
    \sqrt{A_s}r M_{pl}~.
\end{equation}
Of course, the dependence on $r$ can be exchanged for a dependence on $n_s$ through the relation \eqref{FullHouse}.
Analytic expressions for $N_{\rm re}$, $\Gamma|_{\Gamma=H}$, $T_{\rm re}$ and the couplings can be obtained by insertion into \eqref{Nk}, \eqref{Nre}, \eqref{Tre}, and \eqref{GammaConstraint} with \eqref{GammaPerturbativeGeneric}, e.g.,
\begin{align}
    \g=\left(\dfrac{\#\sqrt{3}}{8\sqrt{2}}\right)^{1/2}{\rm exp}\left[3\left[\dfrac{8}{r}-\dfrac{1}{2}+\ln\left(\dfrac{k}{a_0 T_0}\right)+\dfrac{1}{4}\ln\left(\dfrac{40}{\pi^2 g_*}\right)+\dfrac{1}{3}\ln\left(\dfrac{11g_{s*}}{43}\right)\right]\right](\pi \sqrt{ A_s})^{-3/2}
\end{align}
Fig.~\ref{FLUSH} shows the model predictions in the $n_s$-$r$ plane for different values of the inflaton coupling $\g$, 
for the sake of definiteness chosen to be the Yukawa coupling $y$. 
The model is quite strongly disfavoured by observations for any choice of the inflaton coupling.
Fig.~\ref{STRAIGHT} shows the dependence of $\g$ on $r$, indicating that a measurement with an uncertainty $\sigma_r \sim 10^{-2}$ would be sufficient to pin down the order of magnitude of $\g$. However, in 
the models studied in Sec.~\ref{sec:Benchmark models and analytic estimates}
the dependence of $r$ on $\g$ is weaker for smaller values of $r$, meaning that a more accurate measurement will be needed if the true value of $r$ lies within the currently favoured region. In \cite{Drewes:2019rxn} it was estimated that an uncertainty of $\sigma_r \sim 10^{-3}$ will be needed to impose a constraint on $\g$ in plateau models, which can be achieved with LiteBIRD or CMB-S4. In the remainder of the present article we apply different methods to quantify the knowledge gain on $\g$ and $\Treh$ that can be achieved with those missions.


\section{Benchmark models and analytic estimates}
\label{sec:Benchmark models and analytic estimates}

In this section we introduce the specific models we study in this work and discuss their properties. For the illustrative plots in this section we fix the scalar amplitude to its best fit value \cite{Planck:2018vyg},
\begin{eqnarray}\label{AsBestFit}
A_s=10^{-10}e^{3.043}. 
\end{eqnarray}
Note that this is not necessarily required in our approach, 
but represents a simplification which we justify a posteriori in Sec.~\ref{ChorPasseMole}.
In Sec.~\ref{Sec:ForecastMethod} we treat $A_s$ as a parameter that is derived from the forecasts.

\subsection{Mutated hilltop inflation model}
\label{MHI model}

\begin{figure}
    \centering
    \includegraphics[width=0.7\linewidth]{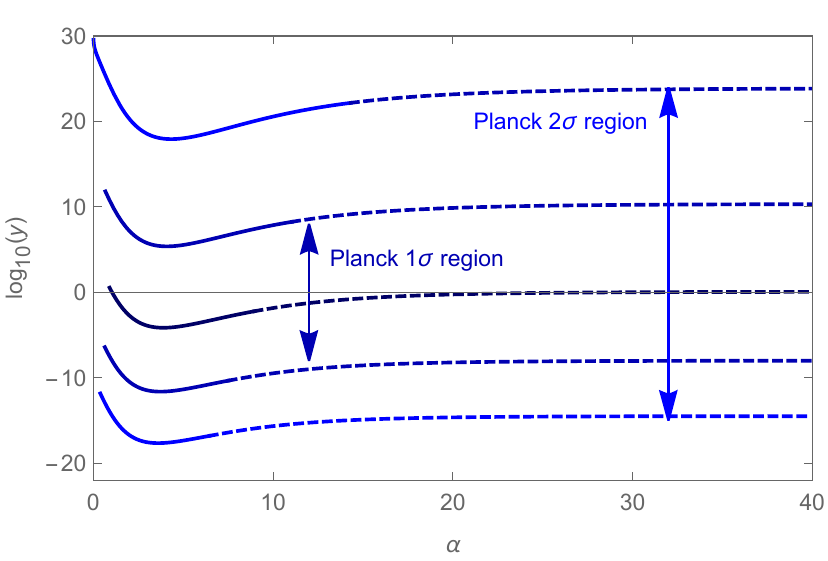}
    \caption{The relation between the inflaton coupling $\g$ and $\alpha$ in the MHI model for 
different choices of $n_s$ marking 
the $1\sigma$ and $2\sigma$ confidence interval from Planck~\cite{Planck:2018vyg}. 
We here chose the Yukawa coupling $\g=y$, corresponding bounds for other couplings can be obtained with the conversion factors in table \ref{ConversionTable}.
   We present examples for different couplings
    in Fig.~\ref{MHI ghy} in the appendix.
The dashed lines correspond values of $r$ that exceed the observational upper bound $r<0.06$ from Planck~\cite{Planck:2018vyg}. 
The relation is non-monotonic for MHI model.
Using the upper bound on $r$
one can in principle impose lower (and upper) bound on the inflaton couplings for with given $n_s$ without $r$ being measured, though the foreseen sensitivities of LiteBIRD and CMB-S4 will not permit this in practice, cf.~Fig.~\ref{fig:my_label}.
We extend the plot to ludicrously large values of the coupling constants for illustrative purposes, a perturbative treatment is only possible below the horizontal line, and \eqref{YippieYaYaySchweinebacke} fulfilled for considerably smaller values (typically of the order of the electron Yukawa coupling, cf.~Figs.~\ref{MHI g}
 - 
 \ref{MHI a}). 
    }
    \label{MHI ghy 1}
\end{figure}

The potential of the MHI model~\cite{Pal:2009sd,Pal:2017bmd} has the form\footnote{The potential of this model is often expressed in terms of a dimensional parameter $\mu=\upmutobealpha M_{pl}$ as
$ \V(\varphi)=M^4\left[1-\frac{1}{{\rm cosh}(\varphi/\mu)}\right]$.
Here we use the dimensionless parameter $\upmutobealpha$ instead for notational consistency with the other models.}

\begin{equation}
\label{MHI V}
 \V(\varphi)
 =M^4\left[1-\frac{1}{{\rm cosh}(\varphi/(\alpha M_{pl}))}\right]~.   
\end{equation}
Here $M$ represents the typical energy scale for hilltop inflation 
and $\alpha$ a parameter that characterises the width of the potential valley, and hence the inflaton mass $m_\phi$.
The potential is shown in Fig.~\ref{VVV}, it is asymptotically flat when $|\varphi|$ is large and has a minimum at  $\varphi =0$.

Defining the end of inflation as the moment when $\epsilon=1$ we find in terms of 
$\upmutobealpha$
\begin{eqnarray}
\varphi_{\rm end} =
 \ \upmutobealpha M_{pl} \ 
{\rm arccosh}
\frac{
3\times2^{2/3} + 2^{1/3} \yy^{2/3}
+ 2 \yy^{1/3} \upmutobealpha
+ 2^{5/3} \upmutobealpha^2
}{6 \upmutobealpha \yy^{1/3}}
\label{phiendMHI}
\end{eqnarray}
with 
$\yy=3\sqrt{6}\xx + 36 \upmutobealpha + 4\upmutobealpha^3$ 
and 
$\xx=\sqrt{
4\upmutobealpha^4 + 22 \upmutobealpha^2 - 1
}$.
The inflaton mass $m_\phi$ and the self-coupling $\lambda_\phi$ of the inflaton can be obtained from the Taylor expansion \eqref{TaylorV} of its potential near the minimum $\varphi=0$, 
\begin{equation}\label{mutatedlamda}
    m_\phi
    =\frac{M^2}{\alpha M_{pl}}~,\quad g_\phi = 0~, \quad \lambda_\phi 
    =-\frac{5M^4}{(\alpha M_{pl})^4}~.
\end{equation}

\begin{figure}
    \centering
    \includegraphics[width=0.69\linewidth]{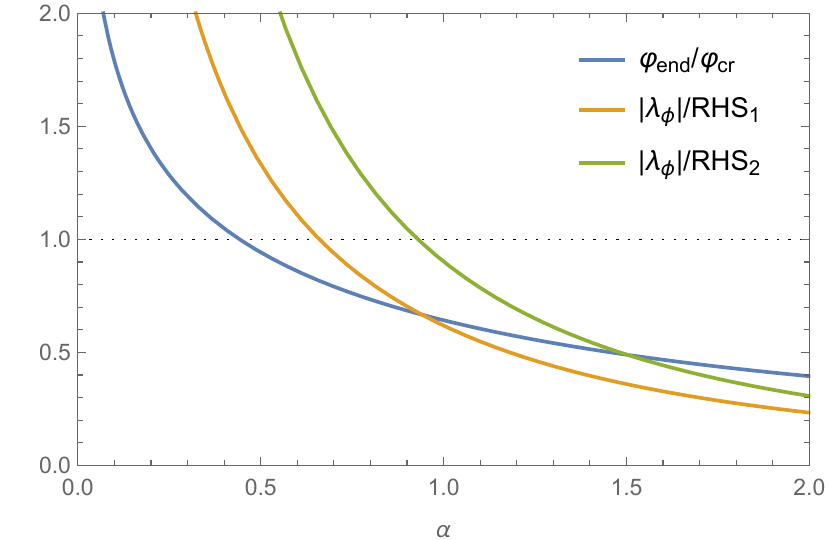}
    \caption{Graphic representation of the conditions on the self-coupling $\lambda_\phi$ in the MHI model, which are functions of $\alpha$. 
    We show $\varphi_{\rm end}/\varphi_{\rm cr}$, $\lambda_\phi/{\rm RHS}_1$ and $\lambda_\phi/{\rm RHS}_2$, where ${\rm RHS}_{1,2}$ represent the RHS of the two inequalities in \eqref{avoid self PR} as well as  \eqref{phi cr}. These conditions roughly require 
    $\alpha>1$.}
    \label{MHI ratios}
\end{figure}

\begin{figure}[!h]
	\centering
	\subfloat[]{
		\includegraphics[width=0.47\linewidth]{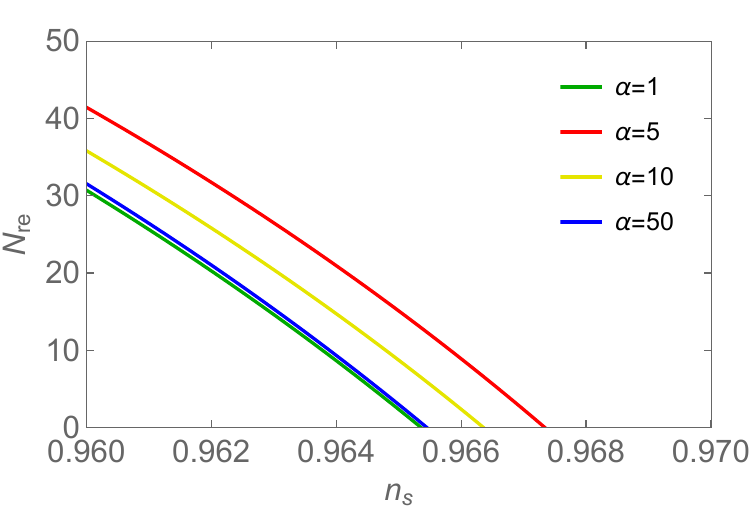}
		\label{MHINre}
	}
\quad
\subfloat[]{
	\includegraphics[width=0.47\linewidth]{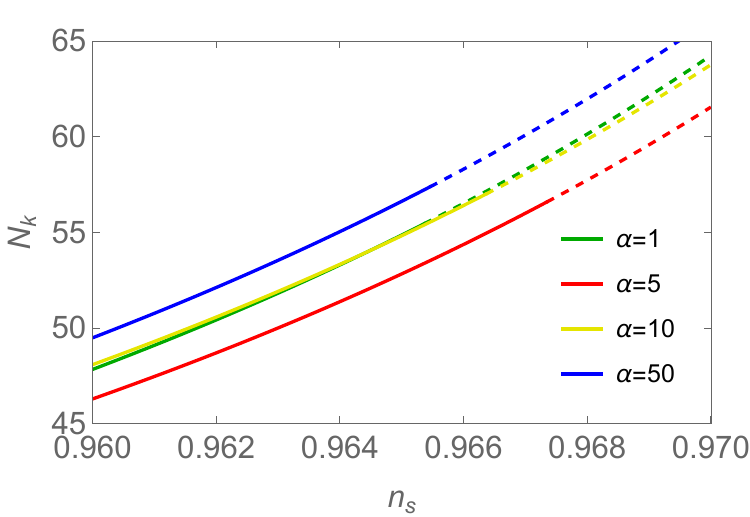}
	\label{MHI Nk}
}
	\quad
	\subfloat[]{
		\includegraphics[width=0.47\linewidth]{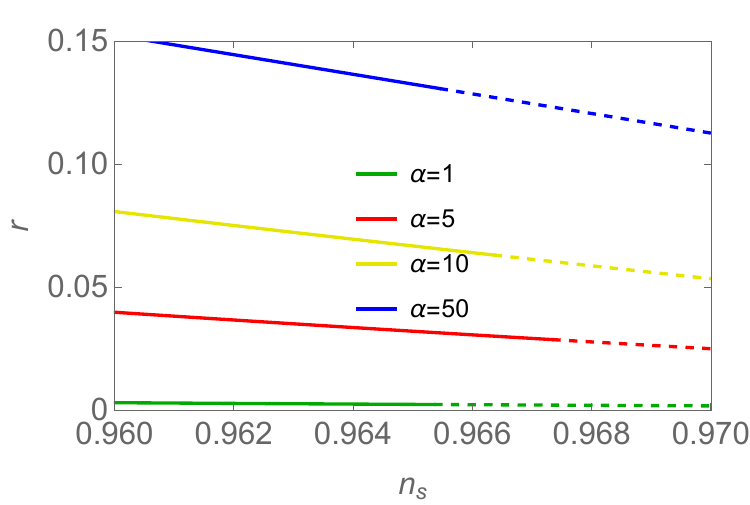}
		\label{MHIr}
	}
\quad
\subfloat[]{
	\includegraphics[width=0.47\linewidth]{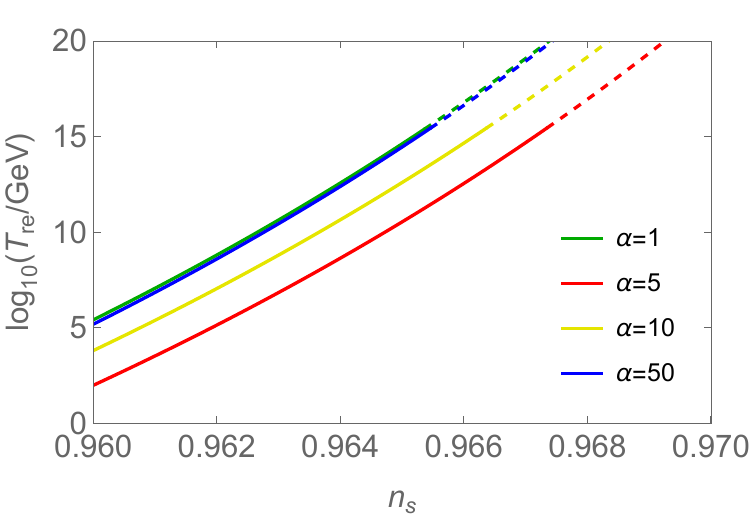}
	\label{MHITre}
}
	\caption{The $n_s$-dependence of $N_{\rm re}$, $N_k$, $r$ and $T_{\rm re}$ in the MHI model for various values of $\alpha$. We take $\bar{w}_{\rm re}=0$ as implied by \eqref{phi cr}
	and fixed $A_s$ to \eqref{AsBestFit}. 
 The colour coding is used to denote different values of $\alpha$. The dashed parts of each line represent the unphysical region where} $N_{\rm re}<0$.
	\label{MHI 4}
\end{figure}

\paragraph{Parametric dependencies.} We shall first investigate whether the set of parameters $\{M,\alpha,\g\}$ can be uniquely determined from the observables $\{n_s, A_s, r\}$.
Using \eqref{H_k} we can express $M$ in terms of other parameters and observables,
\begin{equation}
    M=M_{pl}\sqrt[4]{\frac{3\pi^2r A_s}{2[1-1/{\rm cosh}(\varphi_k/(\alpha M_{pl}))]}},	
    \label{M1}
\end{equation}
consistent with \eqref{InflatonMass}.
We do not find an analytic solution for the set of two equations \eqref{TakaTukaUltras} to express $(\varphi_k,\alpha)$ in terms of $\{r,n_s\}$, but we find numerically that such a solution exists and is unique within the observationally allowed range of $\{n_s,r\}$ and for the values of $\alpha$ that can be made consistent with the conditions
\eqref{YippieYaYaySchweinebacke}.
This permits to numerically express $\alpha$ in terms of $\{n_s,r\}$ and  determine $M$ with \eqref{M1}. 
Hence, the parameters in $\V(\varphi)$ can be fitted without any information on the reheating epoch. 
Using \eqref{Nre}, \eqref{Nk} and \eqref{EquationOfState} one can in addition determine $N_k$ and hence $\Nreh$. Knowledge of $\Nreh$ already permits to determine the reheating temperature \eqref{Tre}.
Finally, one can extract the inflaton coupling from \eqref{GammaConstraint} and \eqref{GammaPerturbativeGeneric} with table \ref{ConversionTable}.
Formally one can always extract a value for $\g$ in this way, as $\g$ obtained from \eqref{GammaConstraint} and \eqref{GammaPerturbativeGeneric} is simply a re-parametrisation of $\Nreh$   
(for fixed inflaton potential there is a bijective mapping between $\Nreh$ and $\g$). 
The parameter $\g$ has a physical meaning as a microphysical coupling constant 
if the use of \eqref{GammaPerturbativeGeneric} on the RHS of \eqref{GammaConstraint} is justified, which is the case when \eqref{YippieYaYaySchweinebacke} are fulfilled.
Hence, it is in principle possible to extract all three parameters $\{M,\alpha,\g\}$ from a measurement of $\{n_s, A_s, r\}$.\footnote{Before moving on, we note in passing that an interesting feature in Fig.~\ref{MHI 4} lies in the non-monotonic dependence of $N_k$ on $\alpha$, which introduces a non-monotonic dependence of $\Treh$ (and therefore the inflaton coupling) on $\alpha$.}

However, in practice this is difficult even when $r$ is determined by future missions like LiteBIRD and CMB-S4.
  Fig.~\ref{MHI ghy 1} shows the dependence of 
 $\alpha$ on $y$ for different values of $n_s$.
The different values of $n_s$ correspond to the range of $1\sigma$ and $2\sigma$ confidence intervals~\cite{Planck:2018vyg}. 
The tensor-to-scalar ratio $r$ changes monotonically along each curve, and the observational upper Planck bound $r<0.06$ \cite{Planck:2018vyg} is shown by the separation of solid and dashed lines. 
Fig.~\ref{MHI ghy 1} also permits to assess the origin of the uncertainties in the inflaton coupling. On one hand, there is the uncertainty in $r$. A measurement of $r$ would restrict the allowed region to a small fraction of each line in Fig.~\ref{MHI ghy 1}. However, the current error bar on $n_s$ translates into a gargantuan uncertainty in the inflaton coupling. Even if $r$ was known with infinite precision (fixing one point along each line in Fig.~\ref{MHI ghy 1}), the uncertainty in $n_s$ covers many orders of magnitude in $\g$.  This shows that the uncertainty in $n_s$ will prohibit a determination of $\g$ with future CMB observations in the MHI model unless further assumptions are made. This is very different from the chaotic inflation model discussed in  Sec.~\ref{ChaosAD} and simply reflects the fact that models with more free parameters offer more freedom.
Hence, in practice 
we have to fix one parameter in the potential in order to determine $\g$. 
We choose this parameter to be $\alpha$. 
This is possible because fixing $\alpha$ does not uniquely fix $n_s$ and $r$, but only restricts predictions to a line in the $n_s$-$r$ plane, the position along which is determined by the value of the inflaton coupling $\g$.
Each choice of $\alpha$ defines a family of inflaton potentials with one remaining free parameter $M$ that can be fitted to data together with $\g$.

\paragraph{Analysis for fixed $\alpha$.} 
The Taylor expansion \eqref{mutatedlamda} yields $\varphi_{\rm cr}/M_{pl}=\sqrt{\frac{12}{5}}\alpha$.
To satisfy the condition \eqref{avoid self PR} $\upmutobealpha$ should roughly be larger than unity, cf.~Fig.~\ref{MHI ratios}.
In the following analysis we choose several values of $\alpha$, corresponding to 
$\alpha=1$, ~$5$, ~$10$ and $50$.
Combining equations~(\ref{Nre})-(\ref{nANDr}) as well as~(\ref{Tre}) we obtain the $n_s$-dependence of four variables in the MHI model: the e-folding number of reheating $\Nreh$, the e-folding number after horizon crossing $N_k$, the tensor-to-scalar ratio $r$ and the reheating temperature $\Treh$, as shown in Fig.~\ref{MHI 4}. These variables have an analytic but complicated and not illuminating dependence on $\alpha$ and $n_s$.

\begin{figure}
    \centering
 	\includegraphics[width=0.7\linewidth]{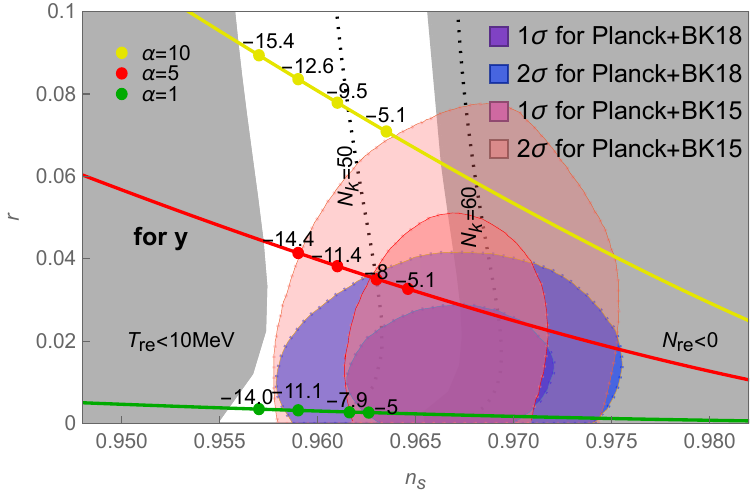}
 	\caption{Impact of the reheating phase on the predictions of the MHI model 
 	for $n_s$ and $r$. Here we assume an Yukawa inflaton coupling $y$;
 	we show results for all couplings in table~\ref{ConversionTable} in Fig.~\ref{MHI R} in the appendix.
 	The red and blue areas display  marginalized joint confidence regions for $(n_s, r)$ at the $1\sigma$ $(68\%)$ CL (dark) and $2\sigma$ $(95\%)$ CL (light) from Planck+BICEP/Keck 2015 and 2018 data, respectively (Fig.~28 in~\cite{Planck:2018vyg} and Fig.~5 in~\cite{BICEP:2021xfz}). 
 The choice of $\alpha$ in the MHI model defines a line in the $n_s$-$r$ plane along which $\Nreh$ changes.
  We show these lines for three values of $\alpha$ as indicated by the colour.
  The discs indicate points corresponding to specific values of $\logy = \log_{10}y \propto \Nreh$  (given by the numbers above the discs) that are consistent with condition \eqref{CrocodileBatidaSpecial}. 
	The grey region on the left side corresponds to the lower bound  $T_{\rm re} < 10$ MeV from BBN in \eqref{TreTBBN}, and the grey area on the right side corresponds to 
 the unphysical region where 
 $N_{\rm re} < 0$. The dotted lines correspond to $\Nk=50$ and $\Nk=60$.}
		\label{MHI R d}
\end{figure}

After eliminating $M$ with \eqref{M1} and fixing $A_s$ to \eqref{AsBestFit}, 
we can compute both $r$ and $\g$ as functions of $n_s$
from \eqref{GammaConstraint} with 
table \ref{ConversionTable}. 
As an example, Fig.~\ref{MHI R d} shows the impact of the reheating era on the observables $n_s$ and $r$ for different values of $\alpha$, assuming a Yukawa coupling. 
Analogue plots for the other interactions in table \ref{ConversionTable} are shown in Fig.~\ref{MHI R} in the appendix. 
The distribution of the discs in Fig.~\ref{MHI g 1 a} shows that a determination of $\g$ by measuring $r$ is considerably easier for larger $\alpha$. 

The relation between the inflaton coupling $y$ and $n_s$ is shown in more detail in
Fig.~\ref{MHI g 1 a}, where we have fixed $\alpha = 5$ and again assumed a Yukawa coupling.
Results for other choices of $\alpha$ and other inflaton interactions are shown in 
Figs.~\ref{MHI g}, \ref{MHI h},  \ref{MHI y} and \ref{MHI a} in the appendix~\ref{Sec:AppendixWithAllPlots}.
The labels on top of the frame
show the value of $r$  
for given $n_s$ and fixed $\alpha$.
The labels on the side of each plot indicate the relation  between ${\rm log}_{10}(T_{\rm re}/{\rm GeV})$ and ${\rm log}_{10}(\g)$ given by \eqref{TRsimple}.
This relation is 
not exactly linear because  the inflaton mass $m_\phi$ changes slightly for different choices of parameters.
In the yellow region condition \eqref{CrocodileBatidaSpecial}
is violated.\footnote{\label{SimpleResonanceFootnote}Given that these conditions are to be read as estimates, we neglect the weak dependence on $n_s$. For example, in Fig.~\ref{MHI g 1} for $\alpha=1$, the condition \eqref{CrocodileBatidaSpecial} gives an upper bound on $\g$ that varies between -5.75 to -5.98 when $n_s$ changes from 0.955 to 0.975. Thus we draw the lower boundary of the yellow region to be horizontal at -5.86, this gives the approximate upper bound for the coupling $\g$ at the magnitude required by perturbative reheating. The situation is similar in other plots.} 
From these figures one can see by eye that a determination of $r$ at the level $\sigma_r \sim  10^{-3}$ would be sufficient to constrain the order of magnitude of the inflaton, consistent with what was found in \cite{Drewes:2019rxn}. 
In the following section \ref{Sec:Methods} we use two methods to quantify the information gain on $\Treh$ and $\g$ that can be achieved with CMB-S4 and LiteBIRD.

Finally, it is instructive to consider the case where all parameters in the potential are fixed from theory. For fixed $M$ and $\alpha$ we can use \eqref{M1} as well as the numerical relation between $n_s$ and $r$ obtained from \eqref{TakaTukaUltras} for fixed $\alpha$ to express both $r$ and $n_s$ in terms of $A_s$.
This permits to express $\g$ in terms of the observable $A_s$ alone. In Fig.~\ref{Horum omnium fortissimi sunt Belgae, propterea quod a cultu atque humanitate provinciae longissime absunt, minimeque ad eos mercatores saepe commeant atque ea quae ad effeminandos animos pertinent important, proximique sunt Germanis, qui trans Rhenum incolunt, quibuscum continenter bellum gerunt.} one can see that the current measurement of $A_s$ is already sufficient to constrain $\g$ if all parameters in the potential are fixed. We do not further pursue the option to constrain $\g$ in this way, but rather focus on the potential of LiteBIRD and CMB-S4 to simultaneously determine $\g$ and the scale of inflation $M$ from data.

\begin{figure}
    \centering
 	\includegraphics[width=0.71\linewidth]{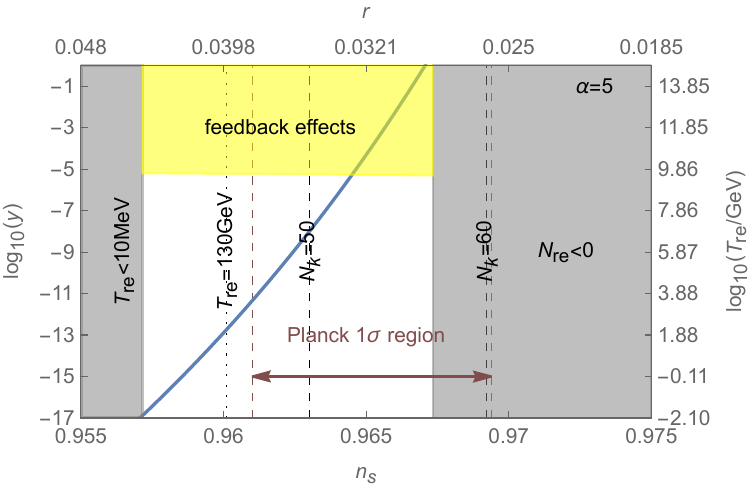}
 	\caption{
 	Dependence of the Yukawa inflaton coupling $y$ on $n_s$ and $r$ in the MHI model for $\alpha=5$, with the Planck error bar on $n_s$ indicated.  
 	Results for other values of $\alpha$ and other couplings are shown in Figs.~\ref{MHI g}-\ref{MHI a} in the appendix. 
 	On the right of the panels we indicate the value of $\Treh$ for given $\g$, assuming instantaneous thermalisation.
	The grey region on the left side corresponds to  $T_{\rm re}<10~{\rm MeV}$ and is excluded by BBN, and the grey area on the right side corresponds to 
 the unphysical region where 
 $N_{\rm re}<0$. The dotted line indicates the electroweak sphaleron freeze-out temperature $T_{\rm re}=131~{\rm GeV}$. The dotted lines correspond to $N_k=50$ and $N_k=60$.
 In the yellow region the condition \eqref{CrocodileBatidaSpecial} is violated for $y$ (we neglected a mild dependence on $r$, cf.~footnote \ref{SimpleResonanceFootnote}); in this regime we can read off $\Treh$ as a function of $r$ or $n_s$, but the conversion to $y$ is questionable. 
	The red dashed lines represent the $1\sigma$ confidence region of the Planck measurement of $n_s$ ($n_s = 0.9652\pm0.0042$) after marginalisation over $r$, which should be understood as illustrative (and not taken literally) because $n_s$ and $r$ are not independent here, and values of $r$ in the panels labeled [d] 
	in Figs.~\ref{MHI g}-\ref{MHI a}
	are in fact already disfavoured by the current constraints on $r$ \cite{Planck:2018vyg,BICEP:2021xfz}.
 	}
		\label{MHI g 1 a}
\end{figure}

\begin{figure}
    \centering
 	\includegraphics[width=0.7\linewidth]{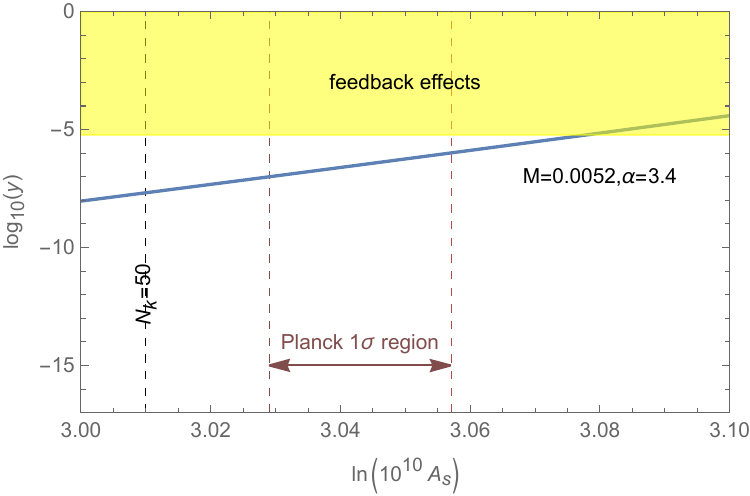}
 	\caption{The Yukawa coupling $y$ as a function of $A_s$ for fixed $M$ and $\alpha$ in the MHI model.}
  \label{Horum omnium fortissimi sunt Belgae, propterea quod a cultu atque humanitate provinciae longissime absunt, minimeque ad eos mercatores saepe commeant atque ea quae ad effeminandos animos pertinent important, proximique sunt Germanis, qui trans Rhenum incolunt, quibuscum continenter bellum gerunt.}
  \end{figure}


\subsection{Radion gauge inflation model}
\label{RGI model}

\begin{figure}
    \centering
    \includegraphics[width=0.7\linewidth]{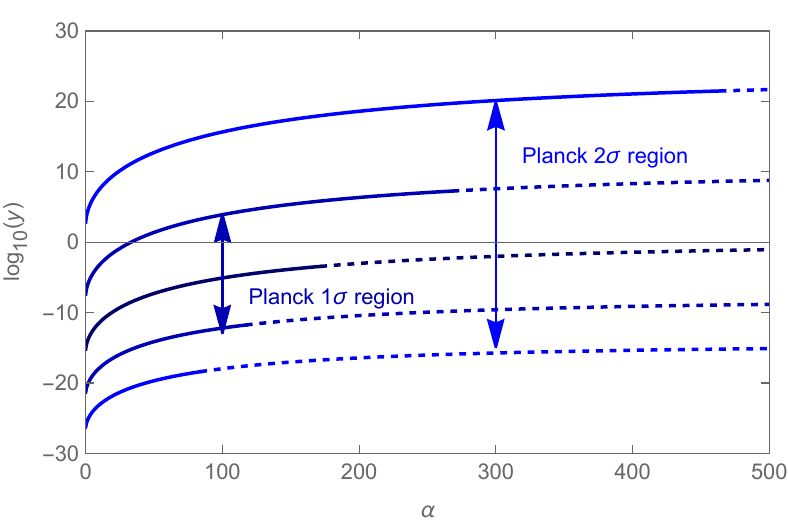}
    \caption{The relation between the inflaton coupling $y$ and $\alpha$ in the RGI model for 
different choices of $n_s$ (marking 
the Planck $1\sigma$ and $2\sigma$ confidence intervals)
with conventions as in Fig.~\ref{MHI ghy 1}. Plot for other couplings are shown in Fig.~\ref{RGI ghy} in the appendix.}
\label{RGI ghy 1}
\end{figure}

The potential of the RGI model has the form~\cite{Fairbairn:2003yx,Martin:2013nzq}
\begin{equation}
 \V(\varphi)=M^4\frac{(\varphi/M_{pl})^2}{\alpha+(\varphi/M_{pl})^2}~.
 \label{RGI V}
\end{equation}
Here $M$ represents the typical energy scale for this model and $\alpha$ is a dimensionless parameter.
The shape of the potential is  shown in Fig.~\ref{VVV}. The scale $M$ can again be expressed in terms of  other parameters through \eqref{H_k},
\begin{equation}
    M =
     M_{pl}
     \left(
     \frac{3\pi^2}{2} r A_s
     \left(1 + \alpha \frac{M^2_{pl}}{\varphi_k^2}\right)
     \right)^{1/4},
    \label{M2}
\end{equation}
consistent with \eqref{InflatonMass}.
Following that, \eqref{TakaTukaUltras} can be solved for $\alpha$,
\begin{equation}\label{珍珠奶茶}
    \alpha=\frac{432r^2}{(8(1-n_s)+r)^2(4(1-n_s)-r)}.
\end{equation}
As in the MHI model, all parameters in the effective potential \eqref{RGI V} can in principle be determined from a measurement of $\{n_s,A_s,r\}$ without knowledge about the reheating phase. 
With the equation of state \eqref{EquationOfState} fixed by the effective potential, 
the relations
\eqref{Nre}, \eqref{Nk}  and \eqref{Tre} can again be used to determine $\Nreh$ and hence $\Treh$. 
This can then finally be converted into a determination of the inflaton coupling $\g$ by means of \eqref{GammaConstraint} and table \ref{ConversionTable},  provided that \eqref{YippieYaYaySchweinebacke} is fulfilled.  
However, Fig.~\ref{RGI ghy 1} shows that the uncertainty in $n_s$ again prohibits a determination of $\g$ even if $r$ was known exactly unless one specifies $\alpha$.

\clearpage\begin{figure}[!h]
	\centering
	\subfloat[]{
		\includegraphics[width=0.47\linewidth]{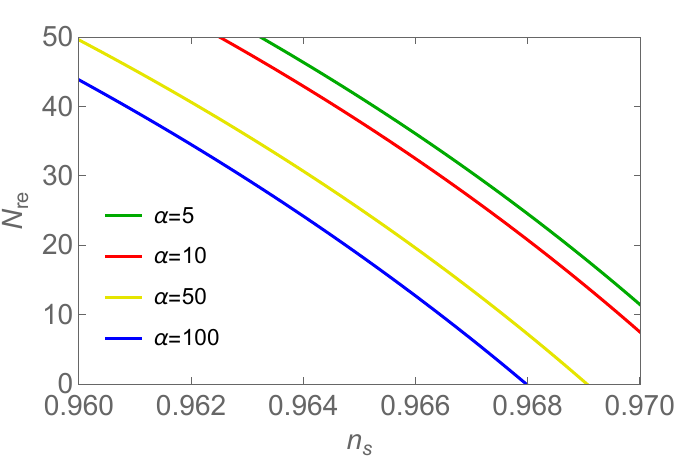}
		\label{RGINre}
	}
	\quad
	\subfloat[]{
		\includegraphics[width=0.47\linewidth]{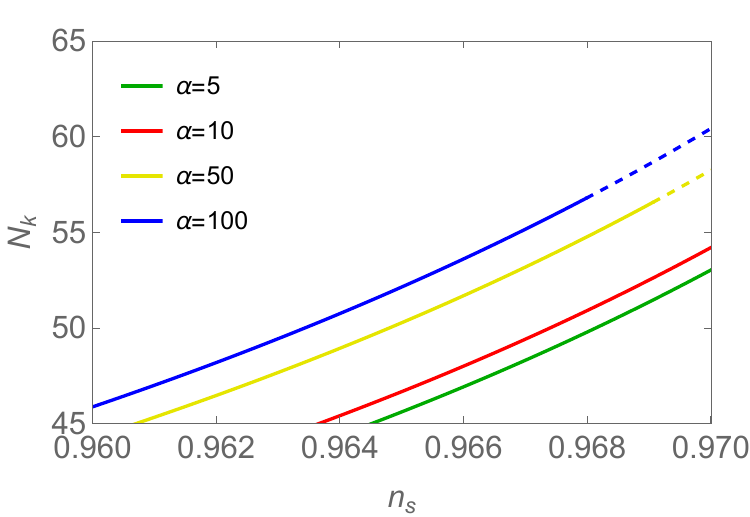}
		\label{RGINk}
	}
	\quad
	\subfloat[]{
		\includegraphics[width=0.47\linewidth]{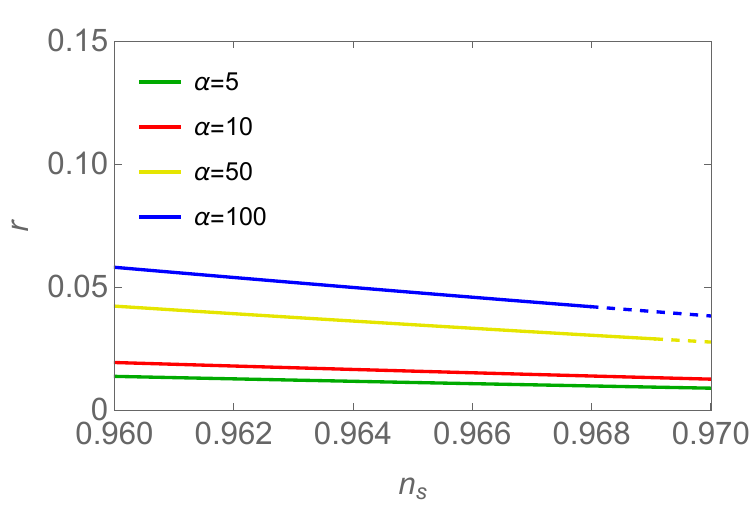}
	\label{RGIr}
    }
    \quad
    	\subfloat[]{
    	\includegraphics[width=0.47\linewidth]{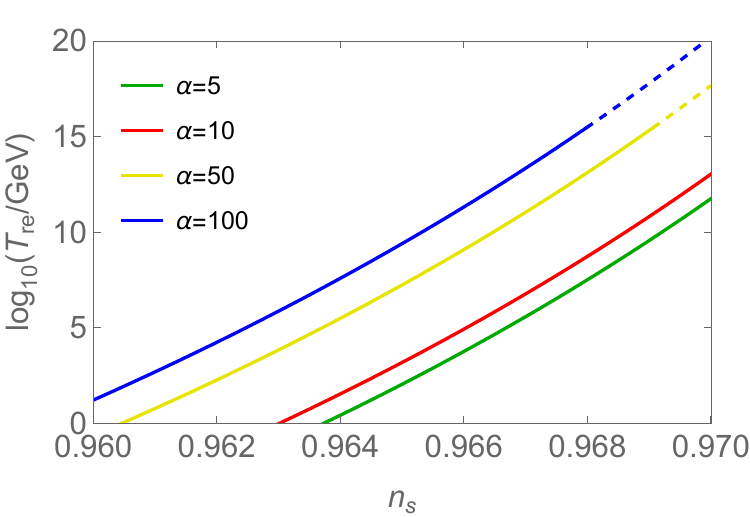}
    	\label{RGITre}
    }
\caption{Dependence of $N_{\rm re}$, $N_k$, $r$ and $T_{\rm re}$ on $n_s$ in the RGI model with different values of $\alpha$, in analogy to Fig.~\ref{MHI 4}.}
\label{RGI 4}
\clearpage\end{figure}

\begin{figure}
    \centering
 	\includegraphics[width=0.7\linewidth]{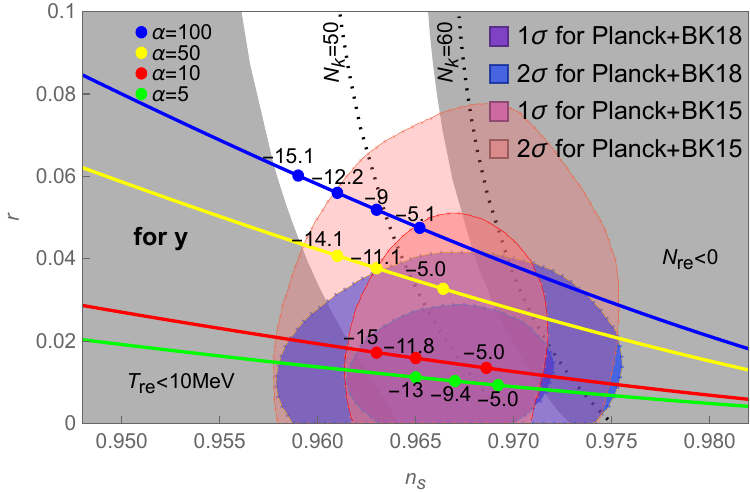}
 	\caption{
  Impact of the reheating phase on the predictions of RGI model for $n_s$ and $r$ for a set of sample points and an Yukawa coupling $y$, with conventions as in Fig.~\ref{MHI R d}. Results for other couplings in table~\ref{ConversionTable} are shown in Fig.~\ref{RGI R} in the appendix.}
		\label{RGI R d}
\end{figure}

Near the minimum  of the potential at $\varphi=0$ the inflaton mass $m_\phi$ and self-couplings are, 
\begin{equation}\label{InflatonMassRGI}
    m_\phi=\sqrt{\frac{2}{\alpha}}\frac{M^2}{M_{pl}}~,\quad 
    g_\phi=0~,\quad
    \lambda_\phi=-\frac{24M^4}{\alpha^2M^4_{pl}}~.
\end{equation}
With \eqref{phi cr} this gives $\varphi_{\rm cr}=\sqrt{\alpha}M_{pl}$. 
Comparing this to 
\begin{eqnarray}
\varphi_{\rm end} = M_{pl}
\frac{\xx^{1/3} - \alpha}{\sqrt{3} \xx^{1/6}}
\label{phiendRGI}
\end{eqnarray}
with $\xx=\alpha^2(27+\alpha+3\sqrt{81+6\alpha})$ 
the condition $\varphi_{\rm end}<\varphi_{\rm cr}$ in \eqref{phi cr} translates into $\alpha>1/2$. 
Condition \eqref{avoid self PR}  implies $\alpha>2.37$. 
In our analysis $\alpha$ is chosen to be four values: $\alpha=5,~10,~50$ and 100.
For fixed $\alpha$ we can get the $n_s$-dependence of $N_{\rm re}$, $N_k$, $r$ and $T_{\rm re}$ in the RGI model, as shown in Fig.~\ref{RGI 4}.
If we fix $n_s$ we can express the inflaton coupling in terms of $\alpha$. 
The dependence is monotonic, as can been seen in Fig.~\ref{RGI ghy 1}. 
The different curves in each panel correspond to different values of $n_s$, and $r$ changes along each curve.

\begin{figure}
    \centering
 	\includegraphics[width=0.71\linewidth]{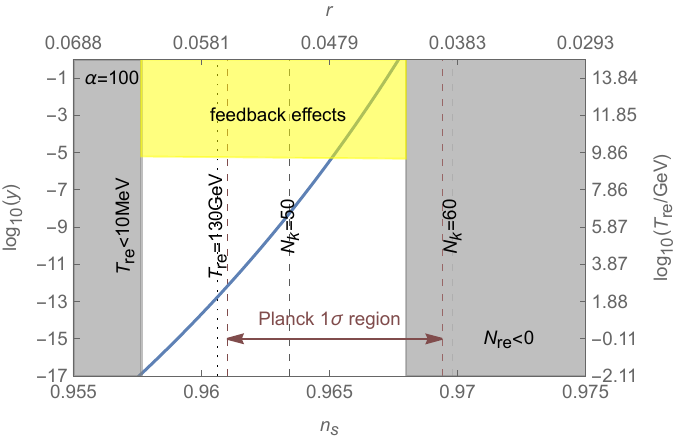}
 	\caption{
 	Dependence of the Yukawa coupling $y$ on $n_s$ and $r$ in the RGI model for $\alpha=100$ with conventions as in Fig.~\ref{MHI g 1 a}. 
  Results for other couplings are shown in Figs.~\ref{RGI g},
  \ref{RGI h},  
  \ref{RGI y},
  \ref{RGI a} in the appendix.
  } 
		\label{RGI g 1 a}
\end{figure}

Fig.~\ref{RGI R d} shows
the impact of the reheating epoch on the model predictions in the $n_s$-$r$ plane
together with
the Planck+BICEP2/Keck 2015 and 2018 results~\cite{Planck:2018vyg,BICEP:2021xfz} ($1\sigma$ and $2\sigma$ CL). 
The corresponding values of ${\rm log}_{10}\g$ are indicated by the numbers above the discs, and different values of $\alpha$ are distinguished by colour coding (same as in  Fig.~\ref{MHI R d}).
Finally, the dependence of the inflaton coupling 
on $n_s$ and $r$ is shown in Fig.~\ref{RGI g 1 a}.

\begin{figure}
    \centering
    \includegraphics[width=0.7\linewidth]{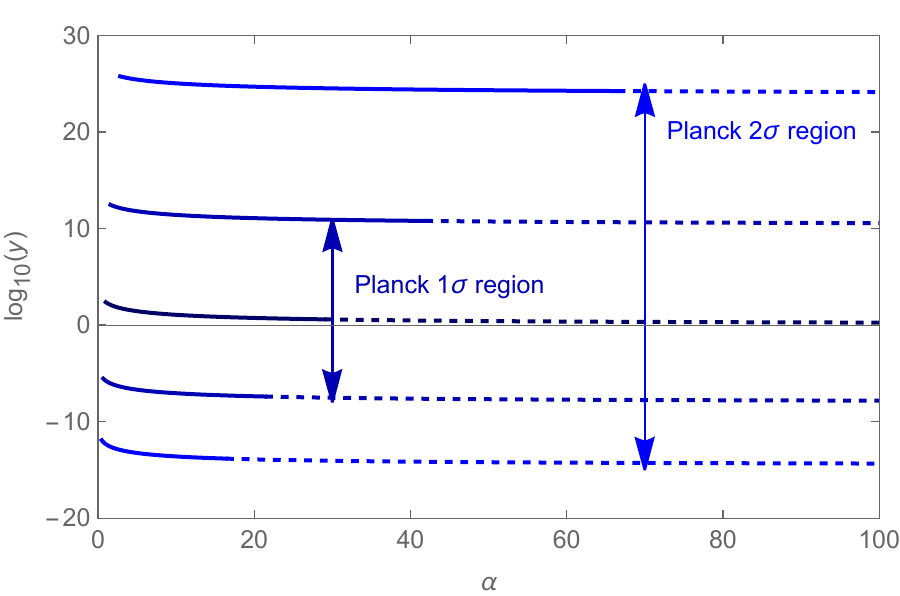}
    \caption{The relation between the inflaton coupling $y$ and $\alpha$ in the $\alpha$-T model for 
different choices of $n_s$ (marking 
the Planck $1\sigma$ and $2\sigma$ confidence intervals)
with conventions as in Fig.~\ref{MHI ghy 1}. Plot for other couplings are shown in Fig.~\ref{RGI ghy} in the appendix.}
\label{alpha T ghy 1}
\end{figure}

\clearpage
\begin{figure}[!h]
\centering
\subfloat[]{
\includegraphics[width=0.45\linewidth]{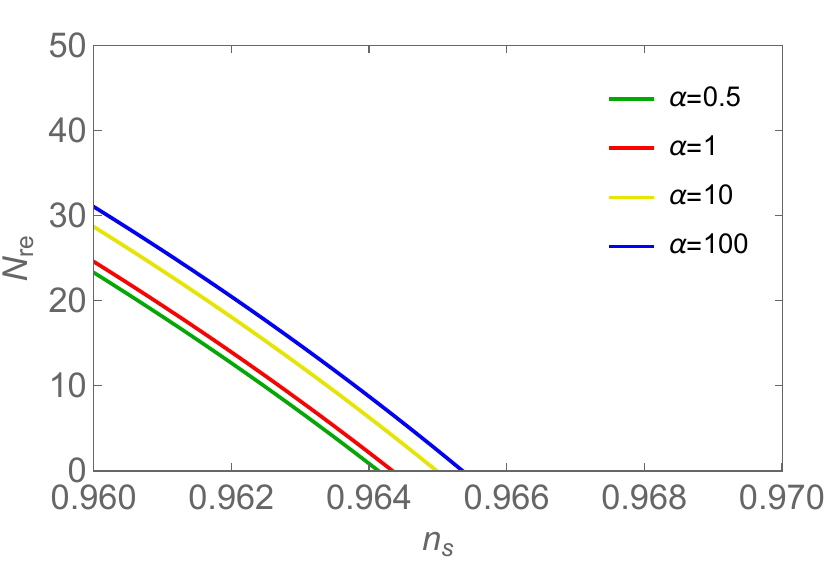}
\label{alpha T Nre}
}
\quad
\subfloat[]{\includegraphics[width=0.45\linewidth]{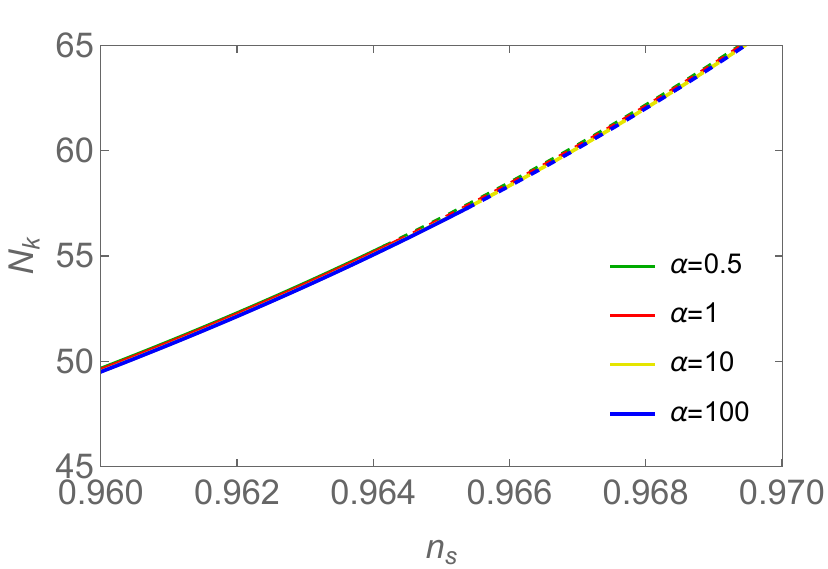}}
\quad
\subfloat[]{\includegraphics[width=0.45\linewidth]{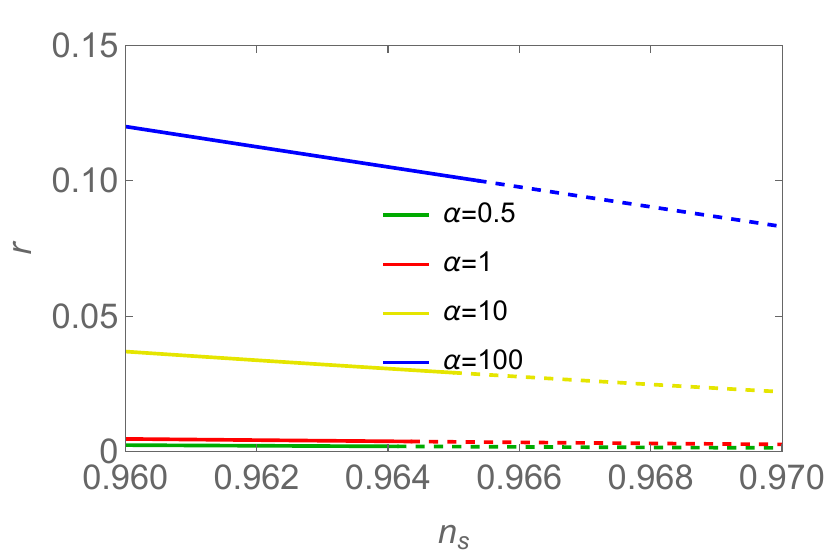}
	\label{alpha T r}
}
\quad
\subfloat[]{
	\includegraphics[width=0.45\linewidth]{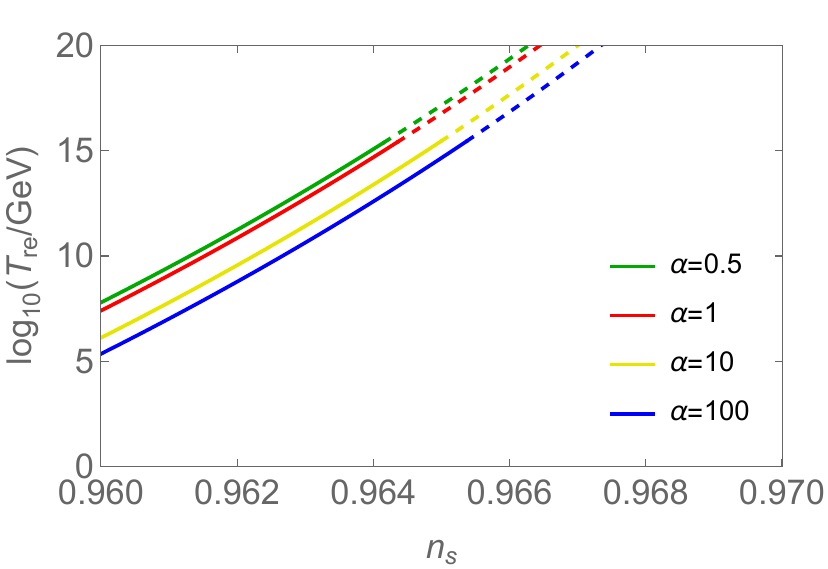}
	\label{alpha T Tre}
}
\caption{Dependence of $N_{\rm re}$, $N_k$, $r$ and $T_{\rm re}$ on $n_s$ in the $\alpha$-T model with different values of $\alpha$, in analogy to Fig.~\ref{MHI 4}.}
\label{alpha T 4}
\clearpage\end{figure}

\subsection{$\alpha$-attractor T-model}
\label{alpa T}
The so-called $\alpha$-attractor models \cite{Kallosh:2013hoa} represent a broad class of popular inflationary models. For some of them constraints on the inflaton coupling have previously been derived in~\cite{Ueno:2016dim,Drewes:2017fmn,Ellis:2021kad}.
Here we focus on the so-called T-type models, which have a symmetric potential. In E-type models it is much more difficult to obtain a bound on $\g$ because self-interactions tend to violate \eqref{avoid self PR}, cf.~appendix \ref{JinURevisited}.
The potential of the $\alpha$-attractor T-models ($\alpha$-T)~\cite{Kallosh:2013maa,Carrasco:2015pla,Carrasco:2015rva} has the form
\begin{equation}
 \V(\varphi)=M^4{\rm tanh}^{2n}
 \left(\frac{\varphi}{\sqrt{6\alpha}M_{pl}}\right) ~.
 \label{alpha V}  
\end{equation}
Here $M$ is the typical energy scale for this model and $\alpha$ is a dimensionless parameter. 
The shape of the potential shown in Fig.~\ref{VVV}.
The scale $M$ can again be expressed in terms of other parameters with the help of \eqref{H_k},
\begin{equation}
    M =
     M_{pl}
     \left(
     \frac{3\pi^2}{2}
     A_s r
     \right)^{1/4}
     \tanh^{-\frac{n}{2}}\left(
     \frac{\varphi_k}{\sqrt{6\alpha}M_{pl}}
     \right),
    \label{M3}
\end{equation}
consistent with \eqref{InflatonMass}.
Again defining the end of inflation as the moment when $\epsilon=1$ we find
\begin{eqnarray}
\varphi_{\rm end} = \frac{M_{pl}}{2}\sqrt{\frac{3\alpha}{2}}{\rm ln}\left(
\frac{3\alpha + 8n^2 + 4n\sqrt{4n^2 + 3\alpha}}{3\alpha}
\right) ~.
\end{eqnarray}
To obtain a parabolic potential near the minimum 
we choose $n=1$.  
With the help of \eqref{TakaTukaUltras} we find
\begin{equation}\label{AlphaInAlphaT}
    \alpha=\frac{4r}{3(1-n_s)(4(1-n_s)-r)}.
\end{equation}
As in the other models, the formalism outlined in Sec.~\ref{Sec:OlleKamellen} can in principle be used to constrain $\Treh$ as a function of $n_s$ or $r$.
However, as in the other examples, in practice the large uncertainty in $n_s$ makes it practically impossible to determine all parameters in the potential from observation (cf.~Fig.~\ref{alpha T ghy 1}), and one has to fix $\alpha$ from model-building considerations.

In order to identify the range of values consistent with the conditions \eqref{avoid self PR} and  \eqref{phi cr} we Taylor expand $\V(\varphi)$ according to \eqref{TaylorV} and identify
\begin{equation}
    m_\phi=\frac{M^2}{\sqrt{3\alpha}M_{pl}} \ , \
    g_\phi = 0 \ , \
    \lambda_\phi=-\frac{4M^4}{9\alpha^2M^4_{pl}}~.
    \label{alphaTlambda}
\end{equation}
The quadratic term in 
\eqref{TaylorV}
dominates over the quartic term at the end of inflation if $\alpha>0.04$.
For $\alpha>0.25$ condition \eqref{avoid self PR} can be fulfilled. 
  Fig.~\ref{alpha T 4} shows the $n_s$-dependence of $N_{\rm re}$, $N_k$, $r$ and $T_{\rm re}$ for these values of $\alpha$.

\begin{figure}
    \centering
 	\includegraphics[width=0.7\linewidth]{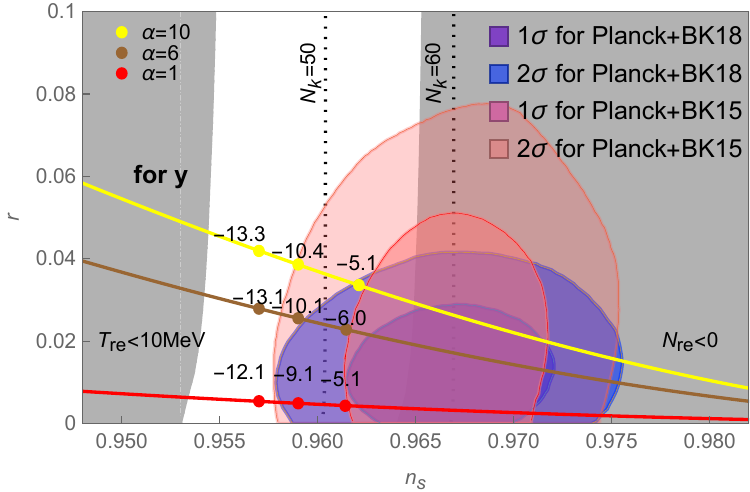}
 	\caption{
  Impact of the reheating phase on the predictions of $\alpha$-T model for $n_s$ and $r$ for a set of sample points and an Yukawa coupling $y$, with conventions as in Fig.~\ref{MHI R d}. Results for other couplings in table~\ref{ConversionTable} in Fig.~\ref{alpha T R} in the appendix.  }
	\label{alpha T R d}
\end{figure}

\begin{figure}
    \centering
 	\includegraphics[width=0.71\linewidth]{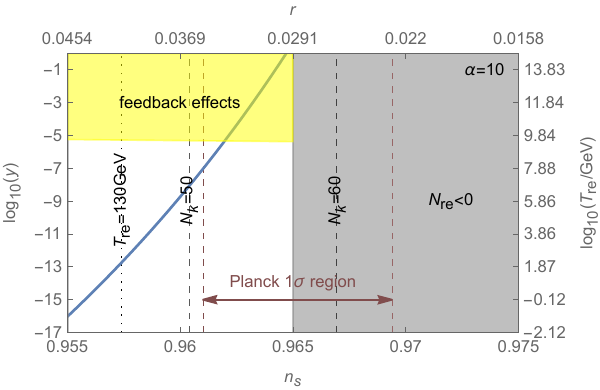}
 	\caption{
 	Dependence of the Yukawa coupling $y$ on $n_s$ and $r$ in the $\alpha$-T model for $\alpha=10$ with conventions as in Fig.~\ref{MHI g 1 a}. 
  Results for other couplings are shown in Figs.~\ref{alpha T g},
  \ref{alpha T h},  
  \ref{alpha T y},
  \ref{alpha T a} in the appendix.
  } 
		\label{alpha T 1 a}
\end{figure}

In our analysis $\alpha$ is chosen to be $0.5,~1,~10,$ and $100$. 
Fig~\ref{alpha T R d} compares the model predictions in the $n_s$-$r$ plane to the Planck+BICEP2/Keck 2015 and 2018 results~\cite{Planck:2018vyg,BICEP:2021xfz} ($1\sigma$ and $2\sigma$ CL). 
The corresponding values of ${\rm log}_{10}y$ are indicated by the numbers over the discs, and different choices of $\alpha$ distinguished by colour of the discs, following the notation of Fig.~\ref{MHI R d}.
We find that observational data slightly disfavours  values of the inflaton coupling where conditions \eqref{YippieYaYaySchweinebacke} are fulfilled.
Finally, the 
dependence of 
the inflaton couplings on $n_s$ is shown in Fig.~\ref{alpha T 1 a}.

\section{Methods for estimating the LiteBIRD and CMB-S4 sensitivities}\label{Sec:Methods}

In the previous section we 
studied the relations between the 
fundamental model parameters $\{M,\alpha,\g\}$,
the physical scales $\{M,m_\phi,\Treh\}$ and the 
properties of the observable CMB perturbations characterised by $\{n_s,A_s,r\}$.
In the following sections we will investigate how much information about the fundamental parameters and physical scales can be gained with observations of LiteBIRD and CMB-S4. 
Our task is straightforwardly phrased in the language of Bayesian statistics. We have some prior knowledge on a set of fundamental parameters $X$ from observation and theoretical consistency, specified in Sec.~\ref{Sec:CurrentKnowledge}, that we can use to define a prior $P(X)$. 
This knowledge will be confronted with new data $\mathcal{D}$, which we can characterise by a likelihood function $P(\mathcal{D}|X)$. Then the posterior distribution for $X$ is given by 
\begin{eqnarray}\label{BayesFormula}
P(X|\mathcal{D})=P(\mathcal{D}|X)P(X)/P(\mathcal{D}),
\end{eqnarray}
with $P(\mathcal{D})=\int dX P(\mathcal{D}|X)P(X)$.  
In principle the set of parameters that we are interested in is $X=\{M,\alpha,\g\}$, but in practice we will always fix $\alpha$.
In the following we use two different methods to evaluate \eqref{BayesFormula}. 
\begin{itemize}
\item In section \ref{Sec:SemiAnalyticMethod} we use the analytic method introduced in \cite{Drewes:2022nhu}.
In this approach $A_s$,
 $n_s$ and $r$ are considered observables, i.e., $\mathcal{D}=\{n_s,A_s,r\}$.
\item In section~\ref{Sec:ForecastMethod} we perform full-fledged forecasts in which $\mathcal{D}$ is a set of generated CMB mock-spectra for given true values of $\g$ and $M$. In addition, the set $X$ then contains the parameters of the concordance model of cosmology, cf.~\eqref{eq:free_parameters}.
\end{itemize}
Before applying these methods we need to specify the sub-classes of models we want to consider, in particular the choice of $\alpha$ in each of the categories MHI, RGI and $\alpha$-T models. 


\subsection{Knowledge prior to observation}\label{Sec:CurrentKnowledge}
 In order to quantify the knowledge gain that can be achieved with LiteBIRD or CMB-S4, we must first quantify the current knowledge, which is needed to define a Bayesian prior. We consider three types of information
\paragraph{Model choice and parametrisation.} 
Even before an explicit prior $P(X)$ is imposed, the parametrisation of $\V$ in terms of the $\{\vv_i\}\subset X$ induces a metric in the mapping between $\mathcal{D}$ and $X$, and $P(\mathcal{D}|X)$ in general is not invariant under changes in the parametrisation. While this does not represent a Bayesian prior in the strict sense, it does reflect prior assumptions about what the fundamental parameters in nature are, and this has an impact on the posteriors.
We adapt the following approach.
\begin{itemize}
    \item For the potentials we use the common parameterisations \eqref{MHI V}, \eqref{RGI V}, \eqref{alpha V}. 
    \item We fix $\alpha$ following the procedure described in Sec.~\ref{sec:choosing_alpha}, in particular imposing \eqref{avoid self PR}.
    \item In Sec.~\ref{Sec:SemiAnalyticMethod} we express $M$ in terms of $\g$, cf.~footnote \ref{SimpleFunctionFootnote}. In Sec.~\ref{Sec:ForecastMethod} we use a flat prior in $M$.
    \item For the inflaton coupling we chose the Yukawa interaction $\g=y$ throughout the main part of this article and use a flat prior in
    \begin{eqnarray}\label{xDef}
\x \equiv \log_{10}y.
\end{eqnarray}
We do not impose \eqref{UBoot} because one can still extract $\Treh$ in the regime where this condition is not fulfilled (in which case $y$ is merely a proxy for $\Treh$).
\end{itemize}

    \paragraph{Overall geometry of the universe.} 
    One of the main motivations for cosmic inflation comes from the facts that the observable universe is homogeneous, isotropic, and spatially flat on large scales. 
    We emphasise that this piece of information does not come from the study of the spectrum of CMB perturbations, but can be inferred from the high level of isotropy of the CMB, which was known \emph{prior} to the missions that study the spectrum of perturbations.
Explaining this observational fact\footnote{Of course, one may argue that values of $N_k$ much smaller than \eqref{NkCondition} are strictly speaking not observationally excluded. In such scenarios inflation simply fails to do its job, and the homogeneity and isotropy of the CMB remain unexplained. We adapt the viewpoint that inflation as a paradigm draws is motivation from the requirement to solve these puzzles, making \eqref{NkCondition} an observational bound on the theory.
    } imposes a lower bound on the total duration of inflation, defined as $N_{\rm inf}=\ln(a_{\rm end}/a_{\rm ini})$ with $a$ the scale factor and the subscripts $_{\rm ini}$ and $_{\rm end}$ referring to values at the beginning and end of inflation, respectively.\footnote{The term \emph{duration of inflation} is somewhat inconsistently used for two different quantities in the literature, namely $\Nk$ and  $N_{\rm inf}$.} 
    The value of the lower bound on $\Ninf$ depends on the specific property that one wants to explain~\cite{Kolb:1990vq,Liddle:2000cg,Mukhanov:2005sc}. 
    \begin{itemize}
    \item One possibility is to demand that inflation must last long enough to explain the \emph{horizon problem}. 
    This requires that the entire current Hubble volume originates from a region that was in causal contact before inflation, implying that the comoving mode $\khor$  
      must have left the horizon during inflation (as otherwise the horizon would have been smaller than $2\pi/\khor$ at all times), i.e.,
     \begin{eqnarray}\label{DurationOfInflation}
         \Ninf > \Nhor \simeq \Nk + \log[10^4 /(2\pi/0.05)] \simeq \Nk  + 5,
     \end{eqnarray}
           with $\Nhor$ the number of $e$-folds between the horizon crossing of the mode $\khor$, $0.05$/Mpc our pivot scale and $\sim 10^4$ Mpc the current particle horizon). 
           The inequality \eqref{DurationOfInflation} is of course not an upper bound on $\Nk$, but a lower bound on $\Ninf$.
           The resulting constraint on the shape of the potential can always be fulfilled in the plateau-type models considered here by choosing an initial (pre-inflationary) value of $\varphi_{\rm ini} > \varphi_{\khor}$,  \footnote{The actual value of $\varphi_{\rm ini}$ cannot be fixed from knowledge of $\V(\varphi)$, but must be evaluated in the UV-complete theory in which the inflationary model is embedded, hence we treat it as a parameter that can be dialed freely here (with the rather weak constraint that $\rho_\varphi < M_{pl}^4$).} 
           as $\V(\varphi)$ becomes asymptotically flat for larger $\varphi$. 
   \item Another possibility is to demand that the amplitude of any pre-inflationary density perturbations shall be reduced sufficiently that they are not seen in the CMB. Assuming that this amplitude is $O[1]$ before inflation, this implies $\dot{a}_{\rm ini}/\dot{a}_{\rm CMB} < 10^{-5}$ \cite{Mukhanov:2005sc}. This ratio can be expressed as
   \begin{eqnarray}
       10^{-5} &>& \frac{\dot{a}_{\rm ini} \dot{a}_{\rm end}}{\dot{a}_{\rm end}\dot{a}_{\rm CMB}} 
       \simeq 
       \frac{a_{\rm ini}}{a_{\rm end}} 
       \frac{a_{\rm reh}}{a_{\rm CMB} } 
       \frac{a_{\rm end}}{a_{\rm reh}} 
       \frac{H_{\rm end}}{H_{\rm CMB}} 
       \simeq
       e^{-N_{\rm inf}}
       \frac{T_{\rm CMB}}{\Treh}  
       e^{-\Nreh}
       \sqrt{\frac{\rhoend}{\rho_{\rm CMB}}}    
   \end{eqnarray}
where we have used $\dot{a}=aH$ and $H_{\rm ini} \simeq H_{\rm end}$. Using \eqref{Tre} and $\wrehbar=0$ implies
\begin{eqnarray}\label{InhomogeneitiesAmplitude}
N_{\rm inf} > 33 + \frac{\ln g_*}{3} + \frac{1}{3}\ln\left(\frac{\Treh}{\rm GeV}\right) + \frac{2}{3}\ln\left(\frac{M}{\rm GeV}\right). 
\end{eqnarray}
For the fiducial values in Tab.~\ref{tab:priors} 
the relation \eqref{DurationOfInflation} implies that we are guaranteed to fulfil \eqref{InhomogeneitiesAmplitude} if 
\begin{eqnarray}\label{NkCondition}
    \Nk > \left\{
\begin{tabular}{c c}
$61$ & MHI\\
$61$ & RGI\\
$75$ & $\alpha$-T
\end{tabular}
    \right.
\end{eqnarray} 
The fact that \eqref{NkCondition} is generally not fulfilled for the fiducial values used in Sec.~\ref{ChorPasseMole} does not indicate a failure of the model, but simply means that the value of $\Nk$ inferred from observation does not guarantee that \eqref{InhomogeneitiesAmplitude} is fulfilled. 
    Even a value of $N_k$ below $50$ is in principle possible and would simply imply a lower value of $M$.\footnote{For lower $M$ (and hence lower $\rhoend$) the physical horizon grew less after inflation, hence a shorter inflationary period is needed to ensure that before inflation it was bigger than the scale corresponding to our current horizon.
    While values as low as $\Nk \sim 30$ are possible in some models (see e.g.~\cite{Daido:2017wwb}), in the scenarios under consideration here such a low value of $\Nk$ would be inconsistent with the observed $A_s$. 
    } 
\end{itemize}
Hence, no strict bound on $\Nk$ can be derived from the requirement to explain the overall geometry of the universe. 
We nevertheless illustrate the range of $\Nk$ in different parts of the parameter space by indicating the interval
   \begin{eqnarray}\label{PlanckNkRange}
        50 < N_k < 60
    \end{eqnarray}
in our plots. This interval is chosen to be consistent with the interval that the Planck collaboration uses to illustrate the uncertainty in model predictions due to the lack of knowledge about the reheating epoch, cf.~Fig.~8 in \cite{Planck:2018jri}.

\paragraph{Previous measurements of the CMB power spectrum.}
The power spectrum of CMB anisotropies has been studied by a number of probes, with the most recent one being the combined analysis of Planck, WMAP, and BICEP/Keck observations in \cite{BICEP:2021xfz}. This data was obtained from the same CMB that will be studied by CMB-S4 and LiteBIRD, hence it strictly speaking does not represent independent information. To avoid complications related to combining correlated data in a Bayesian analysis we refrain from including this information into our priors. Practically this has little consequences; 
though current CMB data already 
contain information about reheating for some models \cite{Martin:2014nya,Martin:2016oyk},
the bounds on $r$ and $n_s$ reported in \cite{BICEP:2021xfz} do not impose a relevant constraint on $\g$ or $\Treh$ in the models considered here.\footnote{
This has been quantified explicitly in terms of posteriors in \cite{Drewes:2022nhu}, and it is also evident from the limits quoted in \cite{Ellis:2021kad}. In fact, one can almost see it by eye when plugging the uncertainties reported in \cite{BICEP:2021xfz} into \eqref{Tre} and \eqref{GammaConstraint}, or when rescaling the results of earlier works in similar scenarios \cite{Ueno:2016dim,Nozari:2017rta,DiMarco:2017zek,Drewes:2017fmn,Maity:2018dgy,Rashidi:2018ois,German:2020cbw,Mishra:2021wkm} with the reduced error bars on $r$ and $n_s$.
}

\paragraph{Requirement to actually reheat the universe.}
Using \eqref{TR}, the lower bound \eqref{TreTBBN} on $\Treh$ from BBN can be converted into a lower bound on $\g$  \cite{Drewes:2019rxn},
\begin{eqnarray}\label{BBNsimple}
\g > g_*^{1/4} \frac{T_{\rm BBN}}{\sqrt{m_\phi M_{pl}}}\sqrt{\#}
\simeq
\frac{T_{\rm BBN}}{M_{pl}} \left(\frac{g_*}{A_s r}\right)^{1/4}\sqrt{\#}.
\end{eqnarray}
The second relation in \eqref{BBNsimple} is approximately valid for plateau models of inflation, as those considered in this work. 
Most baryogenesis scenarios and Dark Matter production mechanisms require temperatures considerably above $T_{\rm BBN}$, cf.~footnote \ref{BBNfootnote}.
However, given our ignorance about these processes, we impose the conservative bound $\Treh > T_{\rm BBN}$. A lower bound $\g$ from any other lower bound on $\Treh$ can be obtained trivially from \eqref{BBNsimple} by replacing $T_{\rm BBN}$ with the respective temperature.

\paragraph{Physical consistency.} Physical consistency obviously requires $\Nreh\geq 0$. Conceptually it may be more appropriate to view this as a condition that defines the boundaries of the range of  values for $\{M,\alpha,\g\}$ (and not as a prior, as it is not derived from any prior observation, but from physical consistency alone).


\subsection{Model class selection: choosing $\alpha$}
\label{sec:choosing_alpha}

If $\{n_s,A_s,r\}$ could be measured without uncertainties, then 
$M$, $\alpha$ and $\Nreh$ 
can be determined unambiguously 
in the class of models considered here. 
This would permit to uniquely identify the physical scales via a mapping
\begin{equation}
    \{n_s,A_s,r\} \leftrightarrow \{M,m_\phi,\Treh\},
\end{equation}
using the general formalism presented in Sec.~\ref{sec:TRconstraints} as well as the relations for each model given in Sec.~\ref{sec:Benchmark models and analytic estimates}. 
If the conditions in appendix \ref{Sec:FeedbaclEffects} are fulfilled, then the formalism summarised in Sec.~\ref{Sec:InflatonCoupling} would further permit to determine $\g$ unambiguously, i.e., there is a mapping
\begin{equation}
    \{n_s,A_s,r\} \leftrightarrow \{M,m_\phi,\Treh\} \leftrightarrow \{M,\alpha,\g\}.
\end{equation}
However, in practice one will not be able to determine $\{n_s,A_s,r\}$ exactly from data.  
One result that is common to all classes of models in Sec.~\ref{sec:Benchmark models and analytic estimates} is that current and near-future observational information on the quantities $\{A_s,n_s,r\}$ will not be sufficient to simultaneously fit all microphysical parameters $\{M,\alpha,\g\}$ because the error bar on $n_s$ is too large. This can e.g.~be seen by comparing Figs.~\ref{MHI ghy 1}, \ref{RGI ghy 1} and \ref{alpha T ghy 1} to Fig.~\ref{fig:my_label}.\footnote{It may be that the situation improves if further properties of the CMB spectra are taken into account, such as the running of $n_s$ or non-Gaussianities. In appendix \ref{app:Next-to-leading} we estimate that the CMB-S4 sensitivity to the running of $n_s$ will not be sufficient to considerably improve the measurement of $\g$ for fixed $\alpha$. We postpone a dedicated study of the possibility to constrain all three parameters $\{M,\alpha,\g\}$ from data to future work. } 
To proceed with the analysis we consider families of inflationary models with a given value of $\alpha$.\footnote{While this can be seen as a pragmatic step, it makes the parameter $\alpha$ fundamentally different from $x$ and $M$: while $x$ and $M$ are treated as free parameters within a given class of models, the choice of $\alpha$ defines this class. If one takes this viewpoint, then $\alpha$ should be fixed from theory, e.g., in a UV-completion such a string theory. At a practical level there is another difference: While we impose \eqref{avoid self PR} when choosing $\alpha$, we do not impose the analogue condition \eqref{UBoot} when choosing the range of $x$ in Sec.~\ref{Sec:Methods}. The reason for this is that the curves in Figs.~\ref{cp_MHI}-\ref{cp_alphaT} remain to be valid if one reads them as posteriors for $\Treh$ even  in the regime where their interpretation as posteriors for $x$ is questionable due to the violation of \eqref{UBoot}.}
Fixing $\alpha$ establishes a relation between the inflaton mass $m_\phi$ and the scale of inflation $M$ 
(cf.~\eqref{mutatedlamda}, \eqref{InflatonMassRGI}, \eqref{alphaTlambda}),
with the latter being the sole free parameter in the potential $\V(\varphi)$ in each family of models.
Since this fixes the inflaton mass as a function of $M$ (cf.~\eqref{mutatedlamda}, \eqref{InflatonMassRGI}, \eqref{alphaTlambda}), we practically determine $\{M,\g\}$ from data, and $m_\phi$ is then fixed by $M$ and the choice of $\alpha$. Further, fixing $\alpha$ also establishes a relation between $n_s$ and $r$, 
implying that the predictions for each family of models are restricted to a line in the $n_s$-$r$ plane (cf.~Figs.~\ref{MHI R d}, \ref{RGI R d} and \ref{alpha T R d}).
$M$ and $\g$ vary along this line, and the accuracy by which they can be constrained depends on the sensitivity of future experiments to $n_s$ and $r$. In practice the  information gain that can be achieved with CMB-S4 and LiteBIRD will primarily be owed to their sensitivity to $r$.\footnote{If one fixes $A_s$ to \eqref{AsBestFit} as done in \cite{Drewes:2022nhu}, then $M$ and $\g$ are uniquely fixed for given $r$. In practice the error bar on $A_s$ will not introduce a considerable uncertainty, as we will see in Fig.~\ref{cp_RGI}.}

Our primary goal is to quantify the information gain on $\Treh$ and $\g$ that can be obtained with LiteBIRD and CMB-S4.  
For our study, we select values of $\alpha$ for which the following conditions are fulfilled:
\begin{itemize}
    \item The line in the $n_s$-$r$ plane defined by this choice of $\alpha$ in a given family of models crosses the region favoured by current constraints.
    \item The conditions \eqref{avoid self PR} can be fulfilled.
    \item The resulting line in the $n_s$-$r$ plane is steep enough that a sensitivity $\sigma_r\sim 10^{-3}$ can be sufficient to determine the order of magnitude of $\g$. The is generally the case if $r\sim 10^{-2}$ or larger.
\end{itemize}
Our choice of values for $\alpha$ is summarised in table \ref{tab:priors} together with other parameters relevant for the method described in Sec.~\ref{Sec:ForecastMethod}.  In sec.~\ref{Sec:ForecastMethod} we explain our choice for the fiducial values of $x$ and $M$.

\begin{table}
\centering
\begin{tabular}{c||c|c|c|c|c|c}
model & $\alpha$ & $\bar{x}$ & prior range $\logy$ & $\bar{M} [M_{\text{pl}} ]$ & prior range $M [M_{\text{pl}}]$ &  $\bar{n}_s$\\
\hline
\hline
MHI & 3.4 & -6.32 & $ < 0.59 $ & 0.005193 & $\left[ 0.0050, 0.0054 \right]$ & 0.9640\\
\hline
  RGI   & 19 & -6.82 & $ < 0.58 $ & 0.005296  & $\left[0.0051 , 0.0055 \right]$ & 0.9669\\
 \hline 
  $\alpha$-T & 6 & -1.03 & $ < 0.60$ & 0.005182 & $\left[ 0.0050, 0.0054 \right]$ & 0.9641\\
\end{tabular}
\caption{Fiducial values of $x$ and $M$ ($\bar{x}$ and $\bar{M}$) and their prior ranges, for the different models under consideration. 
Note that the value of $\alpha$ is fixed in advance to define a class of models, following the procedure in Sec.~\ref{sec:choosing_alpha}, it  should therefore not be viewed as a fiducial value in the sense of Sec.~\ref{Sec:ForecastMethod}.
}
\label{tab:priors}
\end{table}


\subsection{A simple analytic method}\label{Sec:SemiAnalyticMethod}
\begin{figure}
    \centering
    \includegraphics[width=0.7\linewidth]{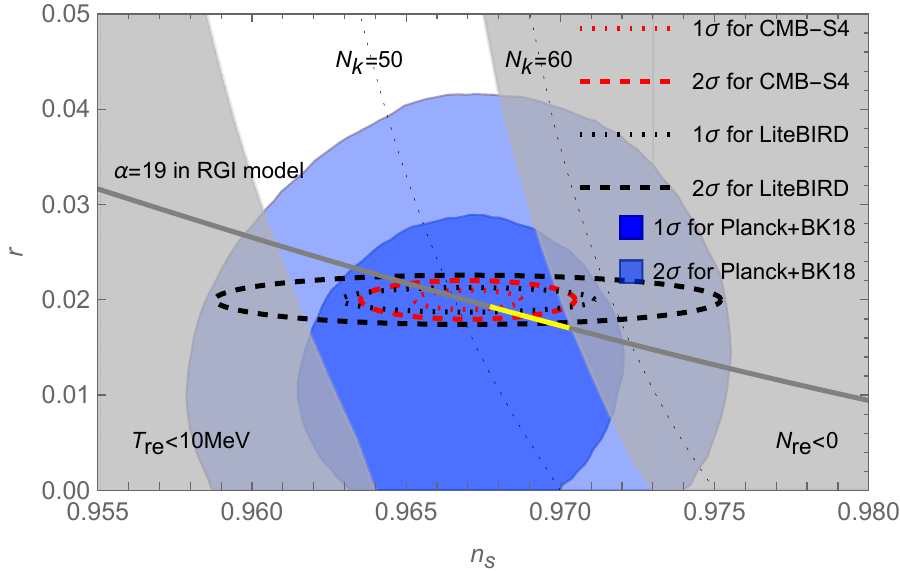}
    \caption{The ellipses indicate the likelihood  \eqref{GaussianLikelihood} as a function of $n_s$ and $r$, with $\bar{n}_s=0.967$, $\bar{r}=0.02$ and 
    $\sigmans=0.0017$, $\sigmar=0.0010$ for CMB-S4 and $\sigmans=0.0041$, $\sigmar=0.0013$ for LiteBIRD (found with the forecast method from Sec.~\ref{Sec:ForecastMethod} with $n_s$ and $r$ as free MCMC parameters).
    The diagonal line is given by \eqref{珍珠奶茶} with $\alpha=19$. The yellow part of this line corresponds to the region where  \eqref{CrocodileBatidaSpecial} is not fulfilled. All other conventions are same as in Fig.~\ref{MHI R d}. }
    \label{fig:my_label}
\end{figure}

In this section we briefly review the method introduced in \cite{Drewes:2022nhu}. 
In this approach the set of observables $\mathcal{D}$ in principle comprises $A_s$, $n_s$ and $r$. 
For illustrative purposes we keep $A_s$ fixed to \eqref{AsBestFit} and determine $M$ from \eqref{H_k}, cf.~\eqref{M1}, \eqref{M2} and \eqref{M3}. 
This is not strictly necessary for this method to be applied, but it conveniently 
reduces \eqref{BayesFormula} to a one-dimensional problem\footnote{\label{SimpleFunctionFootnote}
Using the procedure outlined after \eqref{TakaTukaUltras} one can express $N_k$ and $\Nreh$ and therefore $\{M,\alpha,\g\}$ in terms of $\{A_s,n_s,r\}$.  
For fixed $\alpha$, $n_s$ and $r$ are related by \eqref{TakaTukaUltras} (cf.~\eqref{珍珠奶茶}, \eqref{AlphaInAlphaT}). If $A_s$ is also fixed then $M$ and $\g$ (and also $N_k$ and $\Nreh$) are functions of $r$ alone.
The function $\g(r)$ turns out to be invertible in the observationally allowed range, which permits to express all other quantities ($M$, $N_k$,  $\Nreh$ etc.) in terms of $\g$ alone.
}
with  $X=x={\rm log}_{10}\g$ and $\mathcal{D}=\{n_s,r\}$. 
We a posteriori find that this simplification does not change our conclusions qualitatively, cf.~Fig.~\ref{cp_RGI}.
A prior probability density function 
for $\g$ (more precisely: $\x={\rm log}_{10}\g$) can be constructed from 
\begin{eqnarray}\label{PxdefNew}
P(\x)  
= c_1 
\ \gamma(\x) \ \theta\big(\Nreh(\x)\big) \ \theta\big(\Treh(\x) - T_{\rm BBN} \big),
\end{eqnarray}
with  
$\theta$ the Heaviside function,  
and $\gamma$ a function that allows for a re-weighting of the prior $P(\x)$. 
The constant $c_1$ can be fixed from the requirement
$\int d\x P(\x) = 1$.
To define a likelihood function we use a two-dimensional Gaussian in $n_s$ and $r$ with mean values 
$\nsbar$ and $\rbar$ and variances $\sigmans^2$ and $\sigmar^2$,
\begin{eqnarray}\label{GaussianLikelihood}
\mathcal{N}(n_s,r|\nsbar,\sigmans;\rbar,\sigmar)
=
\frac{1}{2\pi \sigmans\sigmar}{\rm exp}\left(
-\frac{1}{2}\left(\frac{n_s- \nsbar}{\sigmans}\right)^2\left(\frac{r - \rbar}{\sigmar}\right)^2
\right).
\end{eqnarray}
The Gaussian approximation appears to be justified by the shape of the exclusion regions reported by past CMB observations, such as \cite{BICEP:2021xfz}.
In \eqref{GaussianLikelihood} we have in addition assumed a diagonal covariance matrix. While this assumption is not exactly correct, it turns out that it does not lead to a significant loss of accuracy when it comes to the posterior for $\x$.
In \cite{Drewes:2022nhu}  $\sigmans$ and $\sigmar$ were obtained from 
the sensitivities quoted in the literature for both LiteBIRD \cite{LiteBIRD:2022cnt} and CMB-S4 \cite{CMB-S4:2020lpa}.  
This extraction is not trivial because these sensitivities depend on the true values of $n_s$ and $r$ (and also on the assumptions regarding delensing etc), and the complete information about the expected relations is not public. In the present work we obtain $\sigmans$ and $\sigmar$ by performing forecasts as described in Sec.~\ref{Sec:ForecastMethod} with $n_s$ and $r$ as free MCMC parameters. Assuming $\bar{n}_s=0.967$ and $\bar{r}=0.02$ as fiducial values we find $\sigmans=0.0041$ and $\sigmar=0.0013$ for LiteBIRD and $\sigmans=0.0017$, $\sigmar=0.0010$ for CMB-S4 (assuming no delensing). The corresponding contours in the $n_s$-$r$ plane are shown in Fig.~\ref{fig:my_label}. 
The complete likelihood function then reads (recall that $\mathcal{D}=\{n_s,r\}$ here)
\begin{eqnarray}\label{LikelihoodFunction}
P(\mathcal{D}|\x) = c_2\mathcal{N}(n_s(\x),r(\x)|\nsbar,\sigmans;\rbar,\sigmar)\theta(r)
\tilde{\gamma}(\x),
\end{eqnarray}
where $\tilde{\gamma}$ another weighting function and the constant $c_2$ is fixed by normalising $P(\mathcal{D}|\x)$ to unity.
In the following we use $\gamma=\tilde{\gamma}=1$ after checking that the conclusions remain unchanged when using $\gamma=\frac{d\Nk}{d\x}$ or $\tilde{\gamma}=((\frac{d n_s}{d\x})^{2} + (\frac{dr}{dx})^2)^{1/2}$.
The posterior for $\x$ is then obtained from \eqref{BayesFormula} with \eqref{PxdefNew} and \eqref{LikelihoodFunction},
\begin{eqnarray}
P(\x|\mathcal{D})= P(\mathcal{D}|x)P(x)/P(\mathcal{D}).
\end{eqnarray}
The resulting posteriors for the three  models under consideration can be seen in Fig.~\ref{cp_MHI} to \ref{cp_alphaT} together with the results from the next section.

\subsection{Forecast Method}\label{Sec:ForecastMethod}

We also perform a full MCMC based forecast to estimate the sensitivity of LiteBIRD and CMB-S4 on $x$ and $M$. The latter is expected to give a somewhat more realistic estimate of the sensitivities and at the same time serves as a comparison to the method introduced in the previous section. 

As explained in Sec.~\ref{sec:Benchmark models and analytic estimates}, the fundamental parameters $x$, $M$ and $\alpha$ can be mapped into the phenomenological parameters $A_s$, $n_s$ and $r$ which are conventionally used in the context of CMB studies. For the reasons discussed in Sec.~\ref{sec:choosing_alpha}, a simultaneous measurement of
all three parameters $\{M,\alpha,\g\}$ from $\{A_s,n_s,r\}$ 
will not be possible neither with LiteBIRD nor with CMB-S4. Therefore, we fix the value of $\alpha$ according to the criteria described at the end of Sec.~\ref{sec:choosing_alpha} while $x$ and $M$ are allowed to vary freely in our analysis. The corresponding choices for $\alpha$ for the different models considered in this work are summarised in Tab.~\ref{tab:priors}. 
This means that our analysis has in total 6 free MCMC parameters\footnote{We choose the expression \emph{free MCMC parameters} in order to avoid confusion with the free model   parameters of the Lagrangian (an expression that we occasionally used throughout Sec.~\ref{Sec:OlleKamellen} and \ref{sec:Benchmark models and analytic estimates}).}, i.e.  
\begin{equation}
X=   \lbrace \omega_{\mathrm{b}}, \omega_{\mathrm{cdm}}, 100 \theta_{\mathrm{s}}, \tau_{\mathrm{reio}} \rbrace  + x + M \, ,
    \label{eq:free_parameters}
\end{equation}
where the first 4 parameters are the standard $\Lambda$CDM base parameters, namely the baryon density $\omega_{\mathrm{b}}$, the Cold Dark Matter density $\omega_{\mathrm{cdm}}$, the sound horizon at recombination $\theta_{\mathrm{s}}$ and the optical depth to reionization $\tau_{\mathrm{reio}}$. 
We modified the linear Boltzmann-Einstein solver {\sc class} \cite{Blas:2011rf} (which calculates the CMB anisotropy spectrum) such that it takes $x$ and $M$ as free input parameters while the values of $A_s$, $n_s$ and $r$ are inferred from them. This requires to create numerical tables for $A_s$, $n_s$ and $r$ on a grid of $x$ and $M$ values. The step size both of $M$ and $x$ was chosen carefully such that it corresponds to less than $0.2$ of the expected standard deviations in $A_s$, $n_s$ and $r$. The range of the covered $x$ and $M$ values was chosen such that it covers around $10$ of the expected standard deviations in $A_s$, $n_s$ and $r$. We implemented a routine in {\sc class} that interpolates between the $x$ and $M$ values in this table such that the code can interpret any arbitrary value of $x$ and $M$ (inside the covered range)\footnote{We therefore modified and re-used the already existing 2-dim interpolation routine of CLASS that is usually used to derive the helium fraction $Y_p$ as a function of the baryon density $\omega_b$ and the effective number of relativistic degrees of freedom $N_{\mathrm{eff}}$.}. 

\begin{table}[]
    \centering
    \begin{tabular}{c|c|c|c}
        $\bar{\omega}_{\mathrm{b}}$ & $\bar{\omega}_{\mathrm{cdm}}$ & $100\bar{\theta}_{\mathrm{s}}$ & $\bar{\tau}_{\mathrm{reio}}$ \\
         \hline
        0.02237 & 0.1200 & 1.04092 & 0.0544
    \end{tabular}
    \caption{Fiducial values for the 4 base $\Lambda$CDM parameters.}
    \label{tab:fiducial_model}
\end{table}

In order to perform the forecast, we used the built-in forecast tool provided by the Monte Carlo Markov Chain (MCMC)  engine {\sc MontePython} \cite{Audren:2012wb,Brinckmann:2018cvx}. Mock likelihoods both for CMB-S4 and LiteBIRD are as well already built-in (originally developed for \cite{CORE:2016ymi}) and we here assume the same experimental specifications as in \cite{Brinckmann:2018owf}. Note that for CMB-S4 this implies that we assume a sky coverage of $f_{\text{sky}}=0.4$ whereas the latest CMB-S4 science book \cite{CMB-S4:2020lpa} assumes  $f_{\text{sky}}=0.01$. As explained in \cite{CMB-S4:2020lpa}, the optimal choice of $f_{\text{sky}}$ is not only subject to a complicated interplay between raw sensitivity, the ability to remove foregrounds and the ability to delense - but it also depends on the true value of $r$ which is assumed to be $\bar{r}=0$ throughout \cite{CMB-S4:2020lpa}. The CMB-S4 deep survey is therefore designed such that the sky coverage can be increased in case of a detection of $r$. Our choices for the true values of $x$ and $M$ (i.e. the \emph{fiducial} values $\bar{x}$ and $\bar{M}$) in Tab.~\ref{tab:priors} implies $\bar{r}\sim 0.02$, which justifies the choice of a larger sky fraction and we stick to  $f_{\mathrm{sky}}=0.4$ as in \cite{Brinckmann:2018owf}. At the same time, while for small values of $\bar{r}$ and $f_{\mathrm{sky}}$ the sensitivity on $r$ crucially depends on the success of delensing, successful delensing is not that crucial for $f_{\mathrm{sky}}=0.4$ and $r=0.02$ (see fig. 8 in \cite{CMB-S4:2016ple}). In our forecast, we assume both successful delensing\footnote{Note that the delensing procedure as applied in the creation of the mock-likelihoods in {\sc MontePython} is somewhat simplistic: It basically assumes the unlensed B-mode spectrum and adds noise to it. While this may not be the most realistic way to model delensing, we believe that it is sufficient for the purpose of this work.} as well as no delensing at all for the creation of the CMB-S4 mock likelihoods. 
For LiteBIRD in contrast, we make a conservative choice and assume no delensing due to its restricted delensing capabilities (which is also the assumption in \cite{CORE:2016ymi}). According to \cite{LiteBIRD:2023aov}, delensing is expected to improve the constraints on $r$ by about 20~\%. The built-in mock likelihoods in {\sc MontePython} for both experiments are based on the ideal assumption that systematic uncertainties are much smaller than statistical errors. Our analysis is therefore understood to concern noise sensitivities and neglect any foreground effects. 

As described in \cite{Brinckmann:2018cvx}, the {\sc MontePython} forecast method basically consists of 3 steps: 
\begin{itemize}
    \item[i)] Mock data for the fiducial model are created. The fiducial values for the 4 base parameters can be found in Tab.~\ref{tab:fiducial_model} and the fiducial values for $x$ and $M$ in Tab.~\ref{tab:priors}. The 
    values for $\bar{x}$ and $\bar{M}$ are chosen in a way that one can expect the resulting value of $A_s$ near \eqref{AsBestFit} and the values of $n_s$ and $r$ inside the region allowed by current observations, based on the formulae in Sec.~\ref{sec:Benchmark models and analytic estimates}.     It has been shown in \cite{Perotto:2006rj} that scattering of the mock data (due to cosmic variance) can be neglected.
    \item[ii)] The Fisher matrix and its inverse are derived. In case of Gaussian likelihoods, the standard deviations for the parameters in Eq.~\eqref{eq:free_parameters} are directly given by the square roots of the diagonal elements of the inverse Fisher matrix. However, as shown in \cite{Perotto:2006rj}, in general this does not hold for non-Gaussian likelihoods and in particular in presence of parameter degeneracies (as is expected for $x$ and $M$). We therefore proceed with iii).
    \item[iii)] MCMCs are produced, where the inverse Fisher matrix is used as a covariance matrix; the resulting chains are analyzed either with the built-in analysis tool of {\sc MontePython} or with the analysis tool GetDist \cite{Lewis:2019xzd} (which we used to create the posterior plots in figs. \ref{fig:triangle_MHI},\ref{fig:triangle_RGI} and \ref{fig:triangle_alphaT}). As discussed in Sec.~\ref{Sec:CurrentKnowledge}, the condition 
    $N_{\mathrm{re}} \geq 0$ imposes a hard cut on $x$ which can be found in table \ref{tab:priors}. 
    Note that the translation of the condition $\Nreh \geq 0$ 
    into a boundary for $x$ in general depends on the value of $M$ (which is varied freely in our analysis). We have however checked that the prior boundary of $x$ only varies mildly over the prior range of $M$, such that we can simply ignore this dependency in our analysis.
    The prior on $M$ in Tab.~\ref{tab:priors} is of practical nature and has no physical significance (and also does not affect our results), it simply reflects the range of $M$ covered by our numerical table. We assume flat priors for the 4 base parameters.
\end{itemize}

\section{Results and Discussion}\label{ChorPasseMole}

\begin{figure}
    \centering
    \includegraphics[width=\linewidth]{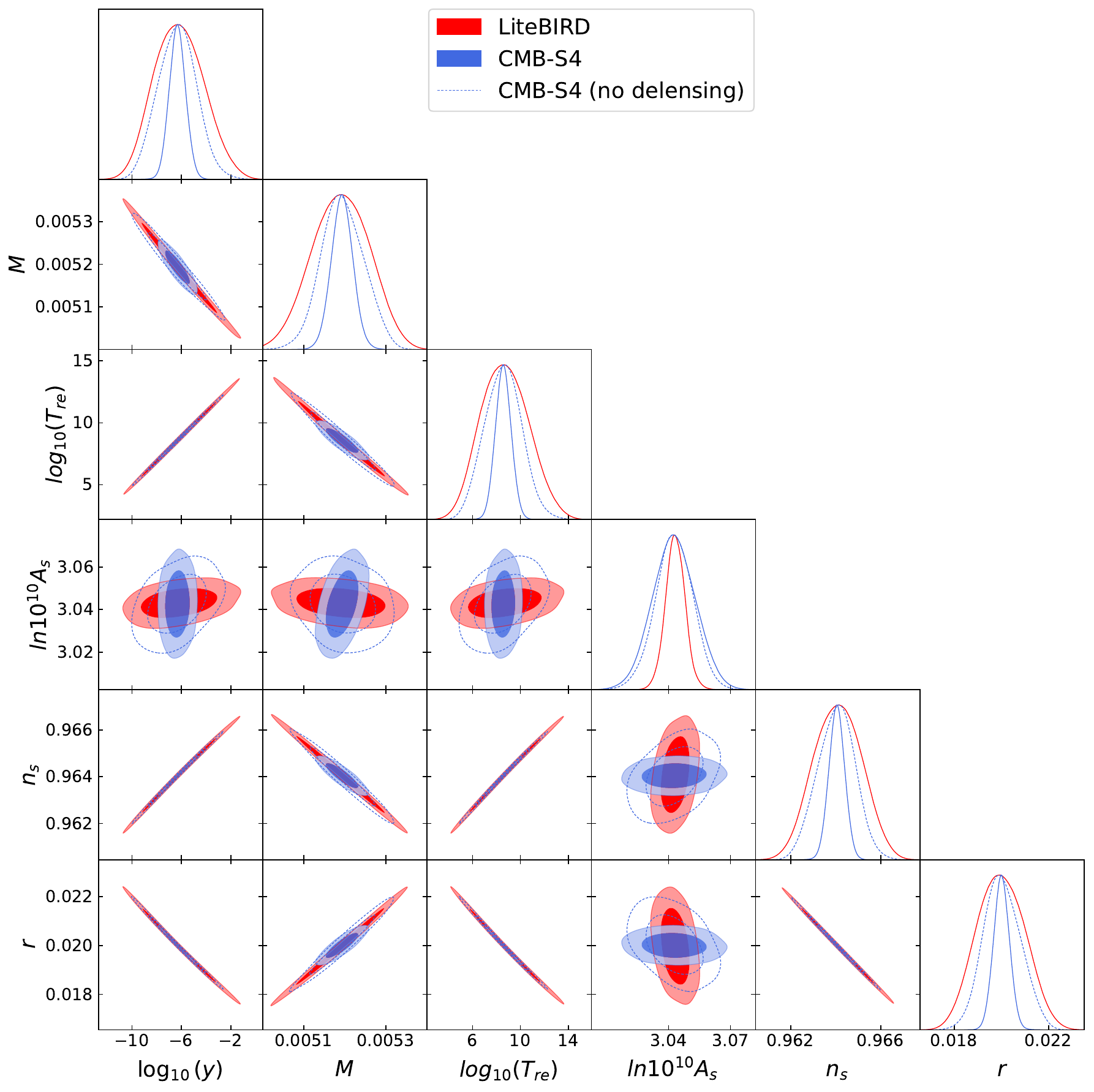}
    \caption{Marginalized 1d- and 2d-posterior distributions for $\logy$, $M$ and the derived parameters $\Treh$, $\ln(10^{10}A_s)$, $n_s$ and $r$ for the MHI model. }
    \label{fig:triangle_MHI}
\end{figure}

\begin{figure}
    \centering
    \includegraphics[width=\linewidth]{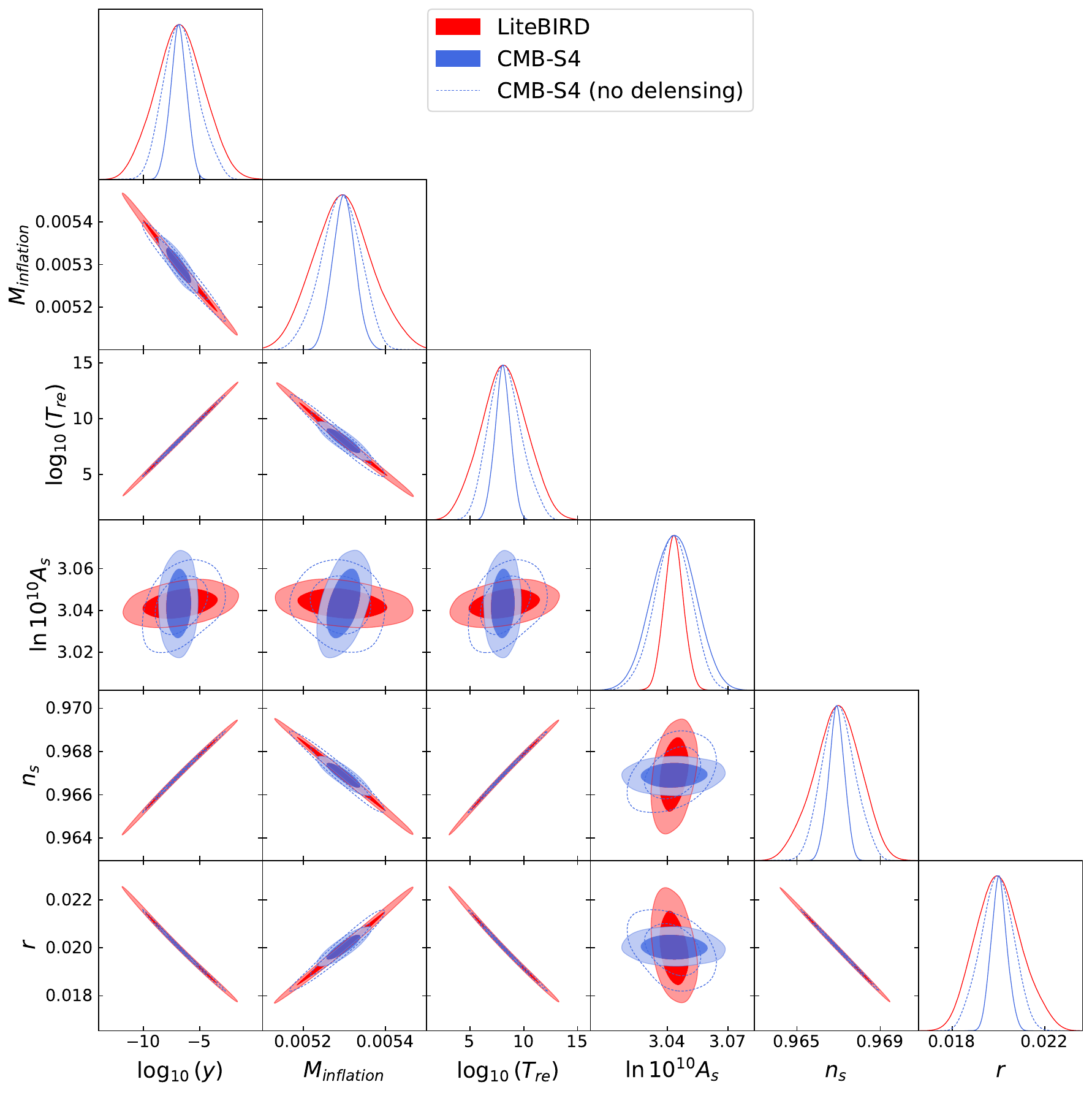}
    \caption{Same as Fig.~\ref{fig:triangle_MHI} but for the RGI model.}
    \label{fig:triangle_RGI}
\end{figure}

\begin{figure}
    \centering
    \includegraphics[width=\linewidth]{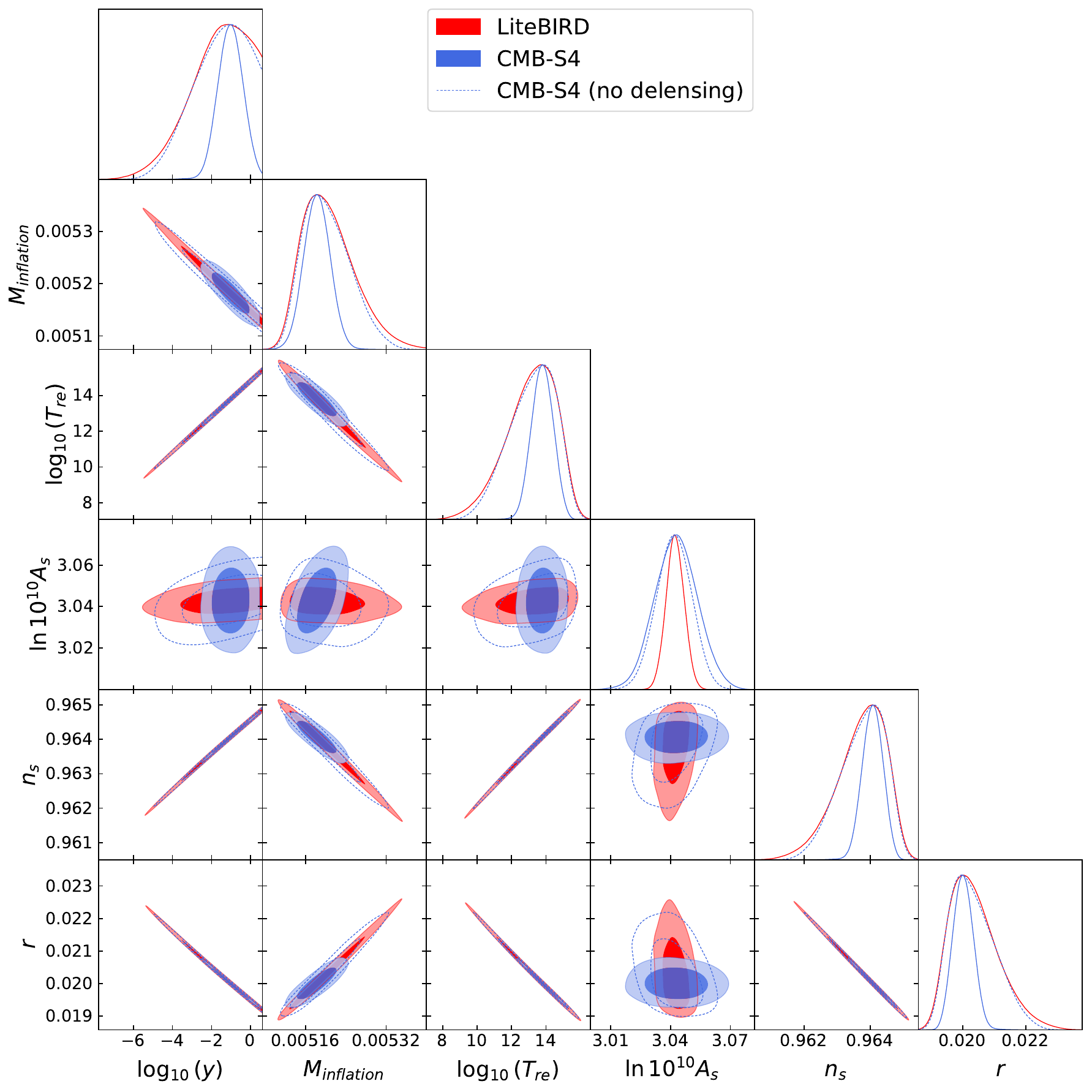}
    \caption{Same as Fig.~\ref{fig:triangle_MHI} but for the $\alpha-T$ model.}
    \label{fig:triangle_alphaT}
\end{figure}

The results of our forecast can be found in Fig.~\ref{fig:triangle_MHI} (MHI model), Fig.~\ref{fig:triangle_RGI} (RGI model) and Fig.~\ref{fig:triangle_alphaT} ($\alpha$-T) in form of 2d- and 1d- marginalized posterior distributions for $x$ and $M$. We also show the resulting posterior distributions for the \textit{derived} parameters ${\rm ln}(10^{10}A_s)$, $n_s$ and $r$, along with the posterior distribution for the reheating temperature $T_{\mathrm{re}}$. The reheating temperature can be calculated from inserting Eq.~\eqref{GammaPerturbativeGeneric} into Eq.~\eqref{TR},
\begin{equation}\label{TreInTermsOfX}
    T_{\rm re}=\left\{
    \begin{aligned}
    && y M \frac{1}{\sqrt{8\pi\alpha}}\left(\frac{90}{\pi^2g_*}\right)^{1/4} \quad {\rm for \ MHI}, \\
    &&y M \frac{1}{\sqrt{8\pi\sqrt{\alpha/2}}}\left(\frac{90}{\pi^2g_*}\right)^{1/4} \quad {\rm for \ RGI}, \\
    && y M \frac{1}{\sqrt{8\pi\sqrt{3\alpha}}}\left(\frac{90}{\pi^2g_*}\right)^{1/4} \quad {\rm for} \ \alpha-T.  
    \end{aligned}
    \right. 
\end{equation}
Figs.~\ref{fig:triangle_MHI}-\ref{fig:triangle_alphaT} show clearly that CMB-S4 as well as LiteBIRD are expected to simultaneously constrain the scale of inflation $M$ as well as the reheating temperature $\Treh$. 

In general, the figures show that CMB-S4 is expected to have higher foreground-free sensitivity on $x$ and $M$ than LiteBIRD -- to which extent however strongly depends on the success of delensing. The only parameter in Figs.~\ref{fig:triangle_MHI}, \ref{fig:triangle_RGI} and \ref{fig:triangle_alphaT} that is more constrained by LiteBIRD than by CMB-S4 is the scalar amplitude $A_s$. At scales below the reionization bump, $\ell \gtrsim 10$, there is a strong degeneracy between $A_s$ and the optical depth to reionization $\tau_{\mathrm{reio}}$.  
This degeneracy is removed in case of LiteBIRD (which is designed to measure exactly this reionization bump) but not in case of CMB-S4 (which will provide no measurement at these largest scales). It is worthwhile noting that due to the different scales observed by LiteBIRD and CMB-S4, the two experiments are considered to be complementary. It is therefore expected that combining data from
both will lead to an improved determination of $\Treh$ and $x$. Due to complications related to partially overlapping data sets, we however do not perform such a combined analysis.

\bgroup
\def\arraystretch{1.5}
\begin{table}
\centering
\begin{tabular}{c||c|c|c}
model & $x$ & $ M [M_{\text{pl}} ]$ & ${\rm log}_{10}(T_{\text{re}}[\text{GeV}])$\\
\hline
\hline
MHI & $-6.17\pm2.08$ & $0.00519\pm0.00007$ & $8.71\pm2.07$ \\
\hline
  RGI   & $-6.75\pm2.19$ & $0.00530\pm0.00007$ & $8.15\pm2.18$ \\
 \hline 
  $\alpha$-T & $-1.67\pm1.42$ & $0.00521\pm0.00005$ & $13.16\pm1.42$ \\
\end{tabular} 
\caption{Expected 1 $\sigma$ foreground-free sensitivities on the inflaton coupling $x$, $M$ and the reheating temperature $\Treh$
for LiteBIRD from the MCMC forecast described in sec.~\ref{Sec:ForecastMethod}. 
Here we considered a Yukawa coupling $y$, corresponding ranges for other types of interactions can be obtained with table \ref{ConversionTable}.}
  \label{tab:sensitivities_litebird}
\end{table}
\egroup

\bgroup
\def\arraystretch{1.5}
\begin{table}
\centering
\begin{tabular}{c||c|c|c}
model & 
$\logY$
& $ M [M_{\text{pl}} ]$ & ${\rm log}_{10}(T_{\text{re}} [\text{GeV}])$\\
\hline
\hline
MHI & \makecell{$-6.31\pm0.69$ \\ \;  $\left( -6.31\pm1.59 \right)$}& \makecell{$0.00519\pm0.00003$ \\ \; $\left( 0.00519\pm0.00005 \right)$ }& \makecell{$8.57\pm0.67$ \\ \; $\left( 8.57\pm1.58 \right)$ }\\
\hline
RGI  & \makecell{$-6.86\pm0.74$ \\ \; $\left( -6.64\pm1.55 \right)$ } & \makecell{$0.00530\pm0.00003$ \\ \; $\left( 0.00530\pm0.00005 \right)$ }&\makecell{ $8.04\pm0.74$ \\ \, $\left( 8.26\pm1.54 \right)$ }\\
\hline
$\alpha$-T & \makecell{$-1.04\pm0.64$ \\ \; $\left( -1.61\pm1.40 \right)$} & \makecell{$0.00518\pm0.00003$ \\ \; $\left( 0.00520\pm0.00005 \right)$ }& \makecell{$13.79\pm0.64$ \\ \; $\left( 13.22\pm1.40 \right)$ }\\
\end{tabular}
\caption{Expected 1 $\sigma$ foregound-free sensitivities on the inflaton coupling $x$, $M$ and the reheating temperature $\Treh$
for CMB-S4 from the MCMC forecast described in sec.~\ref{Sec:ForecastMethod}. The values in brackets are without delensing.
Here we considered a Yukawa coupling $y$, corresponding ranges for other types of interactions can be obtained with table \ref{ConversionTable}. 
}
  \label{tab:sensitivities_cmbs4}
\end{table}
\egroup

\begin{figure}
    \centering
    \includegraphics[width=0.7\linewidth]{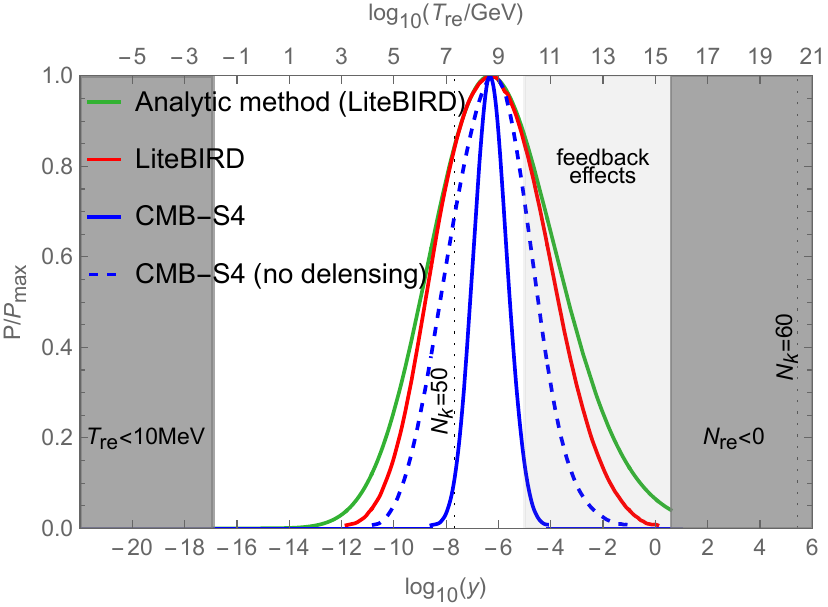}
    \caption{
    The green line is the result of the analytic method from Sec.~\ref{Sec:SemiAnalyticMethod} in the MHI model for LiteBIRD with fixed $A_s$. The red line is the forecast for LiteBIRD, the blue and blue-dashed lines correspond to the forecast for CMB-S4 with or without delensing, respectively (both without fixing $A_s$). 
    In the region labelled as \emph{feedback effects} the conditions outlined in appendix \ref{Sec:FeedbaclEffects} are not fulfilled, implying that the interpretation of the curves as posteriors for the microphysical coupling constant $y$ is questionable in this regime (while they can still be read as posteriors for $\Treh$). Here we assumed a Yukawa coupling $y$, posteriors  for other types of interactions can be obtained with table \ref{ConversionTable}.
    } 
    \label{cp_MHI}
\end{figure}

\begin{figure}
    \centering
    \includegraphics[width=0.7\linewidth]{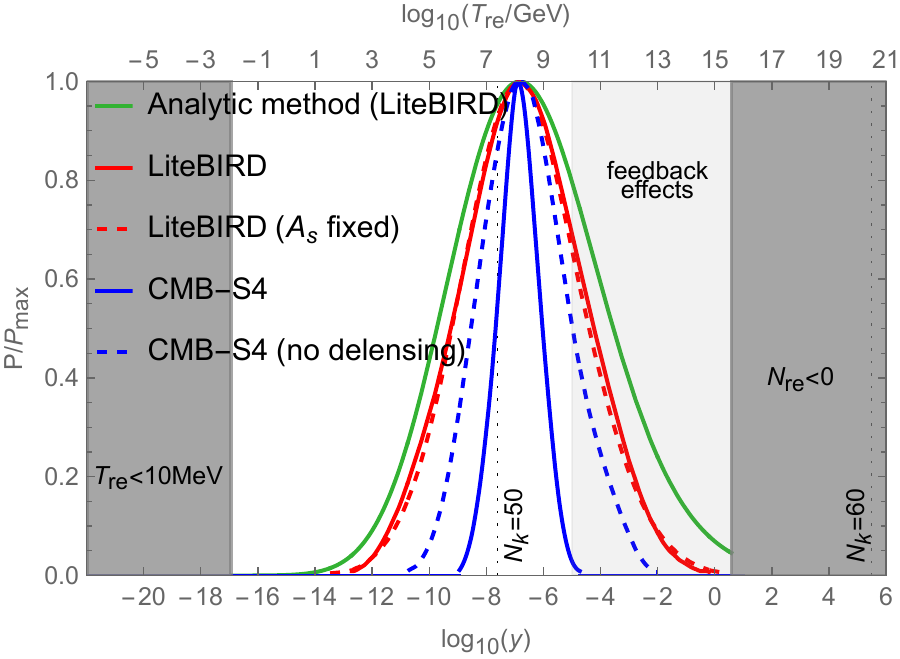}
    \caption{Posteriors for the RGI model with the same conventions as in Fig.~\ref{cp_MHI}.
    In addition to the analytic method with fixed $A_s$ and the forecasts with varying $A_s$, we also show the posterior from a LiteBIRD forecast in which $A_s$ was fixed (red-dashed line). This illustrates that the error bar on $A_s$ does not have a significant impact on the results.}
    \label{cp_RGI}
\end{figure}

\begin{figure}
    \centering
    \includegraphics[width=0.7\linewidth]{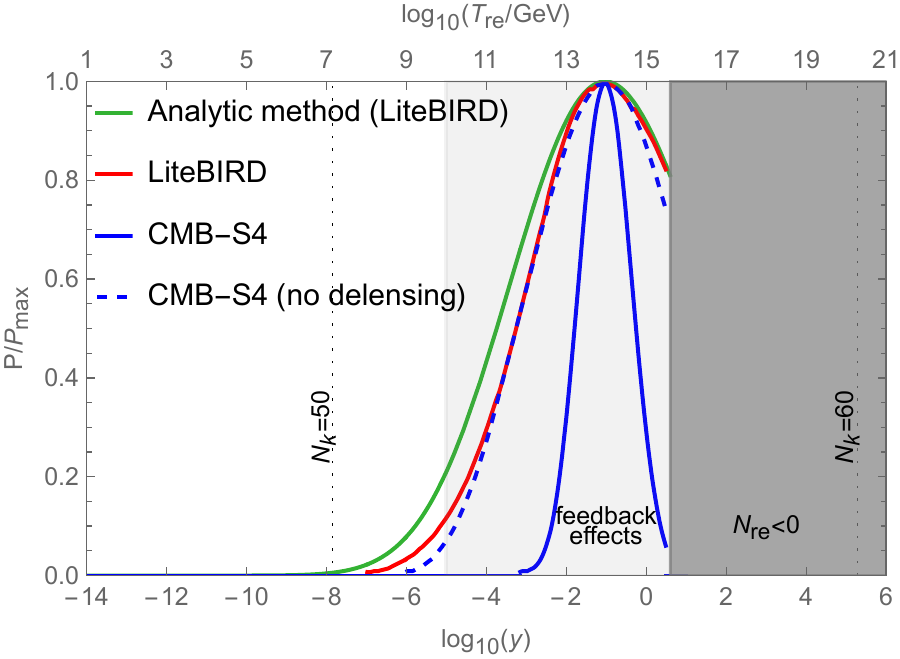}
    \caption{Posteriors for the $\alpha$-T model with the same conventions as in Fig.~\ref{cp_MHI}.
    The peak of the posterior lies in the region where feedback effects invalidate the interpretation in terms of a measurement of the microphysical coupling constant $y$, but one can still measure $\Treh$.}
    \label{cp_alphaT}
\end{figure}

We have imposed flat priors as outlined in Sec.~\ref{Sec:CurrentKnowledge} in order to produce Figs. \ref{fig:triangle_MHI}, \ref{fig:triangle_RGI} and \ref{fig:triangle_alphaT}. 
These figures therefore mainly reflect the experiments' pure foreground-free sensitivity towards $x$ and $M$. 
We also added enlarged figures for the 1-d posterior distribution of $x$ in Figs. \ref{cp_MHI}, \ref{cp_RGI} and \ref{cp_alphaT}. 
The top  x-axis displays the reheating temperature $\Treh$.
In general, $\Treh$ as well as the different prior conditions are all $M$ dependent. We however checked that -- given the very narrow allowed $M$ range -- this $M$ dependence is negligible and we can simply fix $M$ to its fiducial value in Tab.~\ref{tab:priors} in order to produce Figs. \ref{cp_MHI}, \ref{cp_RGI} and \ref{cp_alphaT}. 
Tab.~\ref{tab:sensitivities_litebird} displays 1 $\sigma$ foreground-free sensitivities for $x$, $M$ and $\Treh$. 
 They show that both CMB-S4 and LiteBIRD have the capacity to simultaneously measure the scale of inflation $M$ and the reheating temperature $\Treh$ within class of potentials characterised by a choice of $\alpha$.
For the parameters chosen here the order of magnitude of the inflaton coupling  can also be measured in the MHI and RGI models, as the peaks of the posteriors fall into the white regions in Figs.~\ref{cp_MHI} and \ref{cp_RGI}. 
In the $\alpha$-T model condition \eqref{YippieYaYaySchweinebacke} is violated for the fiducial value of the inflaton coupling $\bar{x}$ that we picked in order to bring the predictions for $n_s$ and $r$ into a regime that is allowed by current observations. Hence, the values for $\g$ given in tables~ \ref{tab:sensitivities_litebird} and \ref{tab:sensitivities_cmbs4} for the $\alpha$-T model cannot be interpreted as measurements of an actual microphysical parameter, but should rather be seen as proxies for $\Treh$.\footnote{\label{PerturbativeTModel}The discs in Fig.~\ref{alpha T R d} indicate that there are parameter choices in $\alpha$-T models for which \eqref{UBoot} 
can be fulfilled while keeping $r$ large enough to enable a measurement of $x$ with LiteBIRD or CMB-S4 and staying within the $2\sigma$ region preferred by current constraints. 
However, since the most popular particle physics of the potential \eqref{alpha V} feature multifield effects that trigger resonant particle production anyway, we decided to choose a fiducial value  $\bar{x}$ that leads to $n_s$ and $r$ inside the current $1\sigma$ region.}


Finally, in Figs.~\ref{cp_MHI}, \ref{cp_RGI} and \ref{cp_alphaT} we also compare the simple analytic method introduced in \cite{Drewes:2022nhu} and discussed in Sec.~\ref{Sec:SemiAnalyticMethod} to the forecast method introduced in Sec.~\ref{Sec:ForecastMethod} for the example of LiteBIRD. 
For the analytic method we therefore set  $\sigmans=0.0041$, $\rbar=0.02$ and $\sigmar=0.0013$ 
in \eqref{GaussianLikelihood} (as in Fig.~\ref{fig:my_label})
 while fixing $A_s$ to \eqref{AsBestFit}. As for the MCMC forecast, we choose the values of $\bar{x}$ and $\alpha$ as in Tab.~\ref{tab:priors}. In both cases we choose the values of $\nsbar$ as in Tab.~\ref{tab:priors}. The resulting posteriors are the green curves in Figs.~\ref{cp_MHI}, \ref{cp_RGI} and \ref{cp_alphaT}. Comparison to the results of Sec.~\ref{Sec:ForecastMethod} shows that the two methods are overall in satisfactory agreement, proving that the simple analytic method gives reliable results when being fed with realistic estimates of $\sigmans$ and $\sigmar$
for given $\nsbar$ and $\rbar$. For the RGI model and for the example of LiteBIRD we have also checked that fixing $A_s$ (as in the analytic method) would not have a significant impact on the results of the MCMC forecast: The red dashed curve in Fig.~\ref{cp_RGI} shows the $x$ posterior for a forecast with $M$ (or respectively $A_s$) and looks almost identical to the one with $M$ being varied freely. However, as can be seen in Fig.~\ref{cp_MHI}, \ref{cp_RGI} and \ref{cp_alphaT} the simple analytic method overestimates the confidence region of $x$ somewhat. This can be understood in the following way: In the analytic method the $n_s$-$r$ 2d-posterior is translated into a 1d-posterior for $x$ basically by inserting parametric expression for $n_s(x)$ and $r(x)$. In practice this method is therefore equivalent to treating $n_s$ and $r$ as free MCMC parameters while $x$ is a derived parameter. The forecasts described in Sec.~\ref{Sec:ForecastMethod} works the other way around, it assumes $x$ (and $M$ respectively) as free MCMC parameters (being the fundamental parameters of the theory) while $n_s$ and $r$ are derived parameters. Assuming $x$ to be the free MCMC parameter however only admits certain combinations of $n_s$ and $r$ values (for given $\alpha$) such that the accessible parameter space is reduced compared to the case of $n_s$ and $r$ as free MCMC parameters. This is ultimately reflected by a wider $x$ posterior for the analytic method.

Throughout this work, we assume a fiducial value of the tensor-to-scalar ratio $\bar{r}\sim0.02$ and we neglect foreground
contamination of the B-mode polarization spectrum.\footnote{
We note in passing that, while foreground estimates cannot be implemented straightforwardly in the software used in Sec.~\ref{Sec:ForecastMethod}, the analytic method used in Sec.~\ref{Sec:SemiAnalyticMethod} can in principle be used to estimate their impact on $\Treh$ and $\g$ if it is fed with estimates for $\sigmans$ and $\sigmar$ that take foregrounds into consideration. This was done in \cite{Drewes:2022nhu}, where estimates for $\sigmans$ and $\sigmar$ based on collaboration forecasts \cite{CMB-S4:2020lpa} were used. In the present work we do not do this because we aim to compare the two methods from Secs.~\ref{Sec:ForecastMethod} and~\ref{Sec:SemiAnalyticMethod} under equivalent assumptions.
}  
In general, smaller values of $\bar{r}$ imply smaller $\sigma_r$ \cite{CMB-S4:2020lpa} and consequently smaller $x$.  
However, the dependence of the couplings (e.g. $x$) on $r$ is extremely strong at small values of $r$ (as can be seen in e.g. Figs.~\ref{MHI R d},  \ref{MHI g 1 a} etc.), i.e., the couplings change by almost 10 orders of magnitude with $r$ changing by an amount of $\sim0.001$. 
We chose $\bar{r}=0.02$ to balance these two tendencies and optimise the sensitivity to $x$ for the purpose of a proof of principle.

\section{Conclusions}\label{Sec:Conclusion}

We studied the foreground-free sensitivity of LiteBIRD and CMB-S4 to  reheating in three classes of inflationary models, 
namely RGI, MHI and $\alpha$-T models.
In each of these classes of models we considered three fundamental parameters: The scale of inflation $M$, a parameter $\alpha$ that basically determines the relation between $M$ and the inflaton mass $m_\phi$, and a microphysical coupling constant $\g$ which controls the interaction of the inflaton with other fields. 
The latter sets the reheating temperature $\Treh$ because it determines the efficiency of the transfer from $\varphi$ to other degrees of freedom during reheating. 
While $M$ and $\alpha$ are parameters in the effective potential $\V(\varphi)$ that defines a model of inflation, 
gaining information on 
$\g$ probes the connection between a given model of inflation and its embedding into a more fundamental theory of particle physics. 
The relation between $\Treh$ and $\g$ 
is independent of further details of the underlying particle physics model (allowing for an agnostic measurement of $\g$) 
if feedback effects  do not considerably modify the duration of the reheating period.
We considered the four types of interactions summarised in table \ref{ConversionTable}, including scalar interactions, a Yukawa interaction and  axion-like coupling.
We find that LiteBIRD and CMB-S4 will be capable of the first ever measurements of 
the reheating temperature $\Treh$ and the inflaton coupling $\g$.

Our work comprises two separate parts. In Sec.~\ref{sec:Benchmark models and analytic estimates} we studied various dependencies between the fundamental parameters $\{M,\alpha,\g\}$ and other quantities, including the duration of 
inflation (characterised by $N_k$),
the duration of reheating $\Nreh$,
the physical scales $\{M,m_\phi,\Treh\}$, and the properties of CMB perturbations characterised by $\{A_s,n_s,r\}$. Many of these relations are given analytically in Sec.~\ref{sec:Benchmark models and analytic estimates}. We further illustrate them in the extensive set of plots given in appendix \ref{Sec:AppendixWithAllPlots}. 
The main  conclusions that can be drawn from this are:
\begin{itemize}
\item[1)] The error bar on $n_s$ will remain too large to fix all three parameters $\{M,\alpha,\g\}$ from data alone without further model-building assumption.
\item[2)] Fixing $\alpha$ defines families of inflationary models with one free parameter $M$ in the potential. In each of these families $M$ and the order of magnitude of $\Treh$ can, as a rule of thumb, be measured from a determination of $r$ at the level $\sigma_r\sim 10^{-3}$ if the true value of $r$ is around $10^{-2}$ or larger. 
This requires $\alpha>1$ in all models.
\item[3)] This can be translated into a measurement of the order of magnitude of the inflaton coupling $\g$ if the values of all involved coupling constants are smaller than a critical value that can be estimated by \eqref{UBoot}.
For the models under consideration here it can be expressed in terms of half-integer powers of $r A_s$ in \eqref{YippieYaYaySchweinebacke}. A  more general discussion is presented in appendix \ref{Sec:FeedbaclEffects}. 
\end{itemize}

In the second part of this work we investigated whether observations with LiteBIRD or CMB-S4 (both of which are expected to have a sensitivity around $\sigma_r\sim 10^{-3}$) will be able to impose meaningful constraints on 
the reheating temperature and the inflaton coupling. 
The main results of this part are presented in Figs.~\ref{fig:triangle_MHI}-\ref{fig:triangle_alphaT} and Figs.~\ref{cp_MHI}-\ref{cp_alphaT}, with the most important quantities summarised in tables \ref{tab:sensitivities_litebird} and \ref{tab:sensitivities_cmbs4}.
They were obtained by performing MCMC forecasts in which $M$ and ${\rm log}_{10}\g$ were treated as free MCMC parameters (along with the other parameters of the cosmological concordance model), as outlined in Sec.~\ref{Sec:ForecastMethod}. We thereby worked under the ideal assumption of negligible foreground contamination of the B-mode polarization spectrum.
The main conclusions are:
\begin{itemize}
\item[4)] Figs.~\ref{fig:triangle_MHI}-\ref{fig:triangle_alphaT} show that both CMB-S4 and LiteBIRD can measure the orders of magnitudes of $M$ and $\Treh$ in all models considered here. 
\item[5)] In the RGI and MHI  models this  permits to measure the order of magnitude of the inflaton coupling $\g$,  
as the peak of the posterior falls into the white regions in Figs.~\ref{cp_MHI} and \ref{cp_RGI}. 
In the $\alpha$-T model this is not possible  for our fiducial value, which was chosen to be in the $1\sigma$-region of current observations, but it may be possible for parameter values within the $2\sigma$-region,
cf.~footnote \ref{PerturbativeTModel}.
\item[6)] CMB-S4 tends to perform better 
in constraining $\Treh$ and $\g$ 
than LiteBIRD in all scenarios we considered. In addition to the moderately higher sensitivity to $r$, this is also related to the considerably better sensitivity to $n_s$.
However,  how large this advantage really is crucially depends on the ability to perform delensing. 
\item[7)] Since LiteBIRD is significantly more sensitive to $A_s$ while CMB-S4 is significantly more sensitive to $n_s$ and moderately more sensitive to $r$ (assuming successful delensing), combining data from both observations will lead to better constraints on $\g$. 
\end{itemize}
Further improvement would be possible if additional quantities are included (e.g.~the running of $n_s$ or non-Gaussianities), and if non-CMB data is added (e.g.~from galaxy surveys and 21cm observations).  
We estimate in appendix \ref{app:Next-to-leading} that measuring the running of $n_s$ will not change our conclusion qualitatively, but 
a more detailed study goes beyond the scope of this work and should be addressed with future studies.

Finally, we compared the  analytic method summarised in Sec.~\ref{Sec:SemiAnalyticMethod} to the forecasts described in Sec.~\ref{Sec:ForecastMethod}. 
We find that the analytic method gives a good estimate of the CMB-S4 and LiteBIRD sensitivities. 
 This is true even if $A_s$ is fixed, so that one practically performs a very simple one-dimensional application of Bayes formula.
 We illustrate this result for the RGI model in Fig.~\ref{cp_RGI}. The agreement between the methods is not specific to the RGI model.
 \begin{itemize}
\item[8)] We conclude that the analytic method from Sec.~\ref{Sec:SemiAnalyticMethod} provides a very useful tool to estimate the sensitivity of future missions to the reheating temperature and the inflaton coupling, provided that  information on the expected uncertainties $\sigmans$ and $\sigmar$ as functions of the true values of $r$ and $n_s$ is available.
\end{itemize}

To summarise, we studied the relations between fundamental model parameters, the physics of the reheating epoch, and CMB observables in plateau-type models of inflation. We performed the first MCMC forecasts to show that LiteBIRD and CMB-S4 can simultaneously measure the scale of inflation and the order of magnitude of the reheating temperature in a given class of models if the true value of $r$ is at least around $10^{-2}$, and assuming a standard radiation-dominated expansion history between reheating and BBN.
Our results confirm earlier findings obtained with the analytic method in \cite{Drewes:2022nhu}.
This would be the first ever measurement of 
$\Treh$
(in the sense that both an upper and lower bound can be obtained). 
For a sizeable fraction of the parameter space in the models considered here this can be translated into a measurement of 
the inflaton coupling $\g$.  
 Obtaining any information on this microphysical parameter will represent an important achievement for both cosmology and particle physics, 
 as it can provide a clue to understand how a given model of inflation can be embedded into a more fundamental theory of nature.
 This perspective clearly adds weight to the science cases for LiteBIRD and CMB-S4. It also motivates further theoretical work to investigate which observables in addition to $\{ A_s, n_s, r\}$ can be used to constrain the inflaton coupling. 

\subsection*{Acknowledgments}
We thank J. Hamann and Y. Y. Y. Wong for helpful discussions on various aspects of this work. We furthermore thank T. Brinckmann for his help with {\sc MontePython}. 
MaD thanks Eiichiro Komatsu for helpful input during the early stage of this project, 
Antonio M.~Soares for cross-checking the illustrative computation in Sec.~\ref{ChaosAD}, and Jin U Kang for many contributions to previous projects that made this line of research possible.
MaD would also like to thank Yann Mambrini for the invitation to the 
AstroParticle Symposium, and him as well as 
Simon Clery,
Marcos Garcia,
Mathieu Gross,
Kunio Kaneta and
Keith Olive
for the discussions there, which helped improving appendix \ref{Sec:FeedbaclEffects} in the revised version of this manuscript.
Computational resources have been provided by the supercomputing facilities of the Université catholique de Louvain (CISM/UCL) and the Consortium des Équipements de Calcul Intensif en Fédération Wallonie Bruxelles (CÉCI) funded by the Fond de la Recherche Scientifique de Belgique (F.R.S.-FNRS) under convention 2.5020.11 and by the Walloon Region. L.M. acknowledges the State Scholarship Fund managed
by the China Scholarship Council (CSC) and the Project funded by China Postdoctoral Science Foundation (2022M723677). I. M. O. acknowledges support by Fonds de la
recherche scientifique (FRS-FNRS).

\begin{appendix}
\section{Pivot scale for $r$}\label{Sec:PivotScale}
 While we use $k_p=0.05$ Mpc$^{-1}$ throughout this work as our pivot scale, in Sec.~\ref{sec:Benchmark models and analytic estimates} we apply the results from BICEP~\cite{BICEP:2021xfz} who instead choose $k_p=0.002$ Mpc$^{-1}$.
In this brief appendix we, for the sake of full transparency, explicitly write out the derivation of the relation \eqref{eq:r_vs_r_002}  that we used to re-scale the bounds presented in~\cite{BICEP:2021xfz} for several of our figures (namely~\ref{FLUSH}, \ref{MHI R d}, \ref{RGI R d}, \ref{alpha T R d}, \ref{MHI R}, \ref{RGI R} and \ref{alpha T R}).

The primordial spectrum for scalar perturbations is well approximated as
\begin{equation}
    P_s(k)=A_s \left( \frac{k}{0.05 \, \text{Mpc}^{-1}} \right)^{n_s-1},
    \label{eq:primoridal_scalar}
\end{equation}
where we explicitly chose $k_p=0.05$ Mpc$^{-1}$ as the pivot scale. This choice of $k_p$ is also the standard convention of e.g. the Planck \cite{Planck:2018jri} and BICEP \cite{BICEP:2021xfz} collaborations. Note that -- by definition -- $A_s$ is the amplitude of the primordial spectrum at the pivot scale, i.e. $P_s(0.05)=A_s$, and as such depends on the choice of the pivot scale. 
The primordial spectrum of tensor perturbations can be written as
\begin{equation}
    P_t(k) = A_t \left( \frac{k}{0.05 \, \text{Mpc}^{-1}} \right)^{n_t} = r A_s \left( \frac{k}{0.05 \, \text{Mpc}^{-1}} \right)^{-r/8} ,
    \label{eq:primordial_tensor}
\end{equation}
where we applied the definition of the tensor-to-scalar ratio
\begin{equation}
    r \equiv \frac{P_t(0.05)}{P_s(0.05)} = \frac{A_t}{A_s}
\end{equation}
and the consistency relation for the tensor tilt $n_t=-r/8$. Just like $A_s$ and $A_t$ the tensor-to-scalar ratio depends as well on the choice of the pivot scale. 
Depending on the context, occasionally also the tensor-to-scalar ratio at the pivot scale $0.002$ Mpc$^{-1}$ is referred to and defined as
\begin{equation}\label{r002def}
    r_{0.002} \equiv \frac{P_t(0.002)}{P_s(0.002)}.  
\end{equation}
Inserting eqs. \eqref{eq:primoridal_scalar} and \eqref{eq:primordial_tensor} gives us finally the following relation between $r$ and $r_{0.002}$
\begin{equation}
    r_{0.002}= r \left( \frac{0.002}{0.05} \right)^{-r/8 - n_s + 1}.
    \label{eq:r_vs_r_002}
\end{equation}

\section{Next-to-leading order corrections}
\label{app:Next-to-leading}
Throughout our analysis we worked at leading order in the slow roll parameters $\epsilon$ and $\eta$ defined in \eqref{slowrollpara}. 
The sensitivity of next-generation CMB observatories in principle requires the inclusion of second order corrections in these parameters. In this appendix we show that the leading-order results used in the main text are sufficient for our purpose. 
While third-order computations in principle exist \cite{Auclair:2022yxs}, we perform the explicit demonstration at second order. 
At this order a deviation from the power law \eqref{eq:primoridal_scalar} that can be characterised by the so-called running of the spectral index 
\begin{eqnarray}\label{RunToTheHills}
n_{\rm run}=\mathrm{d} n_s/\mathrm{d}\ln k. 
\end{eqnarray}
Moreover, a third slow-roll parameter is required, 
\begin{equation}
    \xi^2\equiv M^4_{pl}\frac{\partial_\varphi \V \partial^3_\varphi \V}{\V^2}.
\end{equation}

Truncating at the second order of slow-roll parameters, the observables $\{n_s,A_s,r,n_{\rm run}\}$ should be modified to be~\cite{Liddle:1994dx, Auclair:2022yxs, Planck:2013jfk}
 \begin{align}
    &n_s=1-6\epsilon_k+2\eta_k-\frac{1}{3}\left( 100+72C \right) \epsilon_k^2+(18+16C)\epsilon_k\eta_k+\frac{2}{3}\eta_k^2-\frac{1}{6}(11+12C)\xi_k^2,\label{nsAp}\\
    &A_s=\frac{H_k^4}{4\pi^2\dot{\varphi}^2_k}=\frac{H_k^4}{-8\pi^2M^2_{pl}\dot{H}_k}=\frac{H_k^2}{8\pi^2M^2_{pl}\epsilon_k}\left(1+\frac{4}{3}\epsilon_k-\frac{2}{3}\eta_k\right),\label{AsAp}\\
&r=16\epsilon_k+ \left( 64C-\frac{64}{3} \right) \epsilon_k^2+\left( \frac{32}{3}-32C \right) \epsilon_k\eta_k,\label{rAp}\\
    &n_{\rm run}=16\epsilon_k\eta_k-24\epsilon_k^2-2\xi_k^2,\label{nrunAp}
\end{align}   
where $C=\gamma_E+\ln2-2$ with $\gamma_E$ being the Euler-Mascheroni constant. 
From the Friedmann equation \eqref{Friedmann} with $\rho=\dot{\varphi}^2/2+\V\left(\varphi\right)$ and \eqref{EOM} 
one 
(by substituting the latter into the derivative of the former) 
finds $\dot{\varphi}=-\sqrt{-2M^2_{pl}\dot{H}}$,
where the minus sign appears because we chose $\varphi >0$.
Then \eqref{Nk} becomes
\begin{equation}
\label{Nkk}
\begin{aligned}
       N_k=&\ln\left(\frac{a_{\rm end}}{a_k}\right)=\int_{\varphi_k}^{\varphi_{\rm end}}\frac{H d\varphi}{\dot{\varphi}}=\int_{\varphi_{\rm end}}^{\varphi_k}d\varphi\sqrt{\frac{H^2}{-2M^2_{pl}\dot{H}}}\\ 
       =&\int_{\varphi_{\rm end}}^{\varphi_k}\frac{d\varphi}{\sqrt{2M^2_{pl}\epsilon}}\left(1+\frac{2}{3}\epsilon-\frac{1}{3}\eta\right).
\end{aligned}
\end{equation}
With the help of \eqref{AsAp}, \eqref{H_k} is corrected to be 
\begin{equation}
    \label{Hk1}H_k^2 = 
    8\pi^2M^2_{pl}A_s\left(\epsilon_k-\frac{4}{3}\epsilon_k^2+\frac{2}{3}\epsilon_k\eta_k\right)
\end{equation}
and 
\begin{equation}\label{Hk2}
    H^2 = 
    \frac{\V(\varphi)}{3M^2_{pl}}\left(1+\frac{1}{3}\epsilon-\frac{1}{3}\epsilon^2+\frac{2}{9}\epsilon\eta\right),
\end{equation}
which modifies the dependence of the normalisation $M$ of $\V(\varphi)$ on $A_s$ and $\varphi_k$.

Furthermore, we should also note that $\epsilon|_{\varphi_{\rm end}}=1$ is only a first-order result for the definition of end of inflation. The true endpoint is defined by $\ddot{a}=0$ which can be translated into $2M^2_{pl}(\partial_\varphi H)^2=H^2$, and inserting \eqref{Hk2} will give the corrected expression of $\varphi_{\rm end}$ at second order. One should also change \eqref{RhoEndApprox} to be 
\begin{equation}\label{rhoendnew}
    \rho_{\rm end} =  
    \V_{\rm end}\left(1+\frac{1}{3}\epsilon-\frac{1}{3}\epsilon^2+\frac{2}{9}\epsilon\eta\right)\Big|_{\varphi=\varphi_{\rm end}}.
\end{equation}

From \eqref{nsAp}, \eqref{rAp} and \eqref{Nkk}-\eqref{rhoendnew} we find that for a given set of $\lbrace M, \alpha, x \rbrace$ the difference between first- and second-order in slow-roll parameters in $\lbrace  A_s, n_s, r \rbrace$ is smaller than the expected sensitivities of LiteBIRD and CMB-S4 on these parameters. We have checked that this holds for the entire range of $\lbrace x, M \rbrace$-values covered in the numerical tables that were used for the MCMC forecast in sec.~\ref{Sec:ForecastMethod}. In other words, neglecting terms of second-order in the slow-roll parameters throughout the analysis in Sec.~\ref{Sec:Methods} is fully consistent. 
In the same manner we find that the values for $n_{\text{run}}$ induced by second-order terms in the slow-roll expansion are below the expected sensitivity on $n_{\text{run}}$, i.e., around $\sigma_{n_{\rm run}}=0.002-0.003$~\cite{CMB-S4:2016ple}. Again, this justifies the neglect of $n_{\text{run}}$ throughout this work, i.e., setting $n_{\text{run}}=0$. Note that this conclusion holds \emph{for the models considered in this work} but should not be generalized to other types of models of inflation. An interesting question for future work is to what degree a determination of $n_{\rm run}$ could break this degeneracy.

\section{Conditions for measuring the coupling constants}\label{Sec:FeedbaclEffects}
Reheating is in general a highly complicated nonequilibrium process (c.f.~\cite{Amin:2014eta,Lozanov:2019jxc}), making the computation of $\Gamma$ in terms of microphysical parameters challenging in practice.
When translating \eqref{GammaConstraint} into a meaningful bound on any individual microphysical parameter (such as the inflaton coupling $\g$),  one faces the problem that 
feedback effects introduce a dependence of $\GG$ on a large sub-set of the parameters $\{\SM_i\}$ in the underlying particle physics theory.  
Rather than being a mere problem of limited computational power, this is actually a fundamental restriction, as the dependence on the $\{\SM_i\}$ 
makes it impossible to determine $\g$ from the CMB without having to specify the details of the underlying particle physics model. 
It is, however, possible if reheating proceeds sufficiently slowly that feedback effects can be neglected. The conditions for this have been studies in detail in \cite{Drewes:2019rxn}, here we only quote the results.

The dominant source of $\{\SM_i\}$-dependence lies in the explosive resonant particle production due to non-perturbative effects, sometimes referred to as preheating. 
Preheating leads to very large occupation numbers for the produced particles. The interactions of $\varphi$ with the produced particles and their interactions amongst each other then affect the effective $\GG$. This introduces a dependence of $\GG$ on the properties of the produced particles, and hence the $\{\SM_i\}$. 
In addition to that there are also thermal corrections to the elementary decay rate \eqref{GammaPerturbativeGeneric}, which bring in a dependence on the $\{\SM_i\}$ (e.g.~through the thermal masses of the decay products), cf.~table \ref{ThermalRates}. 
There are two different types of conditions to ensure that such feedback effects do not modify $\Nreh$.

\begin{table}
\begin{small}
\begin{tabular}{c c c c}
 interaction &  process &  contribution to $\GG$ \\
 \hline 
$g\Phi\chi^2$ & $\varphi\to\chi\chi$ &  
$\frac{g^2}{8\pi \M_\phi}\big(1-2\M_\chi/\M_\phi\big)^{1/2}\big(1+2f_B(\M_\phi/2)\big)\theta(\M_\phi-2\M_\chi)$
&   \cite{Boyanovsky:2004dj}\\
$\frac{\hh}{4}\Phi^2\chi^2$ &  $\varphi\varphi\to\chi\chi$ &  
$\frac{h^2\varphi^2}{256\pi \M_\phi}\big(1-\M_\chi/\M_\phi\big)^{1/2}\big(1+2f_B(\M_\phi)\big)\theta(\M_\phi-\M_\chi)$  
& \cite{Mukaida:2013xxa}\\
$\frac{\upalphatosigma}{\Lambda}\Phi F_{\mu\nu}\tilde{F}^{\mu\nu}$ &  $\varphi\to\gamma\gamma$ &  
$\frac{\upalphatosigma^2}{4\pi}\frac{\M_\phi^3}{\Lambda^2}\big(1-2M_\gamma/\M_\phi\big)^{1/2}\big(1+2f_B(\M_\phi/2)\big)\theta(\M_\phi-2M_\gamma)$  
& \cite{Carenza:2019vzg}\\
$y\Phi\bar{\psi}\psi$ &  
\begin{tabular}{c} 
$\varphi\to\psi\bar{\psi}$,\\ 
$\corr{M_\psi \simeq m_\psi}$ 
\end{tabular}
&
$\frac{y^2}{8\pi}\M_\phi\big(1-2m_\psi/\M_\phi\big)^{3/2}\big(1-2f_F(\M_\phi/2)\big)\theta(\M_\phi-2m_\psi)$  & \cite{Drewes:2013iaa}\\
 \ & 
 \begin{tabular}{c} 
$\varphi\to\psi\bar{\psi}$,\\ 
\corr{$M_\psi \gg m_\psi$}
\end{tabular}
 & 
 $\frac{y^2}{8\pi}\M_\phi\big(1-2M_\psi/\M_\phi\big)^{1/2}\big(1-2f_F(\M_\phi/2)\big)\theta(\M_\phi-2M_\psi)$ 
& 
\cite{Drewes:2013iaa}\\
\end{tabular}.
\end{small}
\caption{\label{ThermalRates}
Contributions to $\GG$ from different elementary processes including thermal corrections 
mediated by some of the interactions in table \ref{ConversionTable}. 
For the $\frac{\hh}{4}\Phi^2\chi^2$ interaction only $\varphi$-annihilations are included, contributions from interactions between the condensate $\varphi$ and its own quanta $\phi$ have been considered in \cite{Ai:2023ahr}. 
This interaction  (which leads to non-linearites that can limit the validity of \eqref{EOM} \cite{Ai:2021gtg})  cannot effectively reheat the universe in the regime where the condition \eqref{CrocodileBatidaSpecial} is fulfilled \cite{Drewes:2019rxn,Garcia:2020wiy,Ai:2023ahr}.
Small letters and capital letters denote vacuum masses and effective thermal masses, respectively. 
In practice the quantum-statistical corrections involving the 
Bose-Einstein and  Fermi-Dirac  distributions
and $f_F(\omega)=(e^{\omega/T} + 1)^{-1}$ and
$f_B(\omega)=(e^{\omega/T} - 1)^{-1}$    
can be neglected, as $T\ll M_\phi$ in the regime where \eqref{GeneralConstraintSelf} and \eqref{CrocodileBatidaSpecial} are fulfilled, so that the use of \eqref{GammaPerturbativeGeneric} is justified.
This simplifies the analysis, as one can easily convert the constraints on one coupling $\g$ in constraints in any of the others with table \ref{ConversionTable}.
Thermal corrections to the $\hchi\Phi\chi^3/3!$ interaction have been considered in \cite{Drewes:2013iaa,Drewes:2015eoa}, but no closed analytic form is known. 
For the other interactions it is easy to see that the $T\to 0$ limit reproduces the rate in table \ref{ConversionTable}.
}
\end{table}

\paragraph{Conditions on the inflaton potential.} The first type concerns the inflaton self-interactions $\{\vv_i\}$ and thereby restricts the choice of the model of inflation (specified by a choice of $\V(\varphi)$). 
For the models considered here the minimum of $\V(\varphi)$ is at $\varphi=0$, so that we can Taylor expand
\begin{eqnarray}\label{TaylorVeff}
\V(\varphi) = \sum_\n
    \frac{\vv_{\n}}{\n !} \frac{\varphi^\n}{\Lambda^{\n-4}} 
\end{eqnarray}
with $\vv_3 = g_\phi/\Lambda $ and $\vv_4 = \lambda_\phi$ in \eqref{TaylorV}, where $\Lambda$ is a mass scale that can be chosen for convenience.
If one neglects radiative corrections, the coefficients in this expansion can be identified with the inflaton mass and coupling constants of the self-interactions in the expansion of the potential $V(\Phi)$ in the Lagrangian. 
The inflaton self-interactions can in general lead to parametric or tachyonic resonances well before elementary decays become relevant, invalidating the use of \eqref{GammaPerturbativeGeneric} in \eqref{GammaConstraint}.
This conversion of the coherent condensate $\varphi$ into its own fluctuations ($\phi$-particles) is sometimes referred to as fragmentation.
Early fragmentation can be avoided if \cite{Drewes:2019rxn}
\begin{eqnarray}
 |\vv_\n| \ll \left(\frac{\omega}{\varphi}\right)^{\n-\frac{5
}{2}}
{\rm min}\left(
\frac{\V^{1/4}}{\sqrt{M_{pl} \varphi}}
,
\sqrt{\frac{\omega}{\varphi}}
\right) \ 
\left(\frac{\omega}{\Lambda}\right)^{4-\n},\label{PreGeneralConstraintSelf}
\end{eqnarray}
with $\omega$ the frequency of the inflaton oscillations.
This condition can be fulfilled in approximately parabolic potentials in which the oscillating inflaton has a matter-like equation of state \eqref{EquationOfState}.\footnote{\label{MarcosFootnote}
The condition \eqref{GeneralConstraintSelf} can also be fulfilled for potentials $\V\propto \varphi^\n$ with $\n >4$ if one assumes a sufficiently large cutoff scale (e.g.~$\Lambda \sim M_{pl}$), as for $\Lambda > \varphi_{\rm end}$ it becomes weaker when increasing $\n$ beyond $4$. This was already noticed in \cite{Drewes:2019rxn} and has been explicitly verified in \cite{Garcia:2023dyf}.
The main issue then is that the inflaton mass then vanishes at tree-level, so it is not immediately clear what to use for $m_\phi$ in \eqref{GammaPerturbativeGeneric}. The results in \cite{Garcia:2023dyf} suggest that the use of a time-dependent effective inflaton mass given by $\partial_\varphi^2\V$ evaluated at the evolving value of $\varphi$ (and not at the minimum of $\V$) gives reasonable estimates. 
If this is confirmed, one may apply the approach used here to scenarios with large field elongations in potentials with $\n >4$, in which case one finds $\wrehbar \simeq (\n-2)/(\n+2)$ \cite{Turner:1983he}.
In the present work we restrict ourselves to the case with $m_\phi\neq0$ and 
mildly non-linear regime of field elongations where \eqref{EquationOfState} applies. 
}
Leaving aside sub-dominant non-linear corrections to the equation of motion \eqref{EOM},\footnote{Such corrections can in principle be treated by means of multiple-scale perturbation theory \cite{Ai:2021gtg}.} 
the frequency of these oscillations is given by the effective in-medium inflaton mass $\omega\simeq \M_\phi$.   
In principle $\M_\phi$ depends not only on the couplings to other particles $\{\g_i\}$, but also on their interactions  with each other, and hence on the $\{\SM_i\}$. However, \eqref{PreGeneralConstraintSelf} imposes the strongest constraint when $\varphi$ is maximal, i.e., at the beginning of the reheating phase, and we can replace $\varphi\to\varphi_{\rm end}$ in \eqref{PreGeneralConstraintSelf} to obtain a conservative bound. 
At that moment the universe is empty. Hence, the frequency of the inflaton oscillations is in good approximation given by its vacuum mass $\omega\simeq m_\phi$, leading to
\begin{eqnarray}
 |\vv_\n| \ll \left(\frac{m_\phi}{\varphi_{\rm end}}\right)^{\n-\frac{5
}{2}}
{\rm min}\left(
\sqrt{\frac{m_\phi}{M_{pl}}}
,
\sqrt{\frac{m_\phi}{\varphi_{\rm end}}}
\right) \ 
\left(\frac{m_\phi}{\Lambda}\right)^{4-\n},\label{GeneralConstraintSelf}
\end{eqnarray}
from which \eqref{avoid self PR} can be obtained.

\paragraph{Conditions on the couplings to other fields.}
Consider an interaction term of the form
\begin{eqnarray}\label{Operator}
    \g \Phi^\n \Lambda^{4-\DD}\mathcal{O}[\{\X_i\}],
\end{eqnarray}
where $\mathcal{O}[\{\X_i\}]$ is an operator of mass dimension $\DD-\n$
that is composed of fields $\X_i$ other than $\Phi$,
and $\g$ a dimensionless coupling constant. 
For $\DD>4$ the scale $\Lambda$ 
marks the cutoff of an effective field theory description in which \eqref{Operator} appears, and we assume that it exceeds all other relevant scales for this effective description to be valid.
For $\DD<4$ we set $\Lambda=m_\phi$ for convenience, in this case $\Lambda$ is simply introduced to make $\g$ dimensionless.
A term of the form \eqref{Operator} introduces a time-varying effective mass $\propto \g \varphi^\n$ for a field $\X$ at tree level if $\mathcal{O}[\{\X_i\}]$ is a bilinear in $\X$. Some examples of this kind are given in table \ref{ConversionTable}. For bosonic $\X$ this implies $\DD = \n+2$, for fermionic $\X$ it implies $\DD = \n + 3$. 
It is generally justified to evaluate the LHS in \eqref{GammaConstraint} with \eqref{GammaPerturbativeGeneric} if \cite{Drewes:2019rxn}
\begin{eqnarray}
|\g| \ll\left(\frac{m_\phi}{\varphi_{\rm end}}\right)^{\n-\frac{1}{2}}
{\rm min}\left(
\sqrt{\frac{m_\phi}{M_{pl}}}
,
\sqrt{\frac{m_\phi}{\varphi_{\rm end}}}
\right) \ 
\left(\frac{m_\phi}{\Lambda}\right)^{4-\DD}. \ \label{CrocodileBatidaSpecial} 
\end{eqnarray}
In \cite{Garcia:2021iag} reheating has been studied numerically for the $\frac{\hh}{4}\Phi^2\chi^2$ and $y\Phi\bar{\psi}\psi$ interactions, including non-perturbative processes. The results displayed in Figs.~2, 3 and 6 in that work are fully consistent with the analytic estimate \eqref{CrocodileBatidaSpecial} for the range of inflaton couplings for which the expansion history (i.e.~$\Nreh$) is unaffected by the feedback effects. 
Moreover, Fig.~4 in \cite{Garcia:2021iag} shows that even the maximal temperature during reheating remains unaffected from feedback for only slightly smaller couplings.\footnote{Thermal corrections were not included in that work. These have been computed in \cite{Drewes:2013iaa}, and their effect on the thermal and expansion history was studied in \cite{Drewes:2014pfa}, where it was found that they can modify the thermal history for rather small values of the coupling constants. However, the analysis in \cite{Drewes:2014pfa} neither treated non-perturbative particle production nor the thermalisation of the produced particles consistently and therefore should be revised before drawing conclusions.}  
The condition \eqref{CrocodileBatidaSpecial} imposes a strong upper bound on interactions that are non-linear in $\Phi$. 
For terms with $\DD>4$ the smallness of $(m_\phi/\varphi_{\rm end})^{\n-\frac{1}{2}}$ can be alleviated by the fact that $m_\phi /\Lambda \ll 1$, but it e.g.~represents a strong restriction on interactions of the type $\frac{\hh}{4}\Phi^2\chi^2$.
This is the reason why we have focused on operators with $\n=1$ in table \ref{ConversionTable}. 
For $\n=1$ and $\omega=m_\phi$ \eqref{CrocodileBatidaSpecial} reduces to \eqref{UBoot}.


\paragraph{Effects that can relax the conditions.} The conditions 
\eqref{avoid self PR} and \eqref{UBoot}
impose very strong constraints on interactions that are not linear in $\Phi$, i.e., with $\n>1$.
Luckily it turns out that the relations  \eqref{GeneralConstraintSelf} and \eqref{CrocodileBatidaSpecial} from which it is derived 
are too conservative in most realistic models.
They guarantee the absence of feedback effects  without relying on any further assumptions about the underlying particle physics model (solely due to redshifting of the produced particles). 
In reality there are a number of mechanisms that suppress feedback which in principle rely on assumptions about the particle physics embedding, but in practice occur quite generically.
One of them is the decay of the produced particles.
The interaction \eqref{Operator} generates a 
time-dependent  contribution
$\propto \g \varphi^\n  \Lambda^{4-\DD}$ to the squared effective mass of the particles $\X$ that couple to $\Phi$.
This causes 
non-perturbative particle production during the zero-crossings of $\varphi$, which can trigger a parametric or tachyonic resonance\footnote{
In regime where conditions \eqref{GeneralConstraintSelf} and \eqref{CrocodileBatidaSpecial} hold the  parametric resonance due to quartic and tachyonic resonance due to cubic interactions behave very similar \cite{Dufaux:2006ee}.
}
that quickly reheats the universe if condition \eqref{CrocodileBatidaSpecial} is violated.
However, the same term makes the produced $\X$-particles very heavy when the elongation of $\varphi$ is maximal, implying that they are very short-lived and can decay between two zero-crossings, which can considerably delay or even avoid the resonant particle production, cf.~e.g.~\cite{Garcia-Bellido:2008ycs}. 
Another effect is the screening of particles into quasiparticles with effective in-medium masses $M_\X$ once the universe is filled with radiation.
This terminates the resonant particle production as soon as 
$\M_\X^2 \sim \g \varphi^\n  \Lambda^{4-\DD}$, which can happen well before the radiation dominates the total energy of the universe.

For instance, consider the $h\phi^2\chi^2/4$ interaction in table \ref{ConversionTable} and add a $\lambda_\chi \chi^4/4!$ self-interaction, which generates a thermal mass $M_\chi^2 \simeq \lambda_\chi T^2/24$.
The thermal mass terminates the resonant non-perturbative particle production for $M_\chi^2 \sim h \varphi^2/2$, which happens at a temperature $T\sim \varphi \sqrt{12 h /\lambda_\chi}$. The ratio between radiation density and $\V$ is given by
$\frac{\pi^2 g_*}{15}\frac{T^4}{m_\phi^2 \varphi^2}\simeq 
\frac{48\pi^{2} g_*}{5}(h/\lambda_\chi)^2 (\varphi/m_\phi)^2$. 
Radiation is sub-dominant at the moment when the resonance is terminated if $h/\lambda_\chi < m_\phi/( 10 \sqrt{g_*} \varphi )$. This condition is parametrically much weaker than \eqref{CrocodileBatidaSpecial}, as it basically requires $h\ll m_\phi/\varphi$, while \eqref{CrocodileBatidaSpecial} requires $h\ll (m_\phi/\varphi)^2$. It becomes even weaker if $\chi$-decays have delayed the resonance so that $\varphi < \varphi_{\rm end}$. 
For the $g\Phi\chi^2$ interaction analogous considerations lead to the condition $\tilde{g} \ll \lambda_\chi$. The assumption that the $\chi$-particles have thermalised does not appear to be crucial, as similar results have been obtained in \cite{Dufaux:2006ee} without that assumption, confirming that effective masses can easily  terminate the resonant particle production before it has converted the inflaton energy into radiation. 

Hence, even if the inflaton couplings violate the conditions \eqref{CrocodileBatidaSpecial} and feedback effects strongly affect the rate of particle production in the initial phase of (p)reheating,
elementary decays may still fix $\Nreh$. 
In this case we can still use \eqref{GammaPerturbativeGeneric} in \eqref{GammaConstraint} to obtain a constraint on $\g$.
The reason is that feedback effects only introduce a $\{\SM_i\}$-dependence in the CMB observables if they modify the duration of reheating i.e., the \emph{expansion history} of the universe. 
This happens if $\Gamma$ depends on the $\{\SM_i\}$ at the moment when it equals $H$, cf.~\eqref{GammaConstraint}.


\section{Revisiting $\alpha$-attractor $E$ models}\label{JinURevisited}

The $\alpha$-attractor models of the $E$-type \cite{Ellis:2013nxa,Kallosh:2013yoa,Carrasco:2015pla,Carrasco:2015rva} 
are characterised by the potential 
\begin{equation}\label{alpha potential}
\V(\varphi)=\M^4\left(1-e^{-\sqrt{\frac{2}{3\alpha}}\frac{\varphi}{M_{pl}}}\right)^{2n}.
\end{equation}
Solving \eqref{slowrollpara} for the condition that $\epsilon=1$ we find 
\begin{eqnarray}
\varphi_{\rm end} = \sqrt{\frac{3\alpha}{2}}M_{pl}{\rm ln}\left(\frac{2n}{\sqrt{3\alpha}}+1\right). \label{alpha phi end}
\end{eqnarray}
From \eqref{nANDr} we further find 
\begin{eqnarray}
\varphi_k=\sqrt{\frac{3\alpha}{2}}M_{pl}{\rm ln}\left(\frac{8n}{\sqrt{3r\alpha}}+1\right)
\end{eqnarray}
In order to ensure that $\V(\varphi)$ is parabolic near the minimum we require that $n=1$,  leading to a Starobinsky-like potential \cite{Starobinsky:1980te}. 
Taylor expansion yields  
\begin{eqnarray}\label{alphaEexpansion}
m_\phi^2 = \frac{4\M^4}{3 M_{pl}^2\alpha} \ , \
g_\phi = \frac{2\M^4}{3M_{pl}^3}\sqrt{\frac{2}{3\alpha^3}} \ , \ \frac{\lambda_\phi}{4!} = \frac{7}{27\alpha^2}\frac{\M^4}{M_{pl}^4}.
\end{eqnarray}
Plugging 
\eqref{alphaEexpansion}
into 
\eqref{phi cr}
we find that indeed the quadratic term dominates over the cubic and quartic at the end of inflation for any $\alpha \geq 1$.   
However, condition \eqref{avoid self PR} is violated for practically all values of $\alpha$ used in \cite{Ueno:2016dim,Drewes:2017fmn,Ellis:2021kad} (and possibly all values of $\alpha$ that can be made consistent with the current observational bound on $r$), 
meaning that the fragmentation due to a self-resonance 
puts the interpretation of CMB observables in terms of the microphysical inflaton coupling in these works 
into question (while the determination of the effective $\Treh$ remains valid).\footnote{Note, however, that this conclusion may be too conservative, as already stated in Sec.~\ref{Sec:FeedbaclEffects}. 
The more detailed analysis in \cite{Lozanov:2017hjm} suggests that a self-resonance may still be avoided for values of $\alpha$ or order unity.} A main problem is that the potential of $E$-type models is not symmetric with respect to $\varphi \leftrightarrow -\varphi$, which leads to a  $g_\phi\neq 0$ that tends to violate \eqref{avoid self PR}. For this reason, and in view of the fact that one of our main goals is to quantify the observational sensitivity to $\g$, we refrain from studying $E$-type models here, in spite of the fact that they would otherwise fit naturally into the present work.

\begin{figure}
	\subfloat[]{\includegraphics[width=0.45\linewidth]{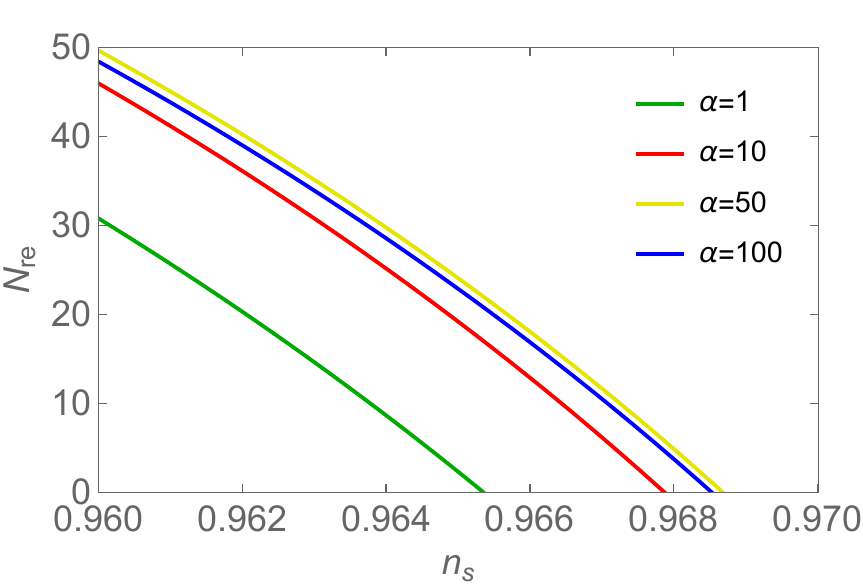}}
	\subfloat[]{\includegraphics[width=0.45\linewidth]{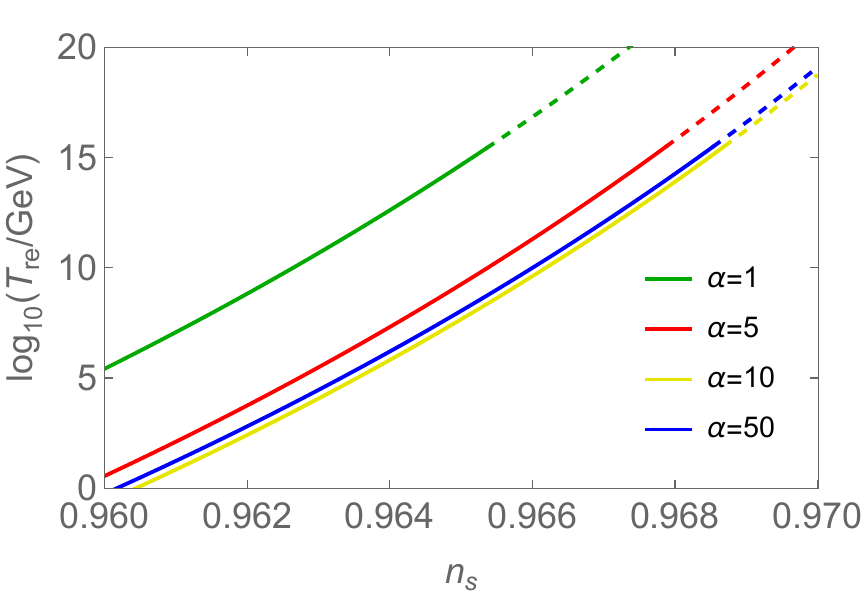}}
	\caption{Corrected version of Fig.~3 in~\cite{Drewes:2017fmn}, describing the $n_s$-dependence of $N_{\rm re}$ and $T_{\rm re}$ in the $\alpha$-attractor E-model with $n=1$ for various values of $\alpha$, in analogy to Fig.~\ref{MHI 4}.}
	\label{alpha NT}
\end{figure}

Before closing we remark that Fig.~3 in~\cite{Drewes:2017fmn}, which describes the $n_s$-dependence of $N_{\rm re}$ and $T_{\rm re}$ in the $\alpha$-attractor E-model with $n=1$ for various values of $\alpha$, needs minor modifications. 
When $n_s$ is fixed, the value of $\Nreh$ and $\Treh$ do not change monotonically in the parameter range $1\leq\alpha\leq100$, as can be seen in Fig.~\ref{alpha NT}. This effect also exists in the mutated hilltop model discussed in section~\ref{MHI model}.

\section{Illustrative plots for various parameter choices}\label{Sec:AppendixWithAllPlots}
In this appendix we show the dependence of $\Treh$, $N_k$, $\Nreh$ and the inflaton coupling constants 
$\g$
on various parameters in the models considered in this work. 
We consider a broader range of choices for $\alpha$, but keep $A_s$ fixed to \eqref{AsBestFit}.

\clearpage
\begin{landscape}
\begin{figure}[h!]
\centering
\subfloat[]{
\includegraphics[width=0.45\linewidth,height=0.4\textheight]{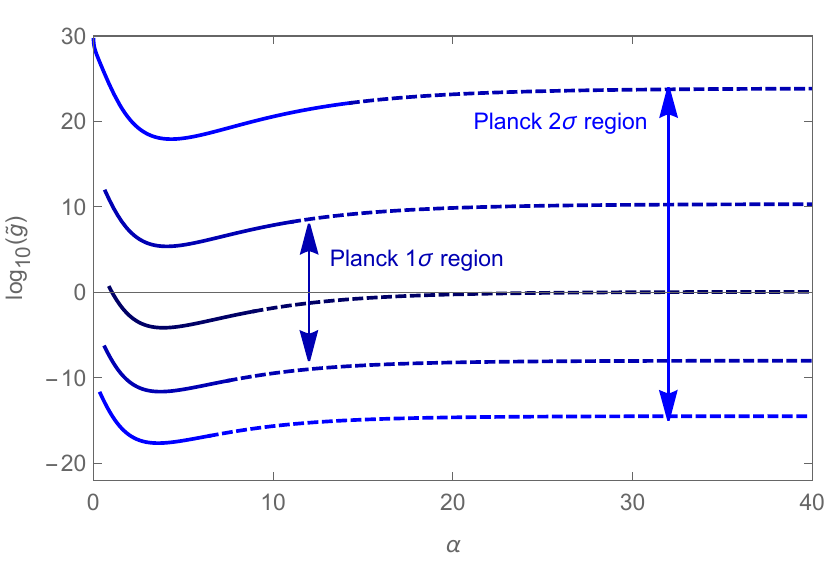}
\label{MHI ghy g}
}
\quad
\subfloat[]{
\includegraphics[width=0.45\linewidth,height=0.4\textheight]{
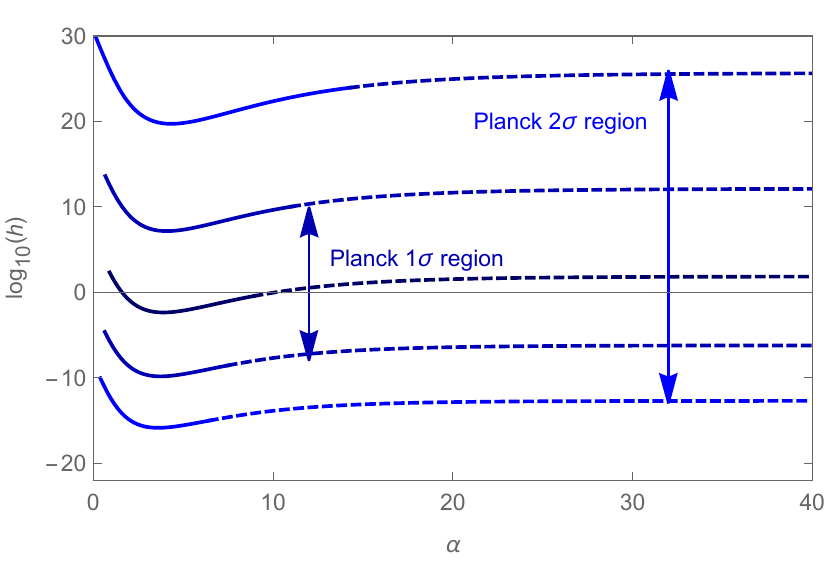}}
\quad
\subfloat[]{
	\includegraphics[width=0.45\linewidth,height=0.4\textheight]{
		mutated_hilltop_inflation/MHI_yru.pdf}}
\quad
\subfloat[]{
	\includegraphics[width=0.45\linewidth,height=0.4\textheight]{
		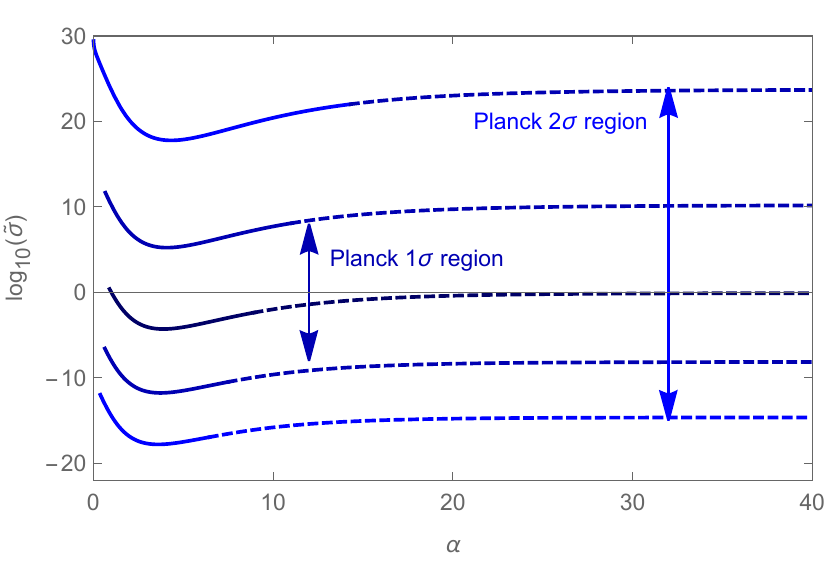}}
\caption{The relation between the inflaton couplings and $\alpha$ in the MHI model for 
different choices of $n_s$, with conventions as in Fig.~\ref{MHI ghy 1}.
}
\label{MHI ghy}
\clearpage\end{figure}
\clearpage

\clearpage\begin{figure}
\centering
\subfloat[]{\includegraphics[width=0.45\linewidth]{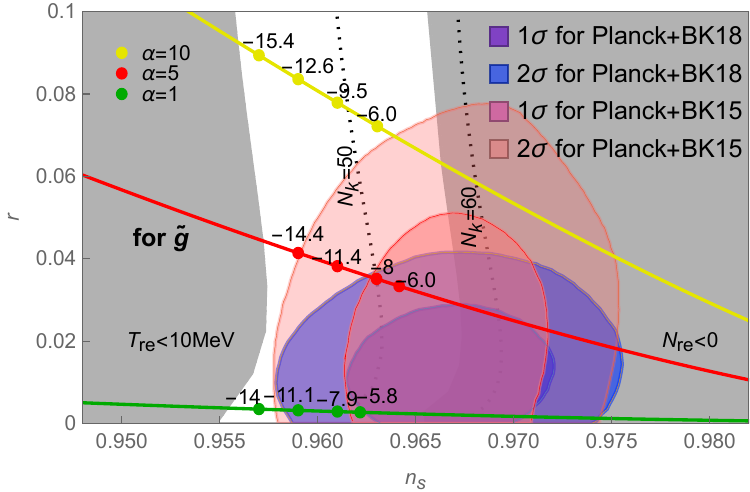}}
\quad
	\subfloat[]{\includegraphics[width=0.45\linewidth]{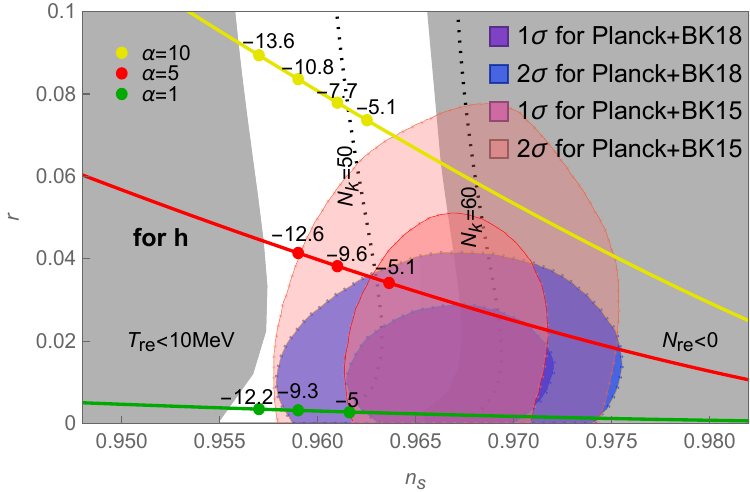}}
	\quad
	\subfloat[]{\includegraphics[width=0.45\linewidth]{mutated_hilltop_inflation/MHI_Ry.pdf}}
	\quad
	\subfloat[]{\includegraphics[width=0.45\linewidth]{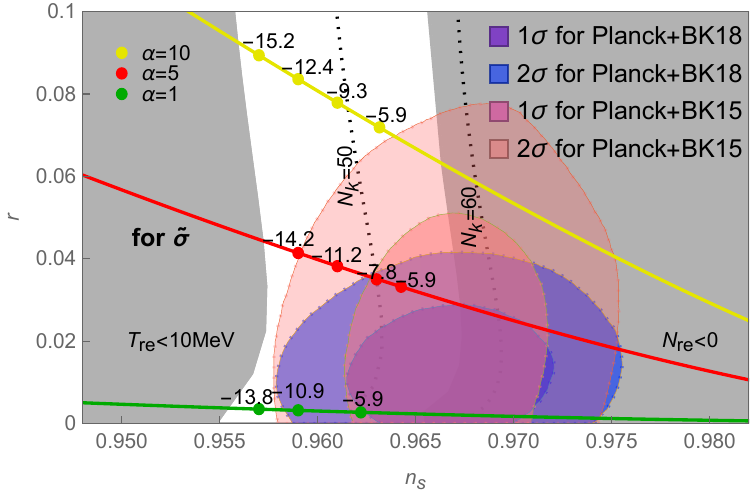}}
	\caption{Impact of the reheating phase on the predictions of the MHI model for $n_s$ and $r$ for a set of sample points.
		In panels [a]-[d] we assume that reheating is primarily driven by an interaction in table~\ref{ConversionTable}.
		The notations are the same with Fig~\ref{MHI R d}.	} 
\label{MHI R}
\clearpage\end{figure}

\clearpage\begin{figure}[!h]
	\centering
	\subfloat[]{
		\includegraphics[width=0.45\linewidth]{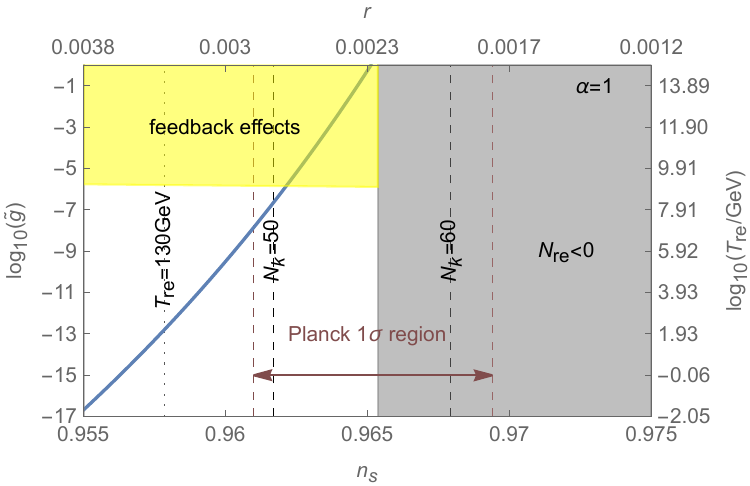}
		\label{MHI g 1}
	}
	\quad
	\subfloat[]{
		\includegraphics[width=0.45\linewidth]{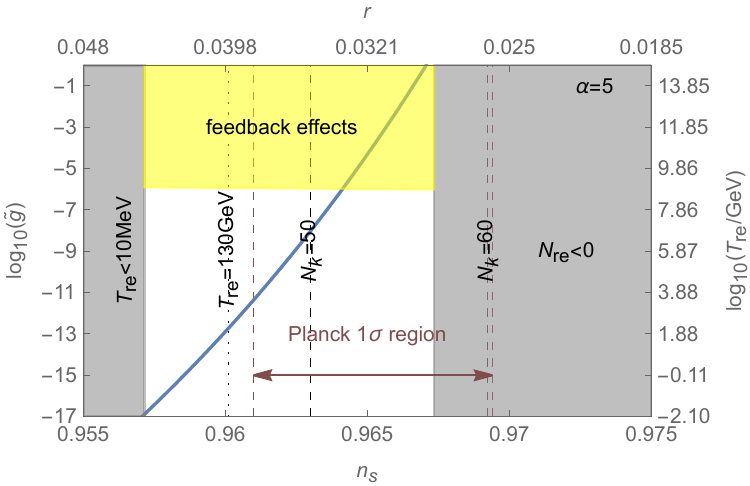}
	}
	\quad
	\subfloat[]{
		\includegraphics[width=0.45\linewidth]{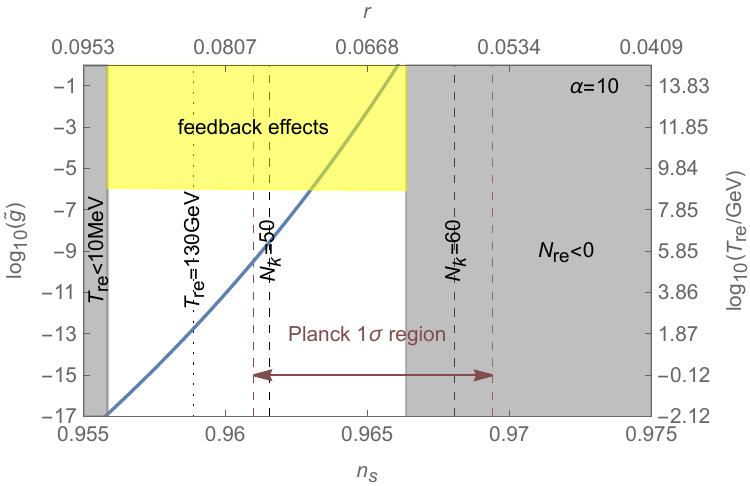}
	}
	\quad
	\subfloat[]{
		\includegraphics[width=0.45\linewidth]{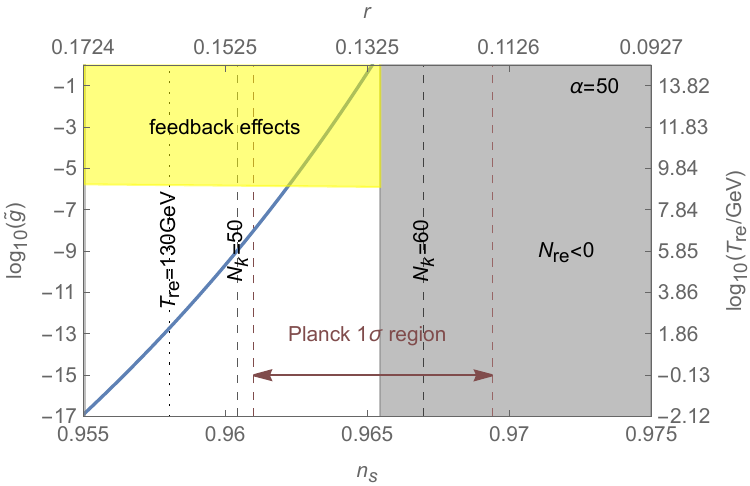}
	}
	\caption{
	Dependence of the scalar coupling $\tilde{g}$ on $n_s$ in the MHI model for different choices of $\alpha$. The notations are the same with Fig.~\ref{MHI g 1 a}.}
	\label{MHI g}
\clearpage\end{figure}
\clearpage\begin{figure}[!h]
	\centering
	\subfloat[]{
		\includegraphics[width=0.45\linewidth]{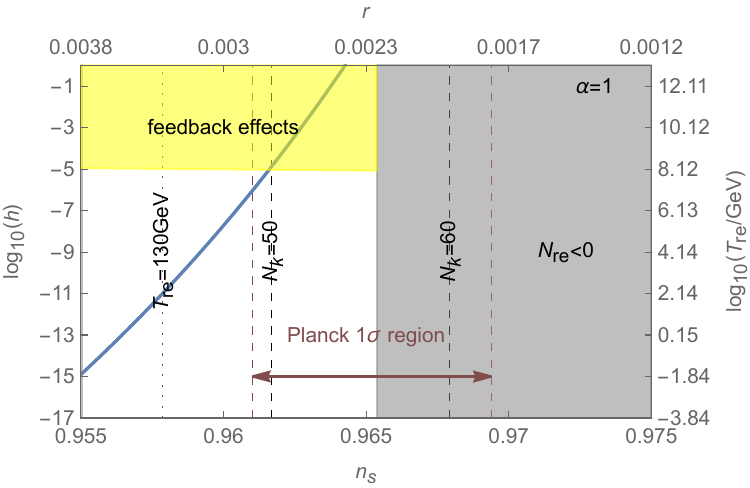}
	}
	\quad
	\subfloat[]{
		\includegraphics[width=0.45\linewidth]{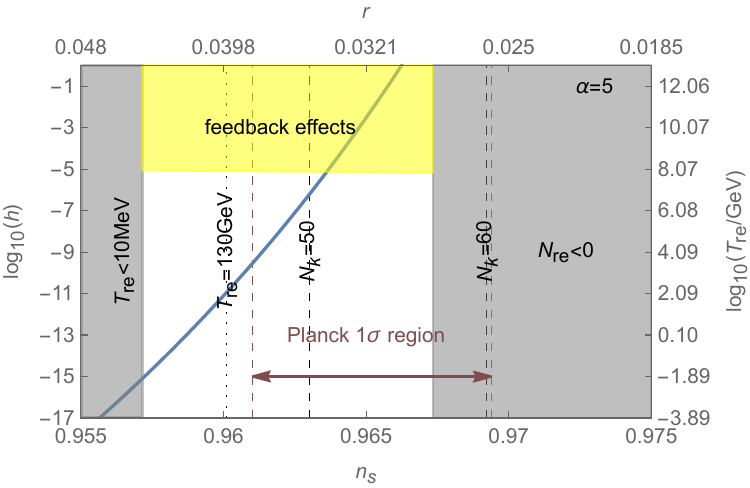}
	}
	\quad
	\subfloat[]{
		\includegraphics[width=0.45\linewidth]{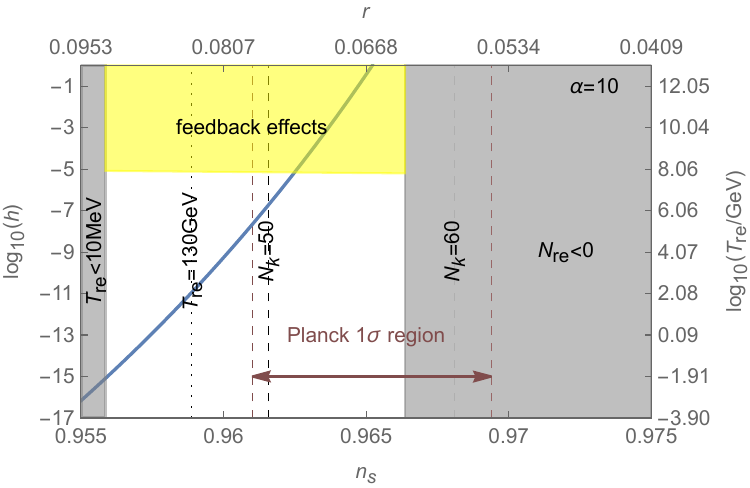}
	}
	\quad
	\subfloat[]{
		\includegraphics[width=0.45\linewidth]{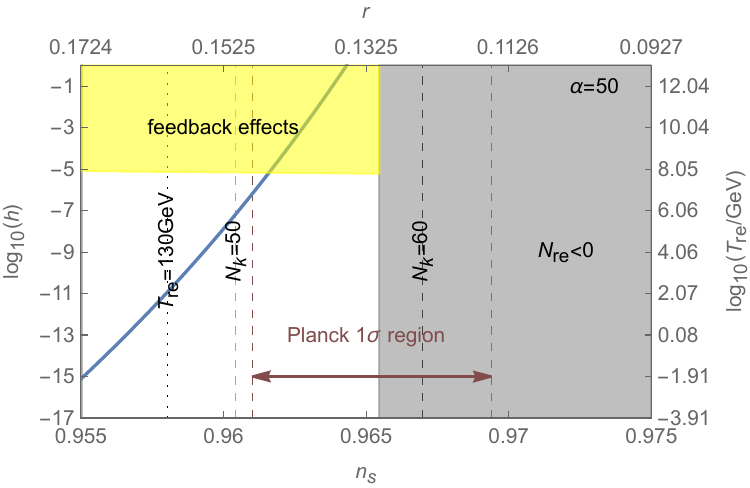}
	}
	\caption{Dependence of the scalar coupling $h$ on $n_s$ in the MHI model for different choices of $\alpha$. 
	The notations are the same with Fig.~\ref{MHI g 1 a}.}
	\label{MHI h}
\clearpage\end{figure}
\clearpage\begin{figure}[!h]
	\centering
	\subfloat[]{
		\includegraphics[width=0.45\linewidth]{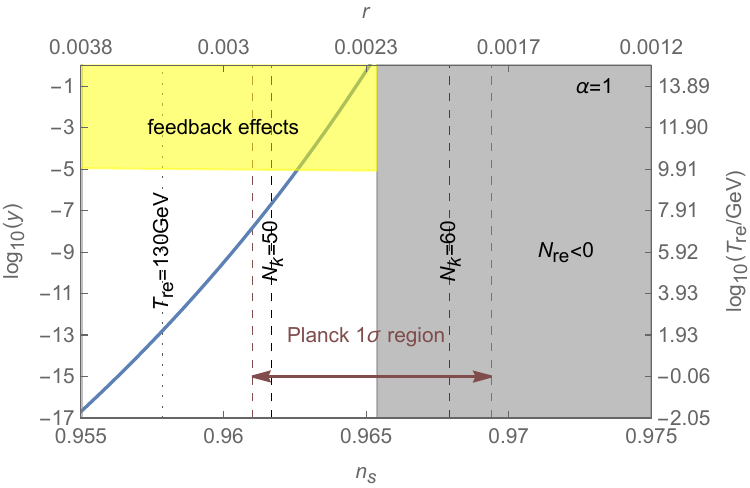}
	}
	\quad
	\subfloat[]{
		\includegraphics[width=0.45\linewidth]{mutated_hilltop_inflation/MHI_y_5.pdf}
	}
	\quad
	\subfloat[]{
		\includegraphics[width=0.45\linewidth]{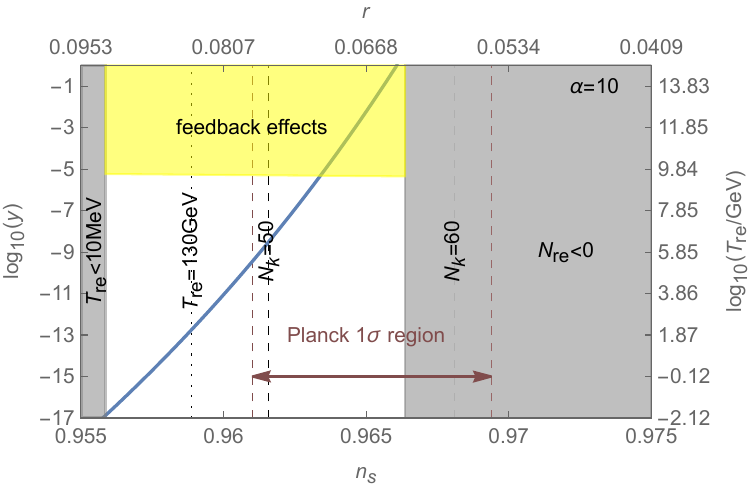}
	}
	\quad
	\subfloat[]{
		\includegraphics[width=0.45\linewidth]{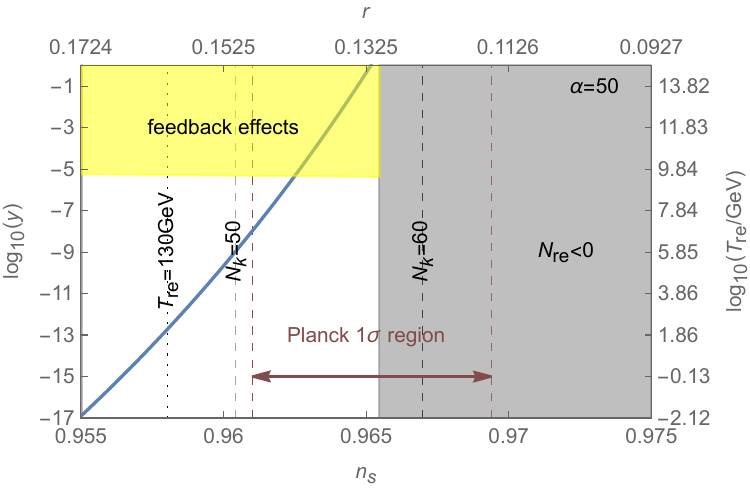}
	}
	\caption{Dependence of the Yukawa coupling $y$ on $n_s$ in the MHI model for different choices of $\alpha$. The notations are the same with Fig.~\ref{MHI g 1 a}.}
	\label{MHI y}
\clearpage\end{figure}
\clearpage\begin{figure}[!h]
	\centering
	\subfloat[]{
		\includegraphics[width=0.45\linewidth]{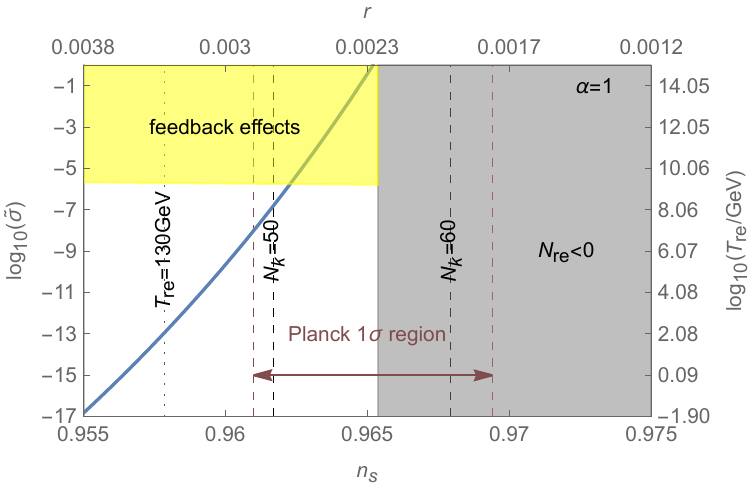}
	}
	\quad
	\subfloat[]{
		\includegraphics[width=0.45\linewidth]{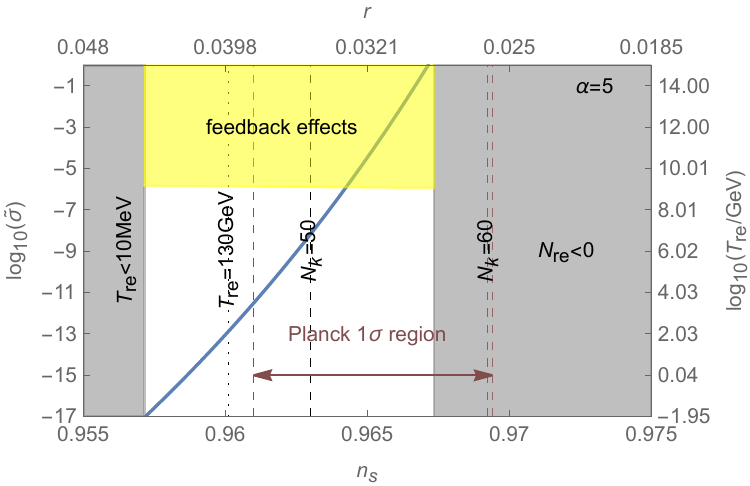}
	}
	\quad
	\subfloat[]{
		\includegraphics[width=0.45\linewidth]{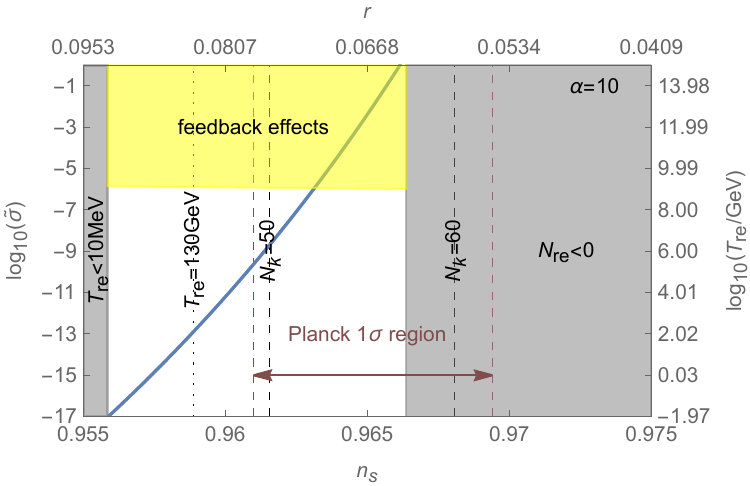}
	}
	\quad
	\subfloat[]{
		\includegraphics[width=0.45\linewidth]{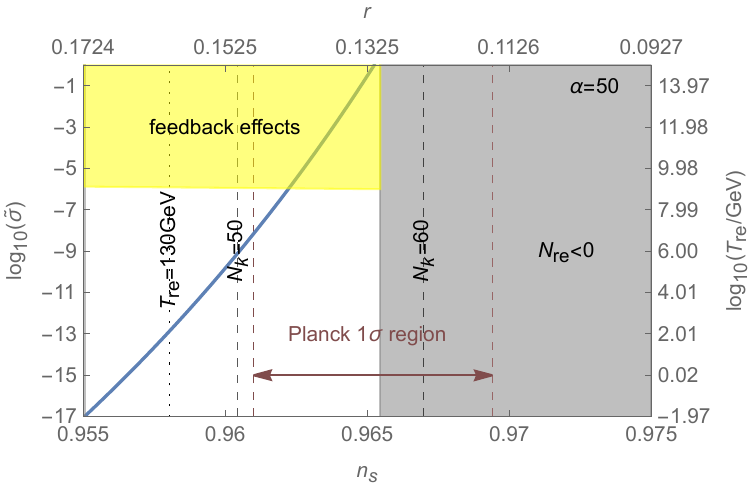}
	}
	\caption{Dependence of the axion-like coupling $\tilde{\upalphatosigma}$ on $n_s$ in the MHI model for different choices of $\alpha$. The notations are the same with Fig.~\ref{MHI g 1 a}.}
	\label{MHI a}
\clearpage\end{figure}



\clearpage\begin{figure}
\centering
	\subfloat[]{\includegraphics[width=0.45\linewidth]{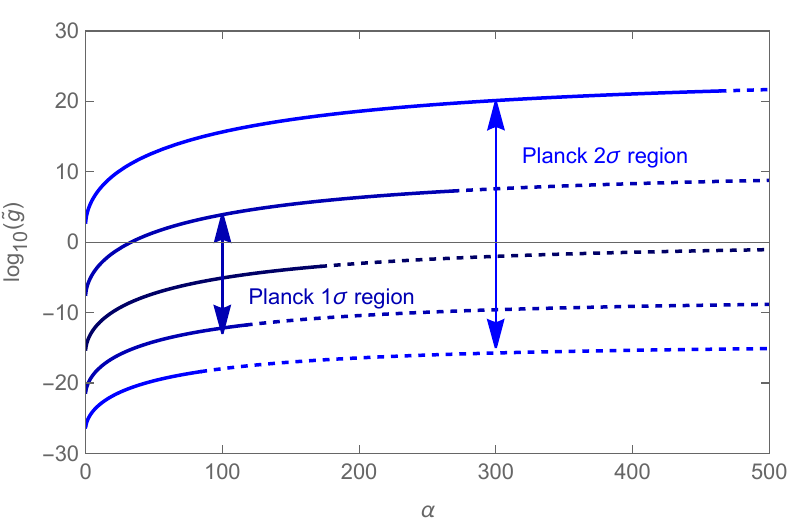}}	
	\quad
	\subfloat[]{\includegraphics[width=0.45\linewidth]{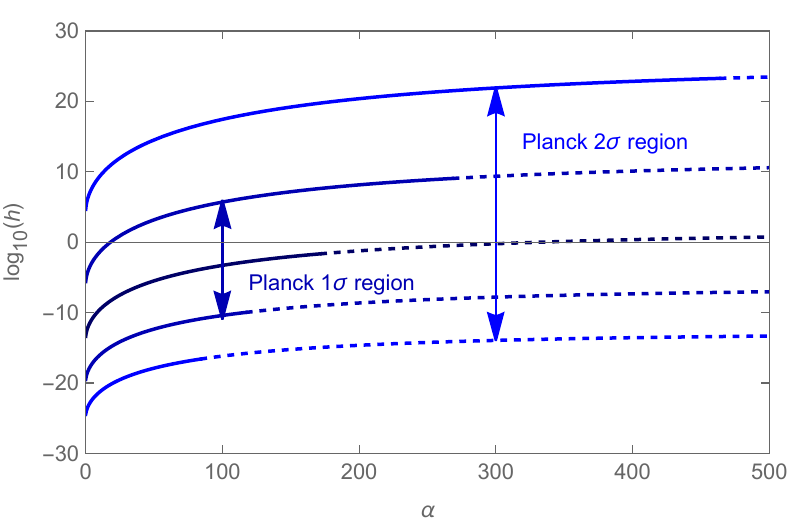}}
	\quad
	\subfloat[]{\includegraphics[width=0.45\linewidth]{radion_gauge_inflation/RGI_yru.pdf}}	
	\quad
	\subfloat[]{\includegraphics[width=0.45\linewidth]{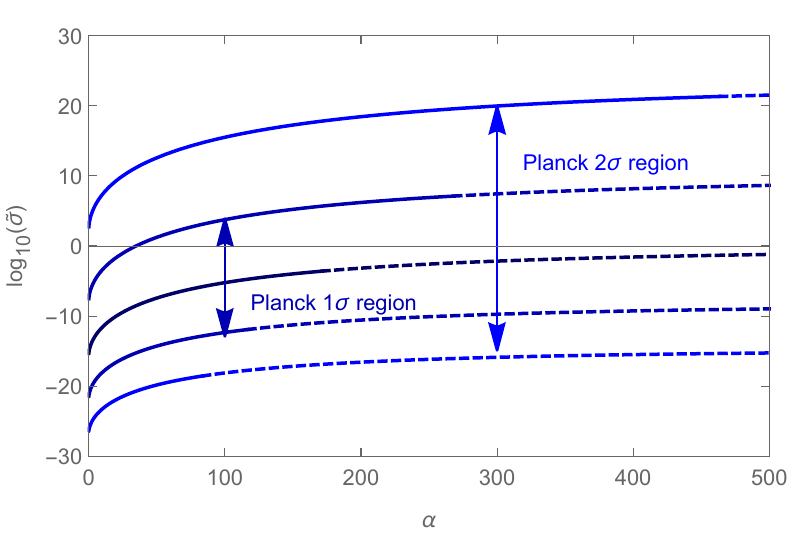}}
	\caption{The relation between the inflaton couplings and $\alpha$ in the RGI model for 
different choices of $n_s$, with conventions as in Fig.~\ref{MHI ghy 1}.}
	\label{RGI ghy}
\clearpage\end{figure}


\clearpage\begin{figure}
	\centering
	\subfloat[]{\includegraphics[width=0.45\linewidth]{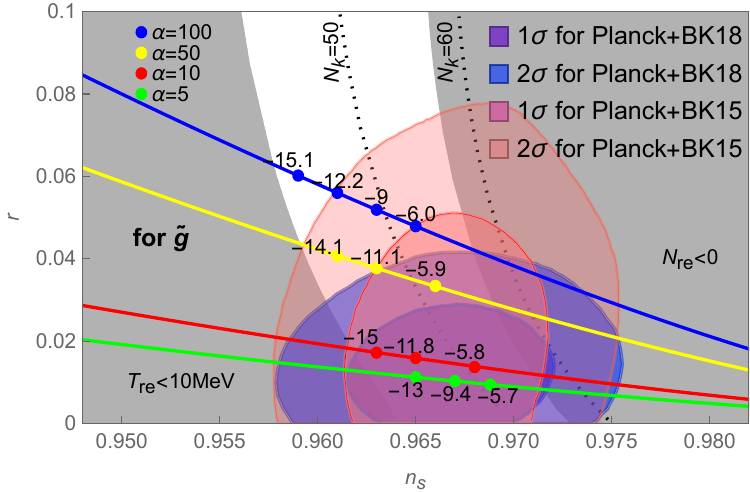}}
	\quad
	\subfloat[]{\includegraphics[width=0.45\linewidth]{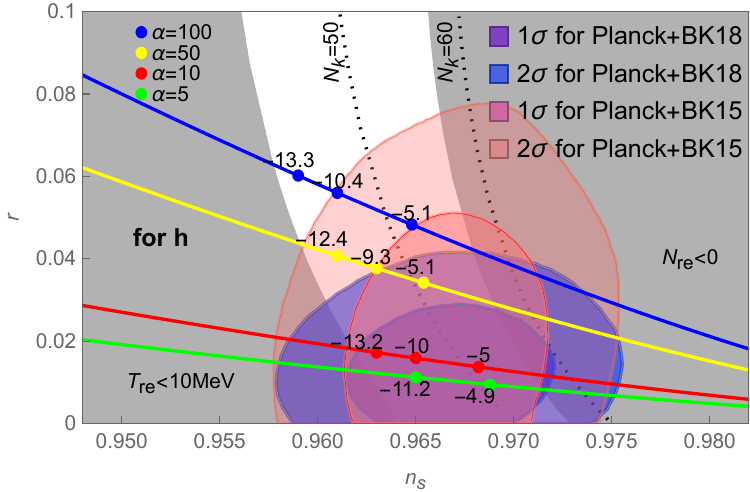}}
	\quad
	\subfloat[]{\includegraphics[width=0.45\linewidth]{radion_gauge_inflation/RGI_Ry.pdf}}
	\quad
	\subfloat[]{\includegraphics[width=0.45\linewidth]{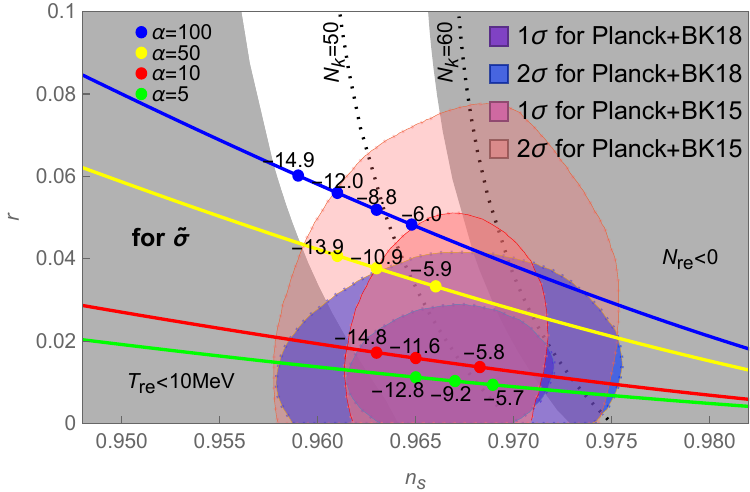}}
	\caption{Impact of the reheating phase on the predictions of the RGI model for $n_s$ and $r$ for a set of sample points.  The notations are the same with Fig~\ref{MHI R d}.} 
	\label{RGI R}
\clearpage\end{figure}


\clearpage\begin{figure}[!h]
\centering
\subfloat[]{
\includegraphics[width=0.46\linewidth]{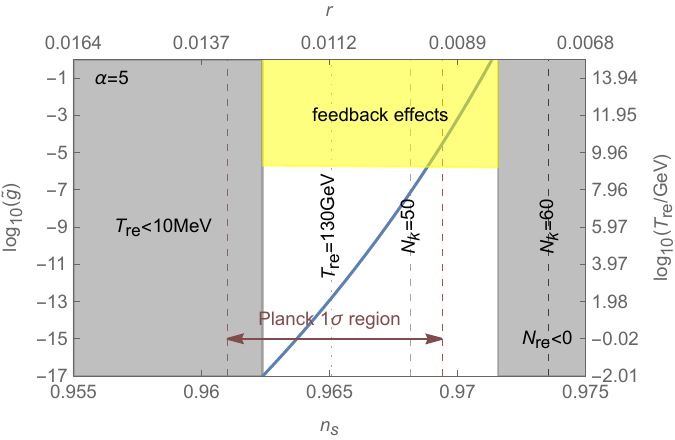}}
\quad
\subfloat[]{\includegraphics[width=0.46\linewidth]{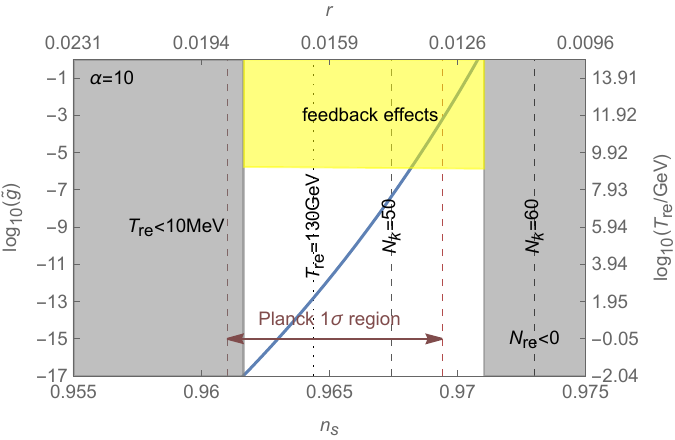}}
\quad
\subfloat[]{\includegraphics[width=0.46\linewidth]{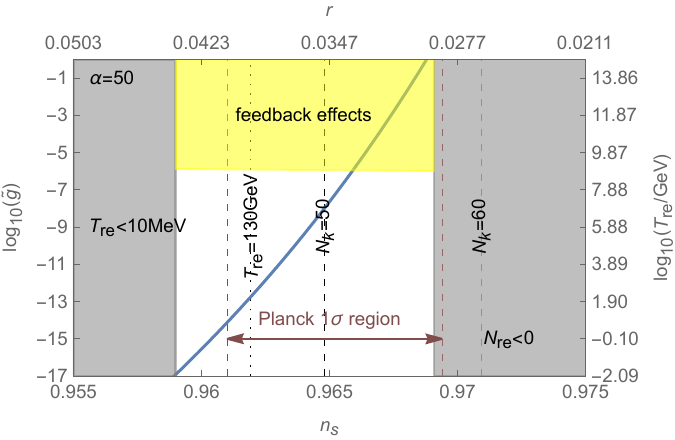}}
\quad
\subfloat[]{\includegraphics[width=0.46\linewidth]{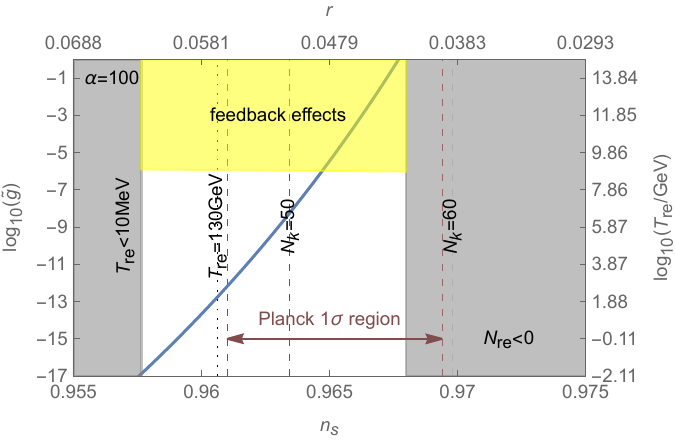}}
\caption{Dependence of the scalar coupling $\tilde{g}$ on $n_s$ in the RGI model for different choices of $\alpha$. The notations are the same with Fig.~\ref{MHI g 1 a}.}
\label{RGI g}
\clearpage\end{figure}
\clearpage\begin{figure}[!h]
	\centering
	\subfloat[]{
		\includegraphics[width=0.46\linewidth]{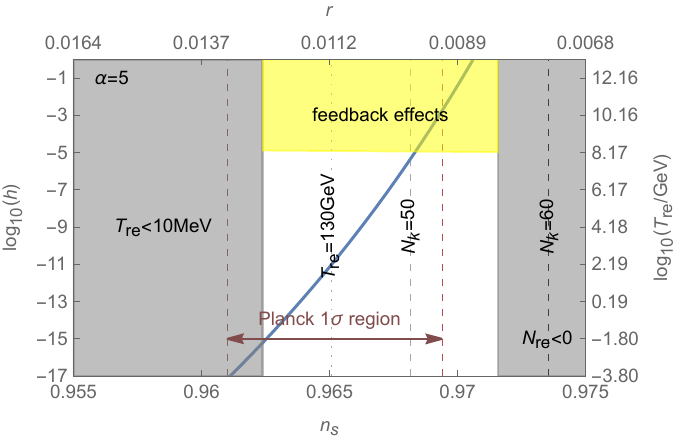}}
	\quad
	\subfloat[]{\includegraphics[width=0.46\linewidth]{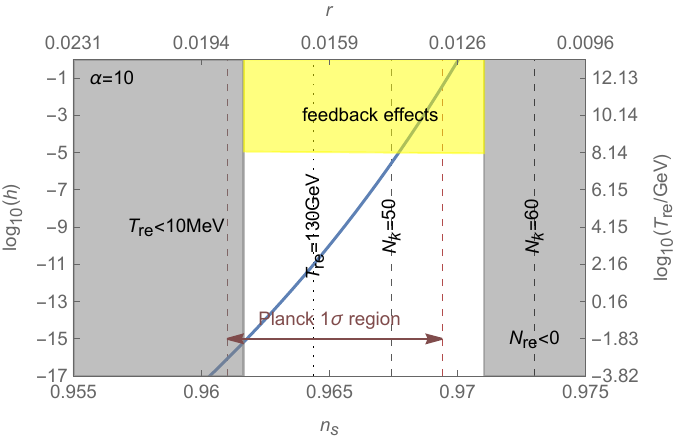}}
	\quad
	\subfloat[]{\includegraphics[width=0.46\linewidth]{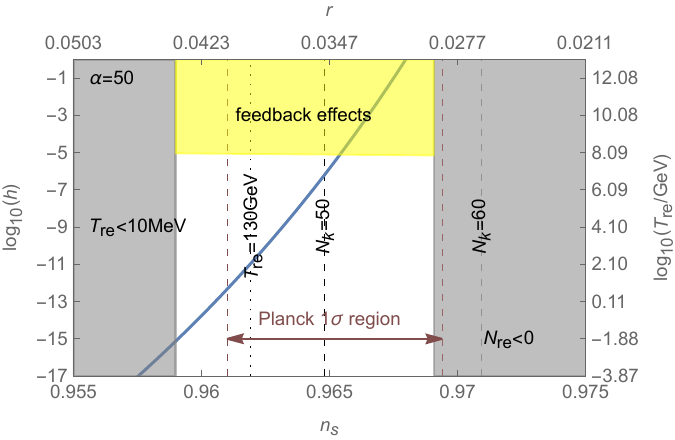}}
	\quad
	\subfloat[]{\includegraphics[width=0.46\linewidth]{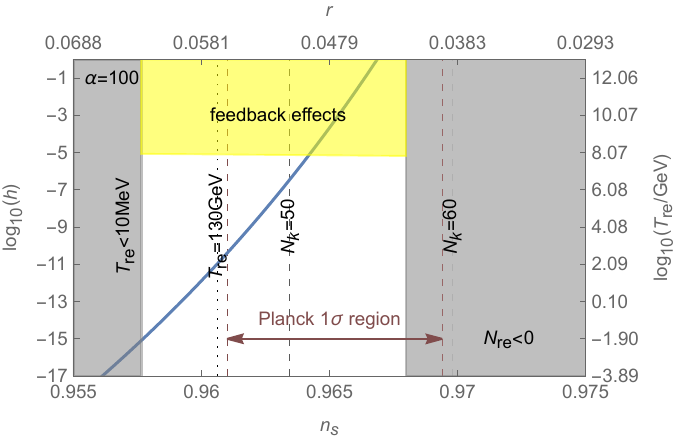}}
	\caption{Dependence of the scalar coupling $h$ on $n_s$ in the RGI model for different choices of $\alpha$. The notations are the same with Fig.~\ref{MHI g 1 a}.}
	\label{RGI h}
\clearpage\end{figure}
\clearpage\begin{figure}[!h]
	\centering
	\subfloat[]{
		\includegraphics[width=0.46\linewidth]{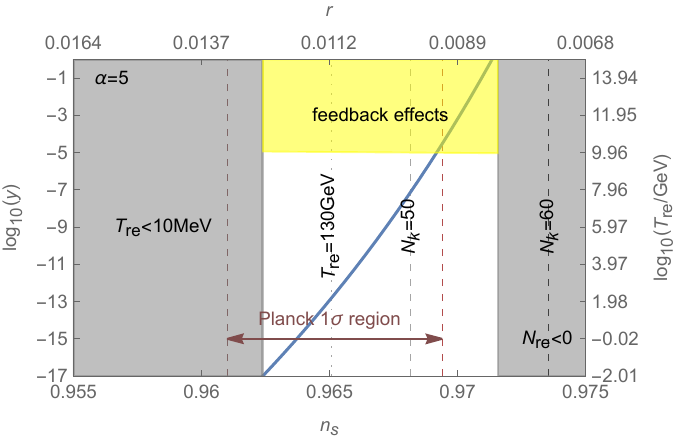}}
	\quad
	\subfloat[]{\includegraphics[width=0.46\linewidth]{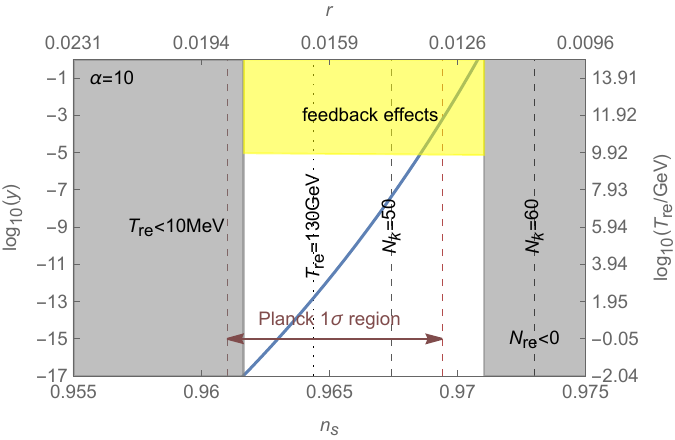}}
	\quad
	\subfloat[]{\includegraphics[width=0.46\linewidth]{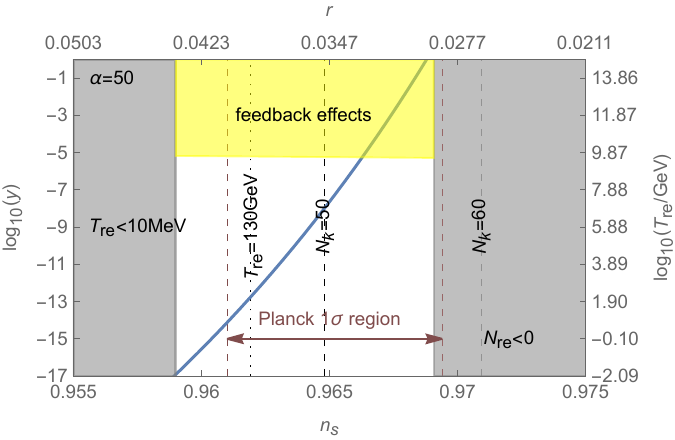}}
	\quad
	\subfloat[]{\includegraphics[width=0.46\linewidth]{radion_gauge_inflation/RGI_y_100.pdf}}
	\caption{Dependence of the Yukawa coupling $y$ on $n_s$ in the RGI model for different choices of $\alpha$. The notations are the same with Fig.~\ref{MHI g 1 a}.}
	\label{RGI y}
\clearpage\end{figure}
\clearpage\begin{figure}[!h]
	\centering
	\subfloat[]{
		\includegraphics[width=0.46\linewidth]{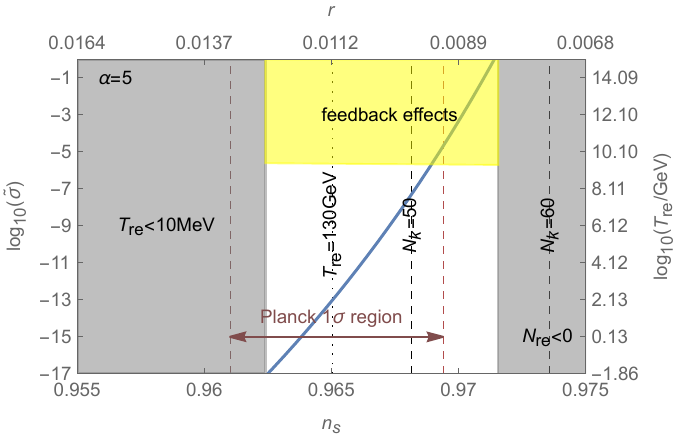}}
	\quad
	\subfloat[]{\includegraphics[width=0.46\linewidth]{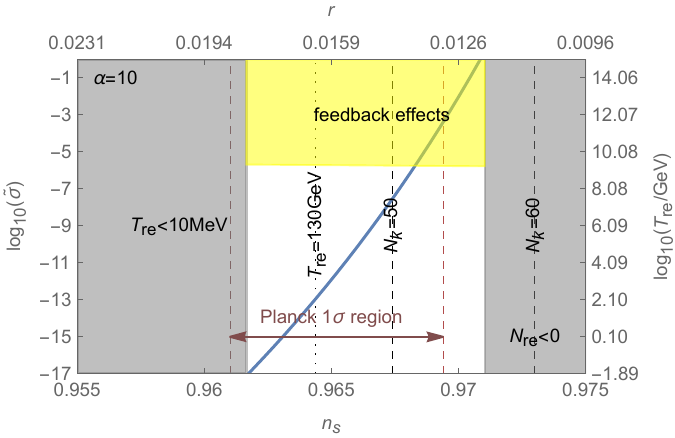}}
	\quad
	\subfloat[]{\includegraphics[width=0.46\linewidth]{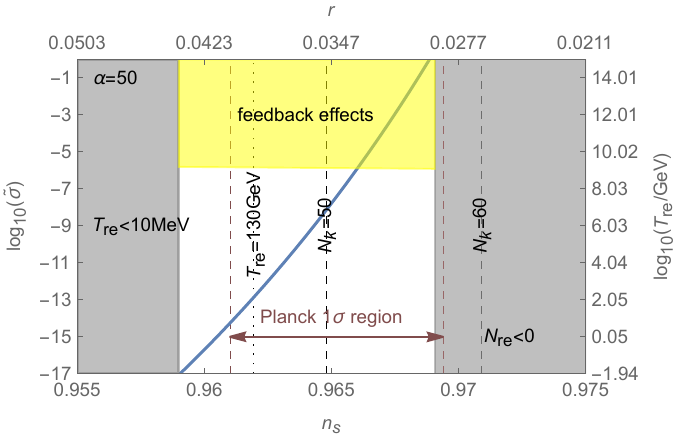}}
	\quad
	\subfloat[]{\includegraphics[width=0.46\linewidth]{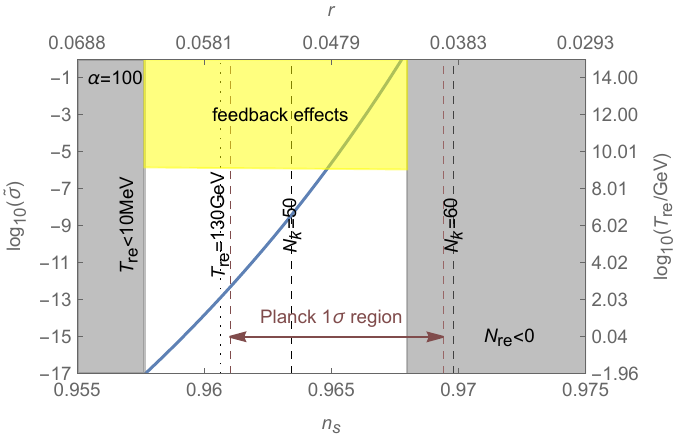}}
	\caption{Dependence of the axion-like coupling $\tilde{\upalphatosigma}$ on $n_s$ in the RGI model for different choices of $\alpha$. The notations are the same with Fig.~\ref{MHI g 1 a}.}
	\label{RGI a}
\clearpage\end{figure}


\clearpage\begin{figure}
\centering
\subfloat[]{\includegraphics[width=0.45\linewidth]{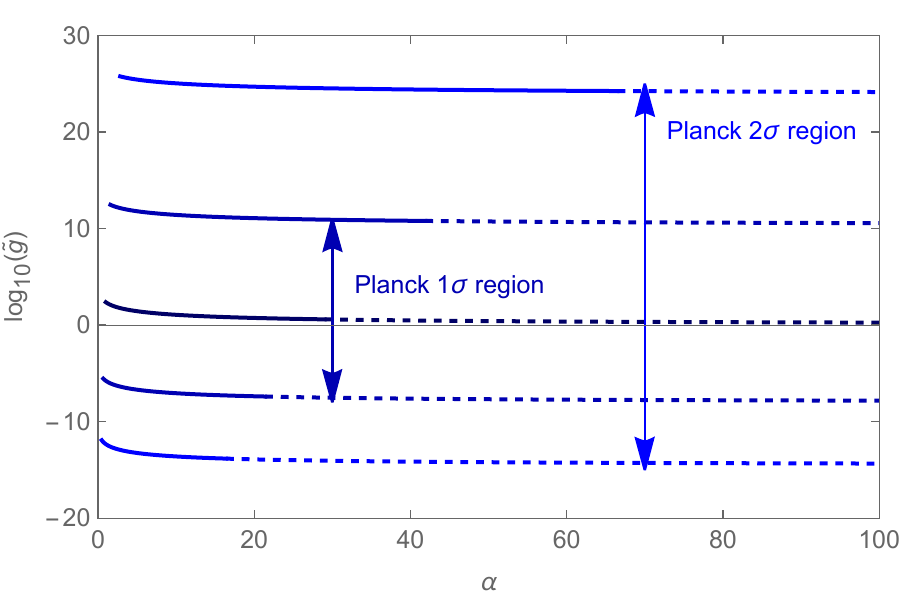}
\label{alpha gru}}
\quad
\subfloat[]{\includegraphics[width=0.45\linewidth]{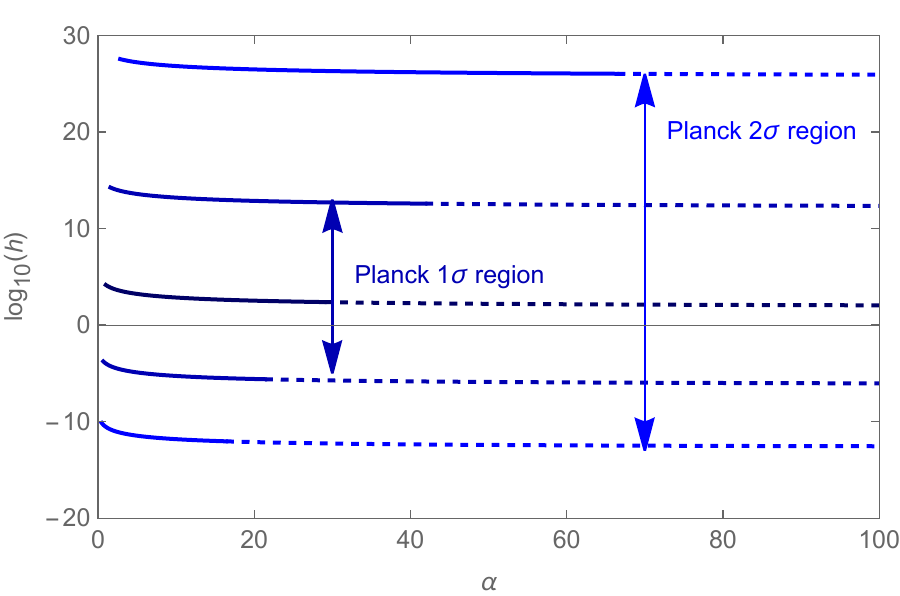}
\label{alpha hru}}
\quad
\subfloat[]{\includegraphics[width=0.45\linewidth]{alpha_T_model/alpha_T_yru.pdf}
	\label{alpha yru}}
\quad
\subfloat[]{\includegraphics[width=0.45\linewidth]{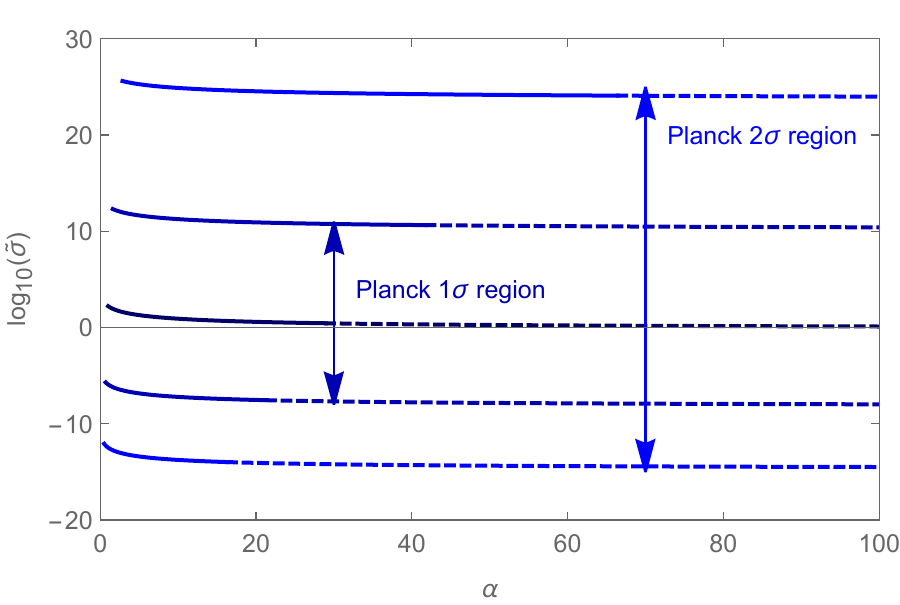}
	\label{alpha xru}}
\caption{The relation between the inflaton couplings and $\alpha$ in the $\alpha$-T model for 
different choices of $n_s$, with conventions as in Fig.~\ref{MHI ghy 1}.
}
\label{alpha ghy}
\clearpage\end{figure}

\clearpage\begin{figure}
	\centering
	\subfloat[]{\includegraphics[width=0.45\linewidth]{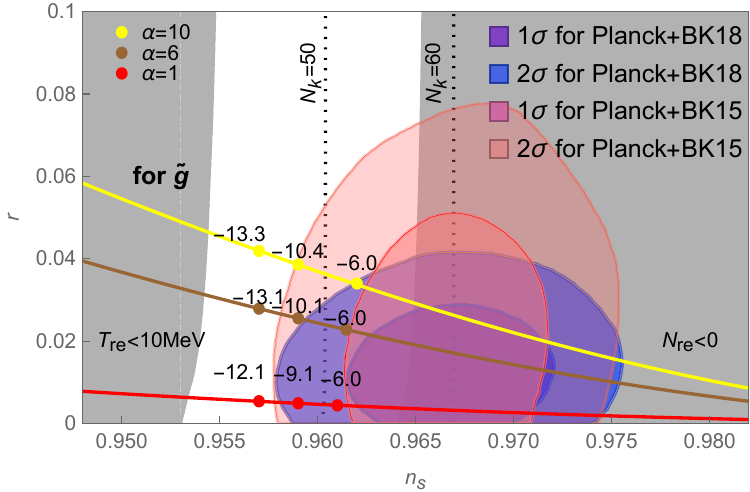}}
	\quad
	\subfloat[]{\includegraphics[width=0.45\linewidth]{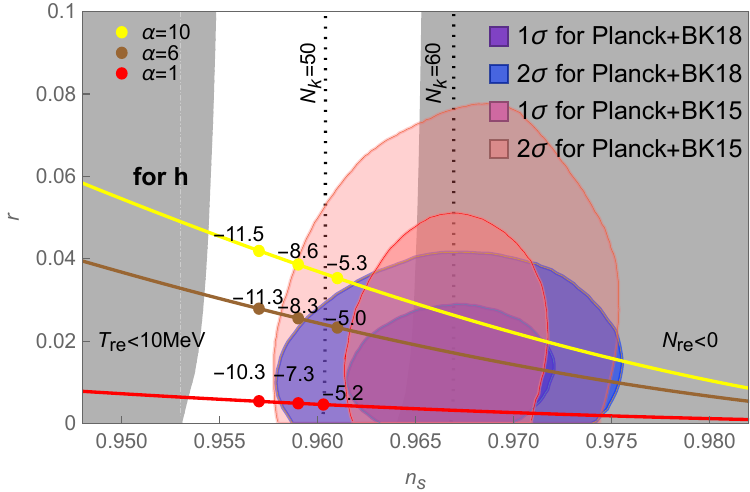}}
	\quad
	\subfloat[]{\includegraphics[width=0.45\linewidth]{alpha_T_model/alpha_T_Ry.pdf}}
	\quad
	\subfloat[]{\includegraphics[width=0.45\linewidth]{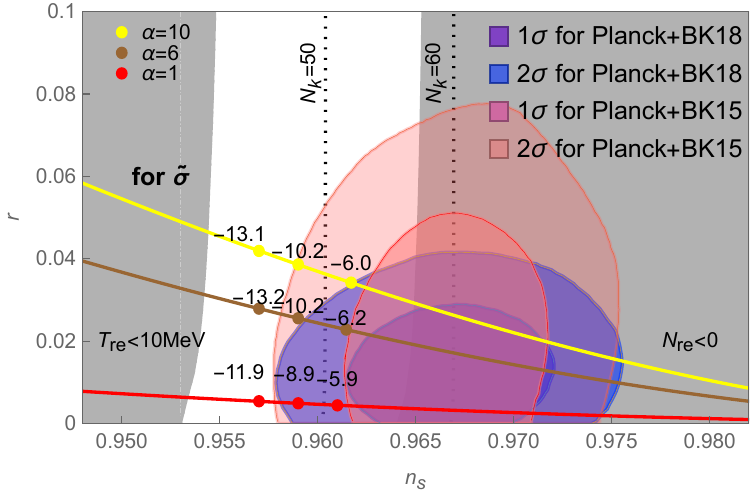}}
	\caption{Impact of the reheating phase on the predictions of the $\alpha$-T model for $n_s$ and $r$ for a set of sample points.  The notations are the same with Fig.~\ref{MHI R d}.} 
	\label{alpha T R}
\clearpage\end{figure}

\clearpage\begin{figure}[!h]
	\centering
	\subfloat[]{
		\includegraphics[width=0.46\linewidth]{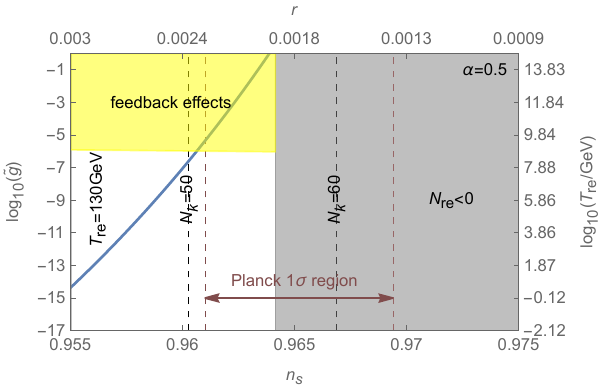}}
	\quad
	\subfloat[]{\includegraphics[width=0.46\linewidth]{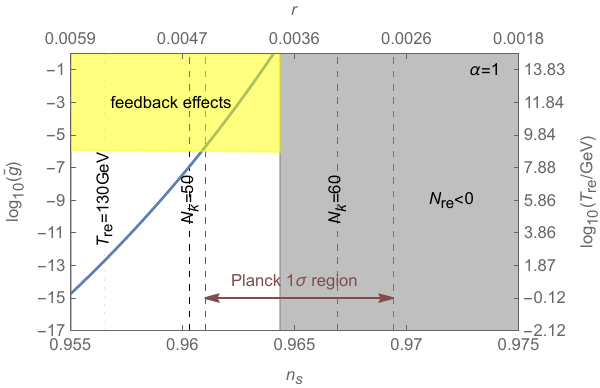}}
	\quad
	\subfloat[]{\includegraphics[width=0.46\linewidth]{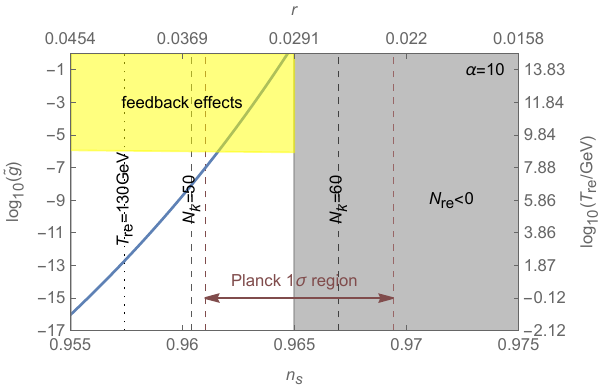}}
	\quad
	\subfloat[]{\includegraphics[width=0.46\linewidth]{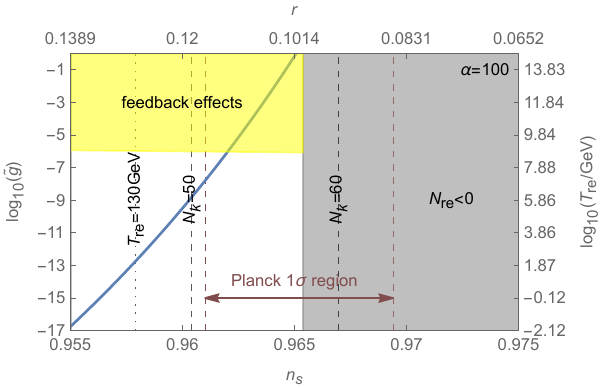}}
	\caption{Dependence of the scalar coupling $\tilde{g}$ on $n_s$ in the $\alpha$-T model for different choices of $\alpha$. The notations are the same with Fig.~\ref{MHI g 1 a}.}
	\label{alpha T g}
\clearpage\end{figure}
\clearpage\begin{figure}[!h]
	\centering
	\subfloat[]{
		\includegraphics[width=0.46\linewidth]{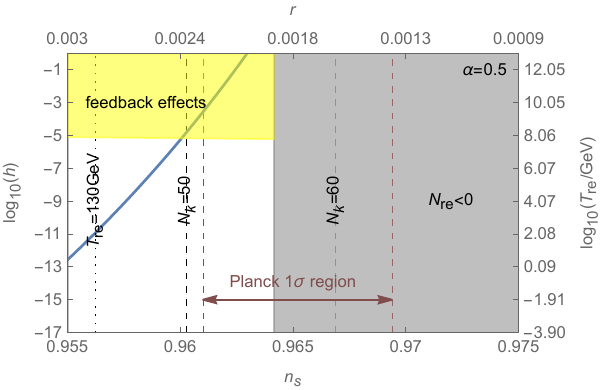}}
	\quad
	\subfloat[]{\includegraphics[width=0.46\linewidth]{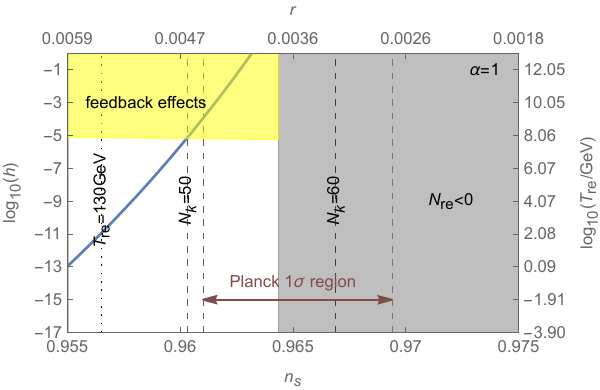}}
	\quad
	\subfloat[]{\includegraphics[width=0.46\linewidth]{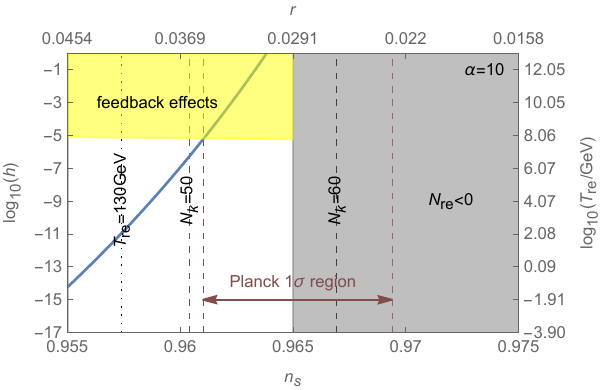}}
	\quad
	\subfloat[]{\includegraphics[width=0.46\linewidth]{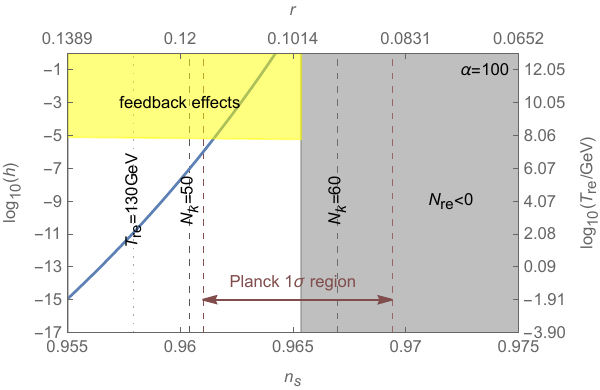}}
	\caption{Dependence of the scalar coupling $h$ on $n_s$ in the $\alpha$-T model for different choices of $\alpha$. The notations are the same with Fig.~\ref{MHI g 1 a}.}
	\label{alpha T h}
\clearpage\end{figure}
\clearpage\begin{figure}[!h]
	\centering
	\subfloat[]{
		\includegraphics[width=0.46\linewidth]{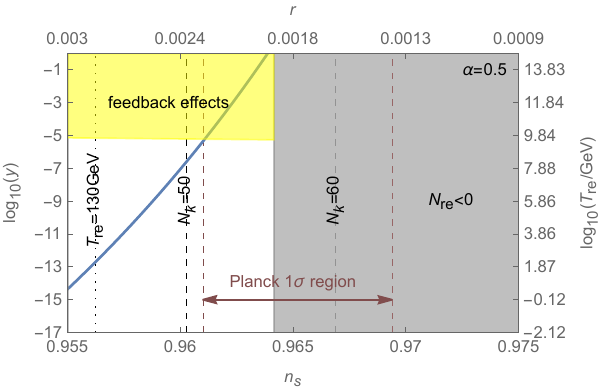}}
	\quad
	\subfloat[]{\includegraphics[width=0.46\linewidth]{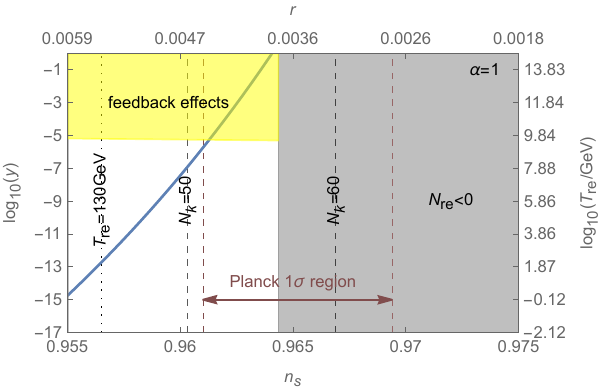}}
	\quad
	\subfloat[]{\includegraphics[width=0.46\linewidth]{alpha_T_model/alpha_T_y_10.pdf}}
	\quad
	\subfloat[]{\includegraphics[width=0.46\linewidth]{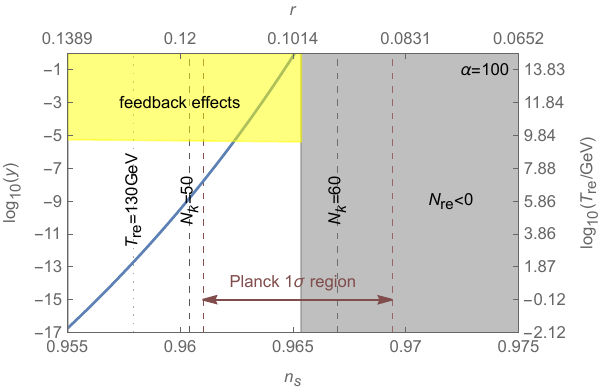}}
	\caption{Dependence of the Yukawa coupling $y$ on $n_s$ in the $\alpha$-T model for different choices of $\alpha$. The notations are the same with Fig.~\ref{MHI g 1 a}.}
	\label{alpha T y}
\clearpage\end{figure}
\clearpage\begin{figure}[!h]
	\centering
	\subfloat[]{
		\includegraphics[width=0.46\linewidth]{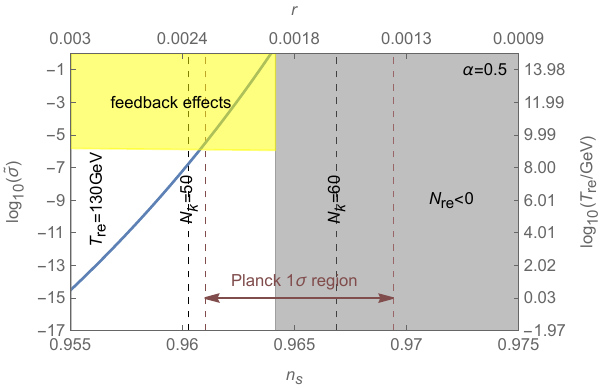}}
	\quad
	\subfloat[]{\includegraphics[width=0.46\linewidth]{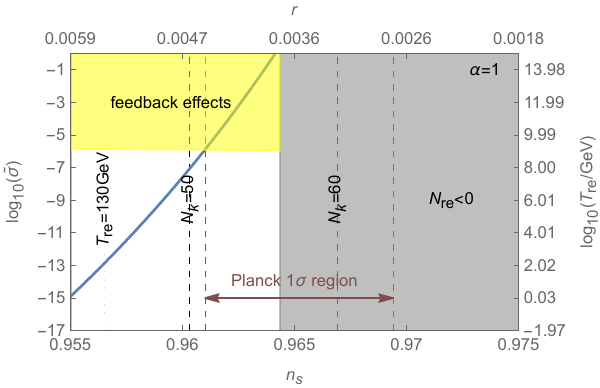}}
	\quad
	\subfloat[]{\includegraphics[width=0.46\linewidth]{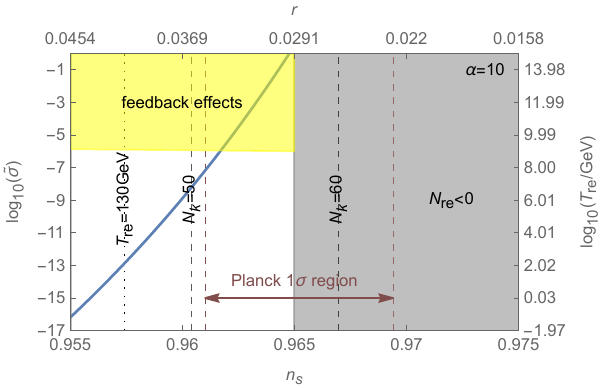}}
	\quad
	\subfloat[]{\includegraphics[width=0.46\linewidth]{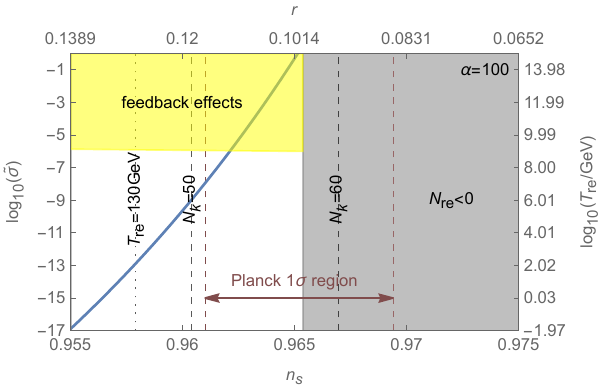}}
	\caption{Dependence of the axion-like coupling $\tilde{\upalphatosigma}$ on $n_s$ in the $\alpha$-T model for different choices of $\alpha$. The notations are the same with Fig.~\ref{MHI g 1 a}.}
	\label{alpha T a}
\clearpage\end{figure}
\clearpage
\end{landscape}

\end{appendix}

\bibliographystyle{JHEP}
\bibliography{main_Mad_2018}{}

\end{document}